\documentclass[epj]{svjour}
\usepackage{placeins}
\usepackage[titletoc,title]{appendix}
\usepackage{authblk}
\usepackage{hyperref}
\usepackage{units}
\usepackage[centertags,intlimits]{amsmath}
\usepackage{amssymb,amscd}
\usepackage{array}
\usepackage{multirow}
\usepackage{xspace}
\usepackage{upgreek}
\usepackage{color}
\usepackage{calc}
\usepackage{mcite}
\usepackage{cite}
\usepackage{graphicx}
\usepackage{subcaption}
\usepackage[export]{adjustbox}

\newcommand{\VERYBROADFIGWIDTH}{1.0\linewidth}
\newcommand{\BROADFIGWIDTH}{1.0\linewidth}

\newcommand{\NARROWFIGWIDTH}{0.42\linewidth}
\newcommand{\VERYNARROWFIGWIDTH}{0.5\linewidth}

\title{Deeply Learning Deep Inelastic Scattering Kinematics}

\author{ Markus Diefenthaler\inst{1}, Abdullah Farhat\inst{2},  Andrii Verbytskyi\inst{3} and Yuesheng Xu\inst{2}}
\authorrunning{M.~Diefenthaler, A.~Farhat, A.~Verbytskyi  and Y.~Xu}
\titlerunning{Deeply Learning Deep Inelastic Scattering Kinematics}
\institute{
Thomas Jefferson National Accelerator Facility, Newport News, VA 23606, USA
\and
Department of Mathematics \& Statistics, Old Dominion University, Norfolk, VA 23529, USA
\and 
Max-Planck-Institut f\"{u}r Physik, D-80805 Munich, Germany
}

\date{Received: date / Revised version: date}
\abstract{
We study the use of deep learning techniques to reconstruct the kinematics of the neutral current deep inelastic scattering (DIS) process in electron-proton collisions. In particular, we use simulated data from the ZEUS experiment at the HERA accelerator facility, 
and train deep neural networks to reconstruct the kinematic variables $Q^2$ and $x$. 
Our approach is based on the information used in the classical construction methods, 
the measurements of the scattered lepton, and the hadronic final state in the detector, 
but is enhanced through correlations and patterns revealed with the simulated data sets. 
We show that, with the appropriate selection of a training set, 
 the neural networks sufficiently surpass all classical reconstruction methods on most of the kinematic range considered. 
Rapid access to large samples of simulated data and the ability of neural networks to effectively extract information from large data sets, 
both suggest that deep learning techniques to reconstruct DIS kinematics can serve as a
 rigorous method to combine and outperform the classical reconstruction methods.

 % end of PACS codes
}
\newcommand{\epjcbreak}[1]{\\#1}
\newcommand{\draftbreak}[1]{}
\newcommand{\arxivbreak}[1]{}
\newcommand{\epjconly}[1]{#1}
\newcommand{\draftonly}[1]{}
\newcommand{\arxivonly}[1]{}

\newcommand{\TABQxbins}{
\begin{table}
\centering
\begin{tabular}{| c | c | c |}
\hline
Bin & $Q^2$ ($\GeV^2$) & $x$ \\ \hline\hline
1 & 120 - 160 & 0.0024 - 0.010 \\
2 & 160 - 320 & 0.0024 - 0.010 \\
3 & 320 - 640 & 0.01 - 0.05 \\
4 & 640 - 1280 & 0.01 - 0.05 \\
5 & 1280 - 2560 & 0.025 - 0.150 \\
6 & 2560 - 5120 & 0.05 - 0.25 \\
7 & 5120 - 10240 & 0.06 - 0.40 \\
8 & 10240 - 20480 & 0.10 - 0.60 \\ \hline
\end{tabular}
\caption{Kinematic bins in $x$ and $Q^2$, see also Fig.~\ref{fig:xq2original}.}
\label{table:bins}
\end{table}
}

\makeatletter
\def\ifabsgreater #1#2{\ifpdfabsnum 
    \dimexpr#1pt>\dimexpr#2pt\relax
    \expandafter\@firstoftwo\else\expandafter\@secondoftwo\fi }
\makeatother

\newcommand{\FIGdis}[2]{
\begin{figure}[htbp]\centering
\includegraphics[width=\BROADFIGWIDTH]{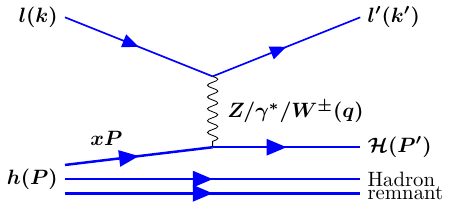}
\caption{#1}
\label{#2}
\end{figure}
}
\newcommand{\FIGdistwo}[2]{
\begin{figure}[htbp]\centering
\begin{subfigure}[b]{\VERYNARROWFIGWIDTH}\adjincludegraphics[width=\VERYNARROWFIGWIDTH]{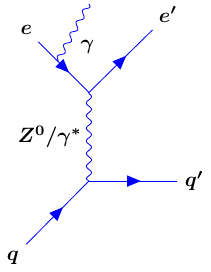}\caption{}\end{subfigure}
\begin{subfigure}[b]{\VERYNARROWFIGWIDTH}\adjincludegraphics[width=\VERYNARROWFIGWIDTH]{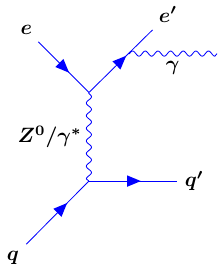}\caption{}\end{subfigure}
\begin{subfigure}[b]{\VERYNARROWFIGWIDTH}\adjincludegraphics[width=\VERYNARROWFIGWIDTH]{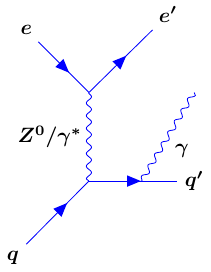}\caption{}\end{subfigure}
\begin{subfigure}[b]{\VERYNARROWFIGWIDTH}\adjincludegraphics[width=\VERYNARROWFIGWIDTH]{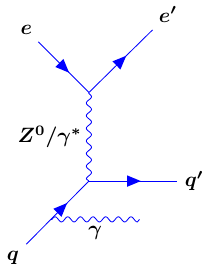}\caption{}\end{subfigure}\\
\begin{subfigure}[b]{\VERYNARROWFIGWIDTH}\adjincludegraphics[width=\VERYNARROWFIGWIDTH]{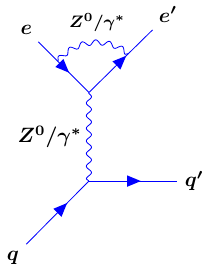}\caption{}\end{subfigure}
\begin{subfigure}[b]{\VERYNARROWFIGWIDTH}\adjincludegraphics[width=\VERYNARROWFIGWIDTH]{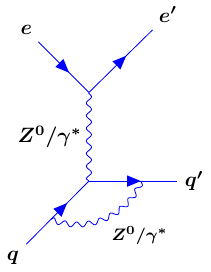}\caption{}\end{subfigure}
\begin{subfigure}[b]{\VERYNARROWFIGWIDTH}\adjincludegraphics[width=\VERYNARROWFIGWIDTH]{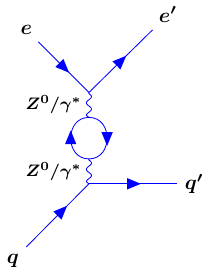}\caption{}\end{subfigure}
\begin{subfigure}[b]{\VERYNARROWFIGWIDTH}\adjincludegraphics[width=\VERYNARROWFIGWIDTH]{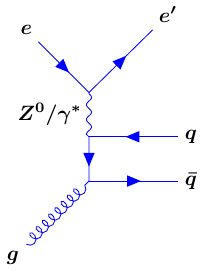}\caption{}\end{subfigure}\\
\begin{subfigure}[b]{\VERYNARROWFIGWIDTH}\adjincludegraphics[width=\VERYNARROWFIGWIDTH]{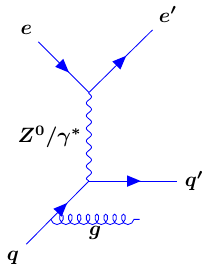}\caption{}\end{subfigure}
\begin{subfigure}[b]{\VERYNARROWFIGWIDTH}\adjincludegraphics[width=\VERYNARROWFIGWIDTH]{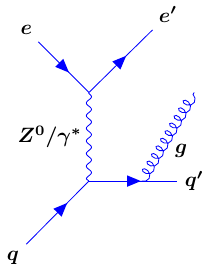}\caption{}\end{subfigure}
\begin{subfigure}[b]{\VERYNARROWFIGWIDTH}\adjincludegraphics[width=\VERYNARROWFIGWIDTH]{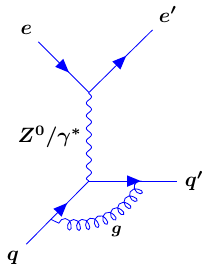}\caption{}\end{subfigure}\\
\caption{#1}
\label{#2}
\end{figure}
}

\newcommand{\FIGdistwoepjc}[2]{
\begin{figure}[htbp]\centering
\begin{subfigure}[b]{0.3\linewidth}\adjincludegraphics[width=1.0\linewidth]{Figures/DIS1a-figure0.pdf}\caption{}\end{subfigure}
\begin{subfigure}[b]{0.3\linewidth}\adjincludegraphics[width=1.0\linewidth]{Figures/DIS2a-figure0.pdf}\caption{}\end{subfigure}
\begin{subfigure}[b]{0.3\linewidth}\adjincludegraphics[width=1.0\linewidth]{Figures/DIS3a-figure0.pdf}\caption{}\end{subfigure}\\
\begin{subfigure}[b]{0.3\linewidth}\adjincludegraphics[width=1.0\linewidth]{Figures/DIS4a-figure0.pdf}\caption{}\end{subfigure}
\begin{subfigure}[b]{0.3\linewidth}\adjincludegraphics[width=1.0\linewidth]{Figures/DIS5a-figure0.pdf}\caption{}\end{subfigure}
\begin{subfigure}[b]{0.3\linewidth}\adjincludegraphics[width=1.0\linewidth]{Figures/DIS6a-figure0.pdf}\caption{}\end{subfigure}\\
\begin{subfigure}[b]{0.3\linewidth}\adjincludegraphics[width=1.0\linewidth]{Figures/DIS7a-figure0.pdf}\caption{}\end{subfigure}
\begin{subfigure}[b]{0.3\linewidth}\adjincludegraphics[width=1.0\linewidth]{Figures/DIS8a-figure0.pdf}\caption{}\end{subfigure}
\begin{subfigure}[b]{0.3\linewidth}\adjincludegraphics[width=1.0\linewidth]{Figures/DIS9a-figure0.pdf}\caption{}\end{subfigure}\\
\begin{subfigure}[b]{0.3\linewidth}\adjincludegraphics[width=1.0\linewidth]{Figures/DIS10a-figure0.pdf}\caption{}\end{subfigure}
\begin{subfigure}[b]{0.3\linewidth}\adjincludegraphics[width=1.0\linewidth]{Figures/DIS11a-figure0.pdf}\caption{}\end{subfigure}\\
\caption{#1}
\label{#2}
\end{figure}
}

\newcommand{\FIGxhistory}[2]{
\begin{figure}[htbp]\centering
\includegraphics[width=\VERYBROADFIGWIDTH,height=0.25\textheight]{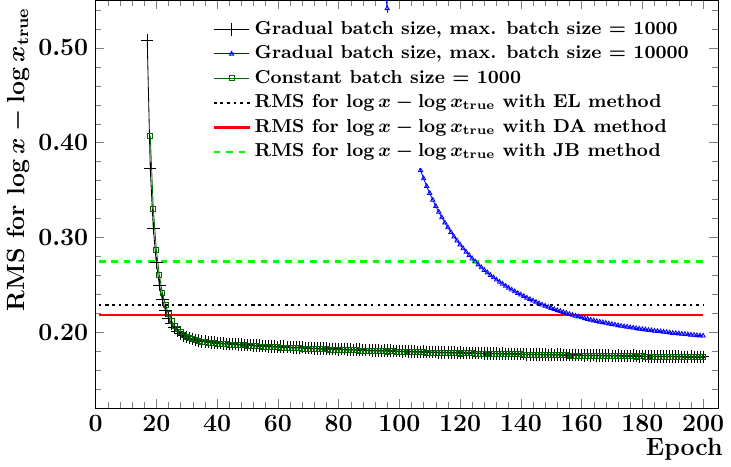}
\caption{#1}
\label{#2}
\end{figure}

}
\newcommand{\FIGtraining}[2]{
\begin{figure}[t]\centering
\includegraphics[width=\VERYBROADFIGWIDTH]{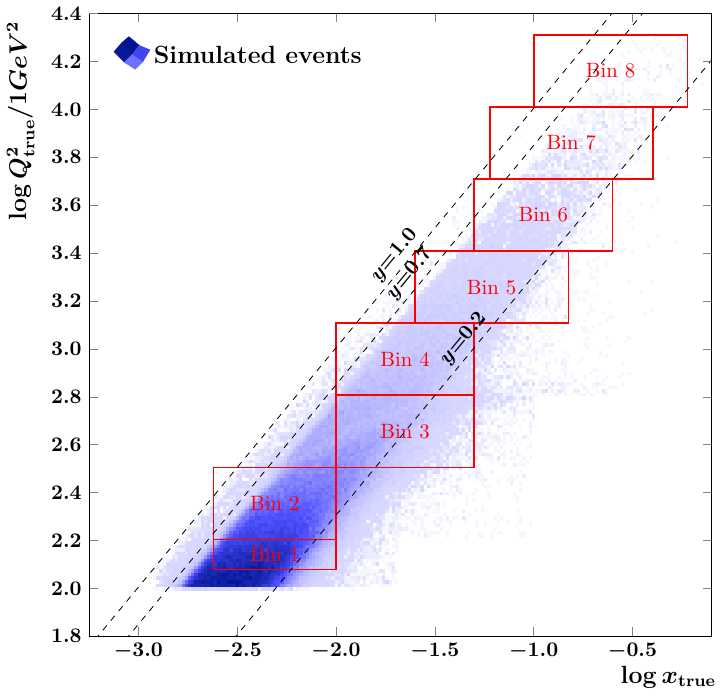}
\caption{#1}
\label{#2}
\end{figure}
}

\newcommand{\FIGqtworesolution}[1]{
\begin{figure}[tbp]\centering
\includegraphics[width=0.5\linewidth,height=0.115\textheight]{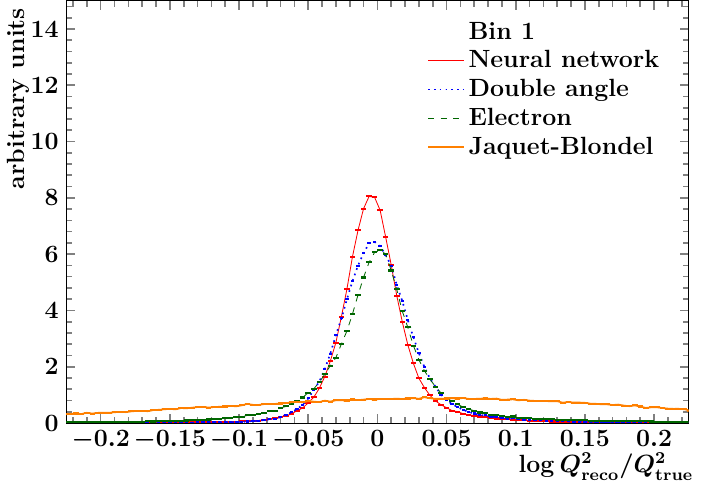}\includegraphics[width=0.5\linewidth,height=0.115\textheight]{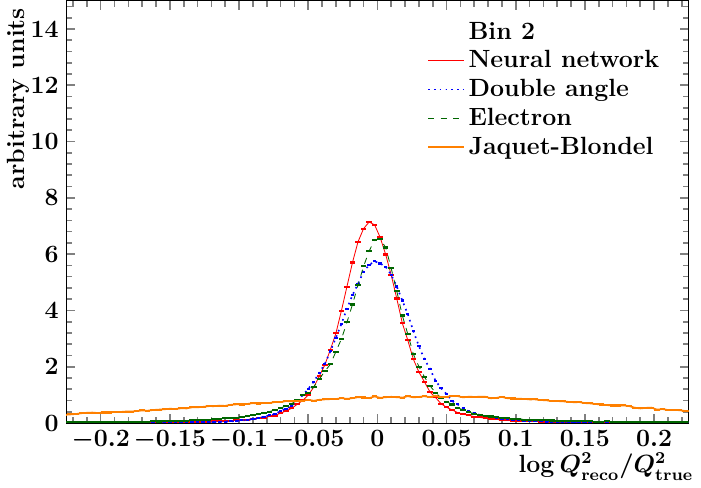}\\
\includegraphics[width=0.5\linewidth,height=0.115\textheight]{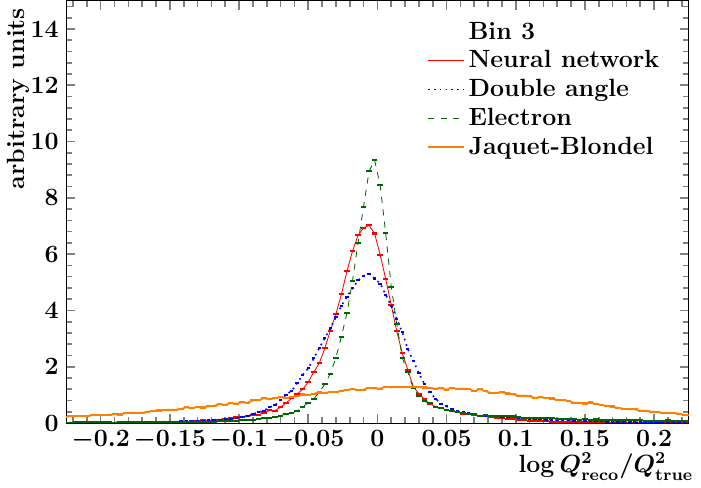}\includegraphics[width=0.5\linewidth,height=0.115\textheight]{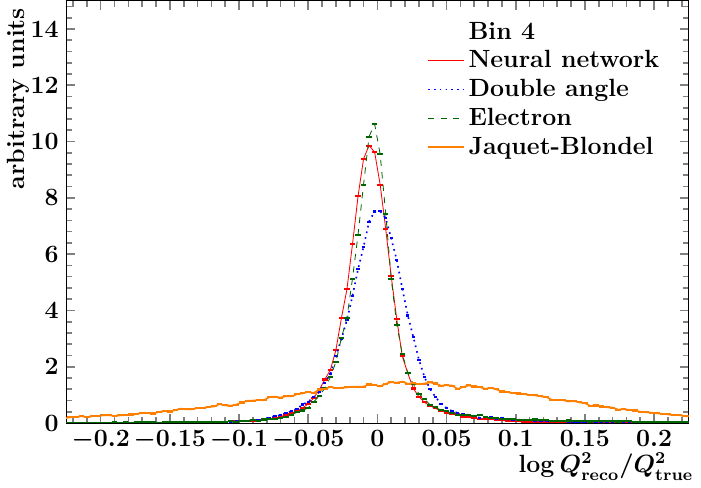}\\
\includegraphics[width=0.5\linewidth,height=0.115\textheight]{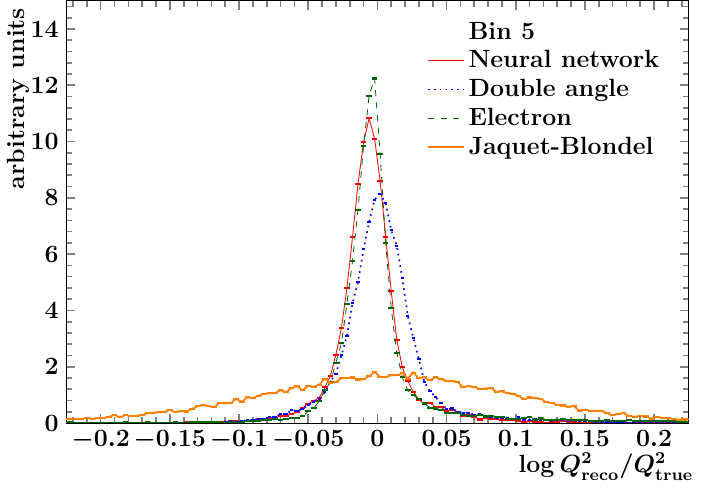}\includegraphics[width=0.5\linewidth,height=0.115\textheight]{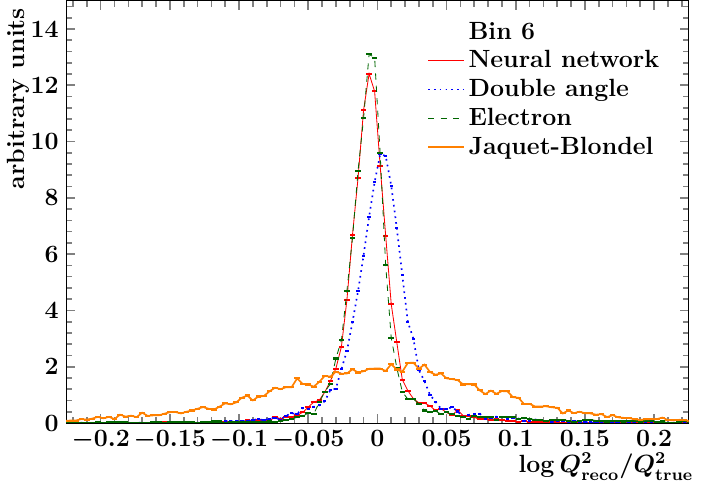}\\
\includegraphics[width=0.5\linewidth,height=0.115\textheight]{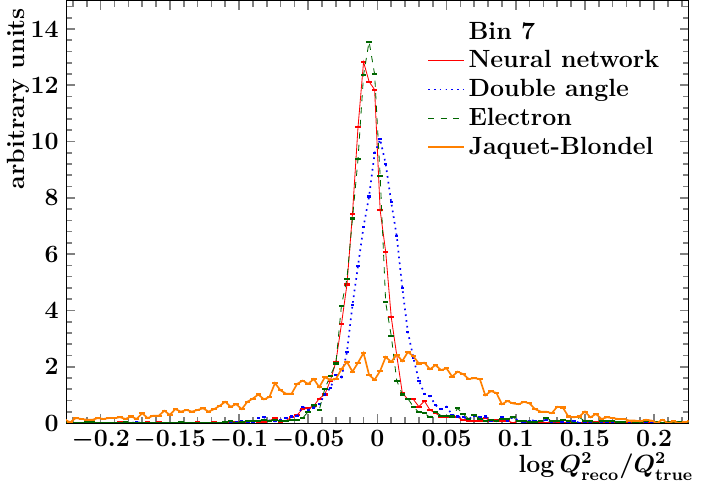}\includegraphics[width=0.5\linewidth,height=0.115\textheight]{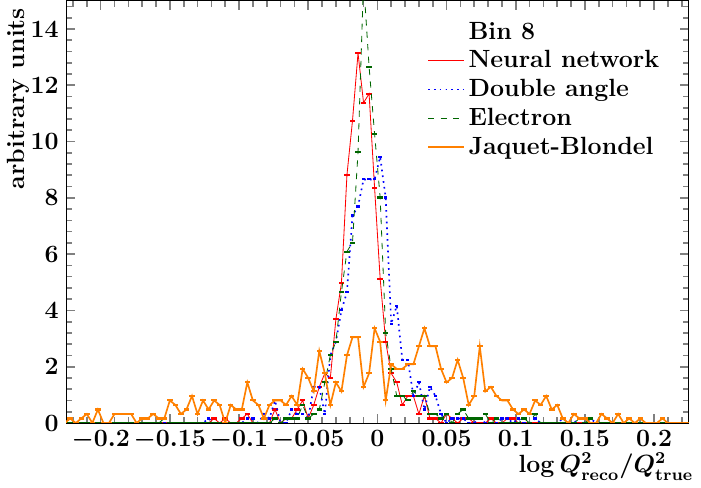}\\
\caption{#1}
\label{fig:qtworesolution}
\end{figure}
}
\newcommand{\FIGxresolution}[1]{
\begin{figure}[tbp]\centering
\includegraphics[width=0.5\linewidth,height=0.115\textheight]{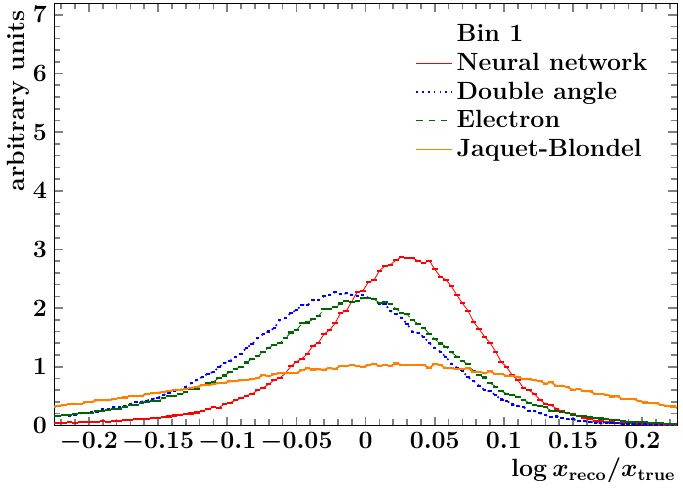}\includegraphics[width=0.5\linewidth,height=0.115\textheight]{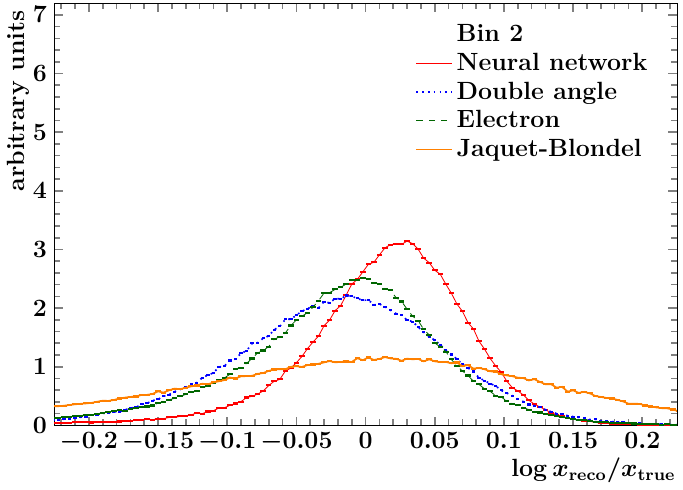}\\
\includegraphics[width=0.5\linewidth,height=0.115\textheight]{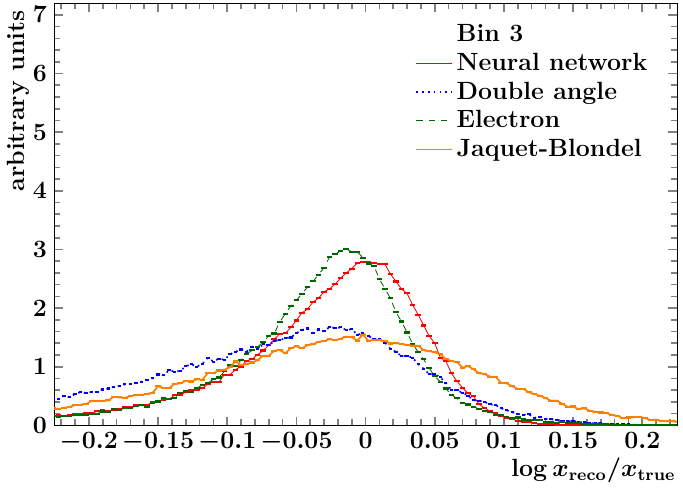}\includegraphics[width=0.5\linewidth,height=0.115\textheight]{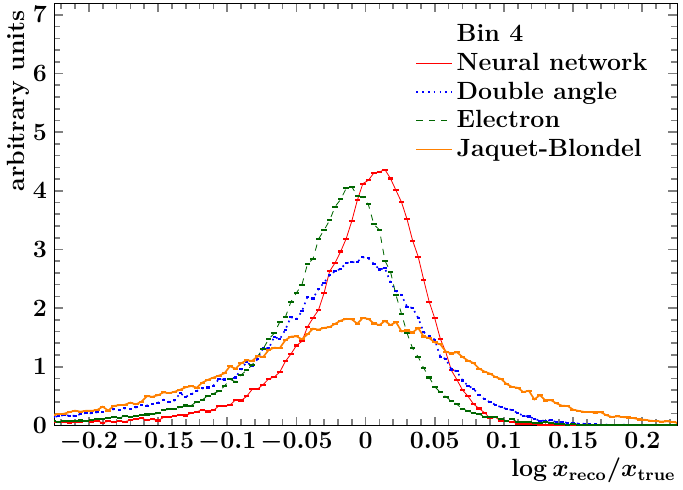}\\
\includegraphics[width=0.5\linewidth,height=0.115\textheight]{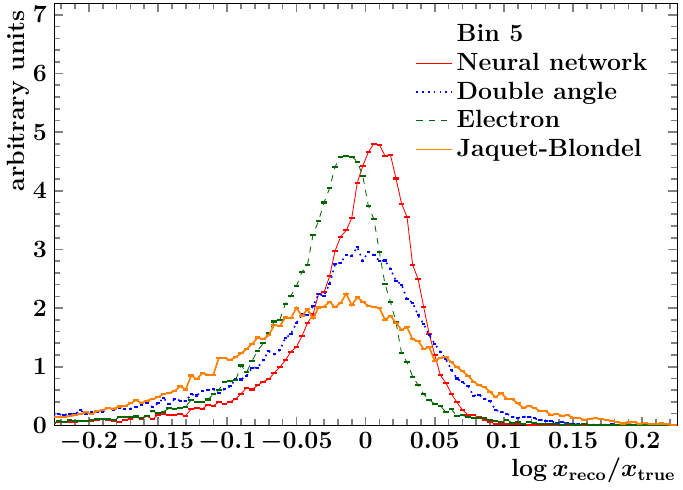}\includegraphics[width=0.5\linewidth,height=0.115\textheight]{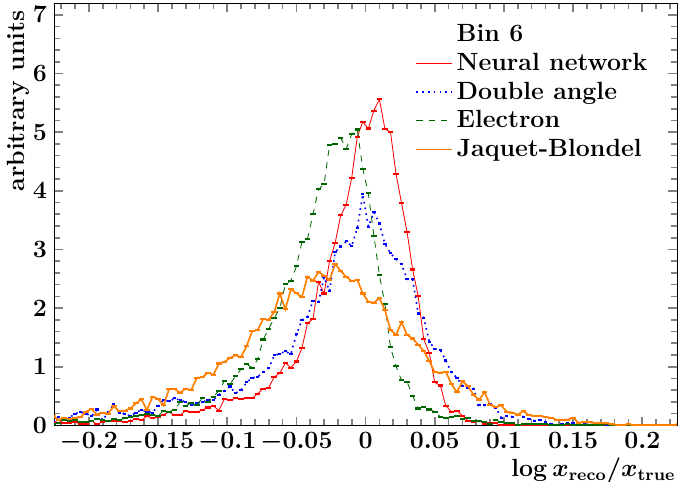}\\
\includegraphics[width=0.5\linewidth,height=0.115\textheight]{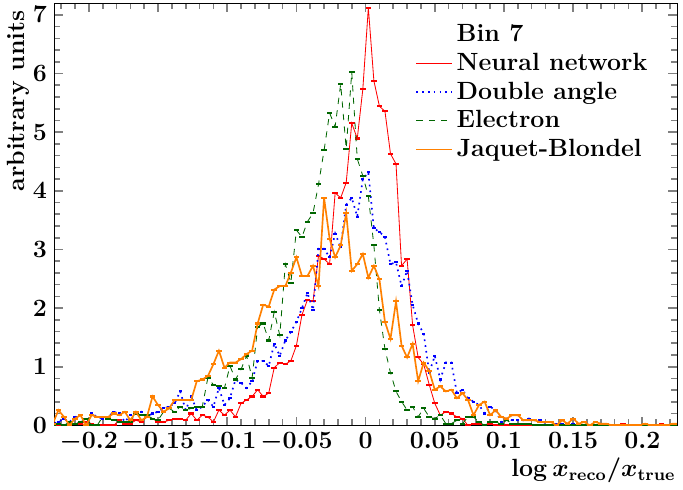}\includegraphics[width=0.5\linewidth,height=0.115\textheight]{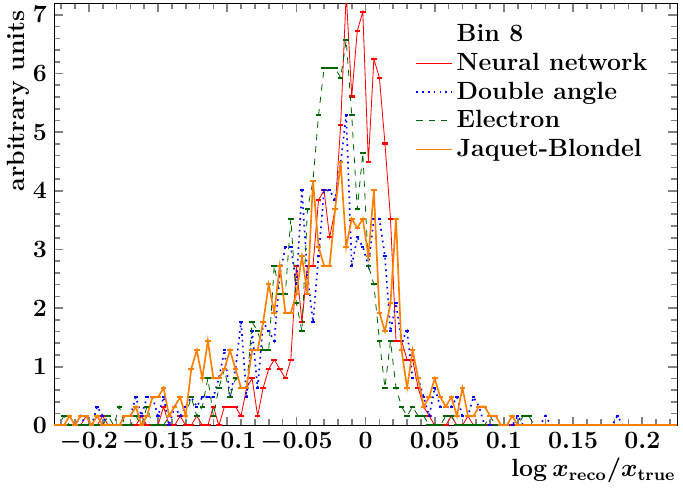}\\
\caption{#1}
\label{fig:xresolution}
\end{figure}
}
\newcommand{\FIGqtwotwod}[1]{
\begin{figure}[htbp]\centering
\includegraphics[width=\NARROWFIGWIDTH]{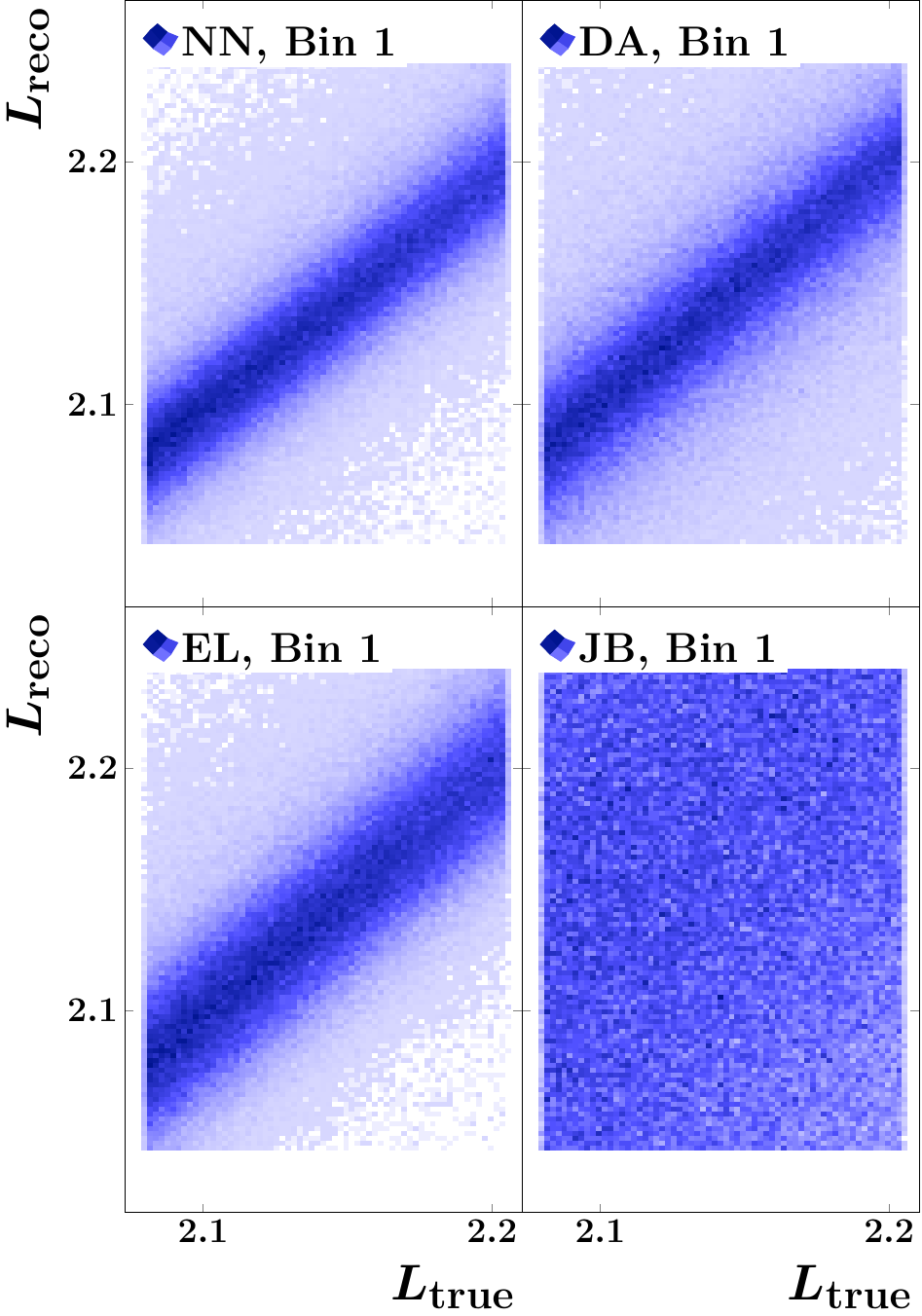}\includegraphics[width=\NARROWFIGWIDTH]{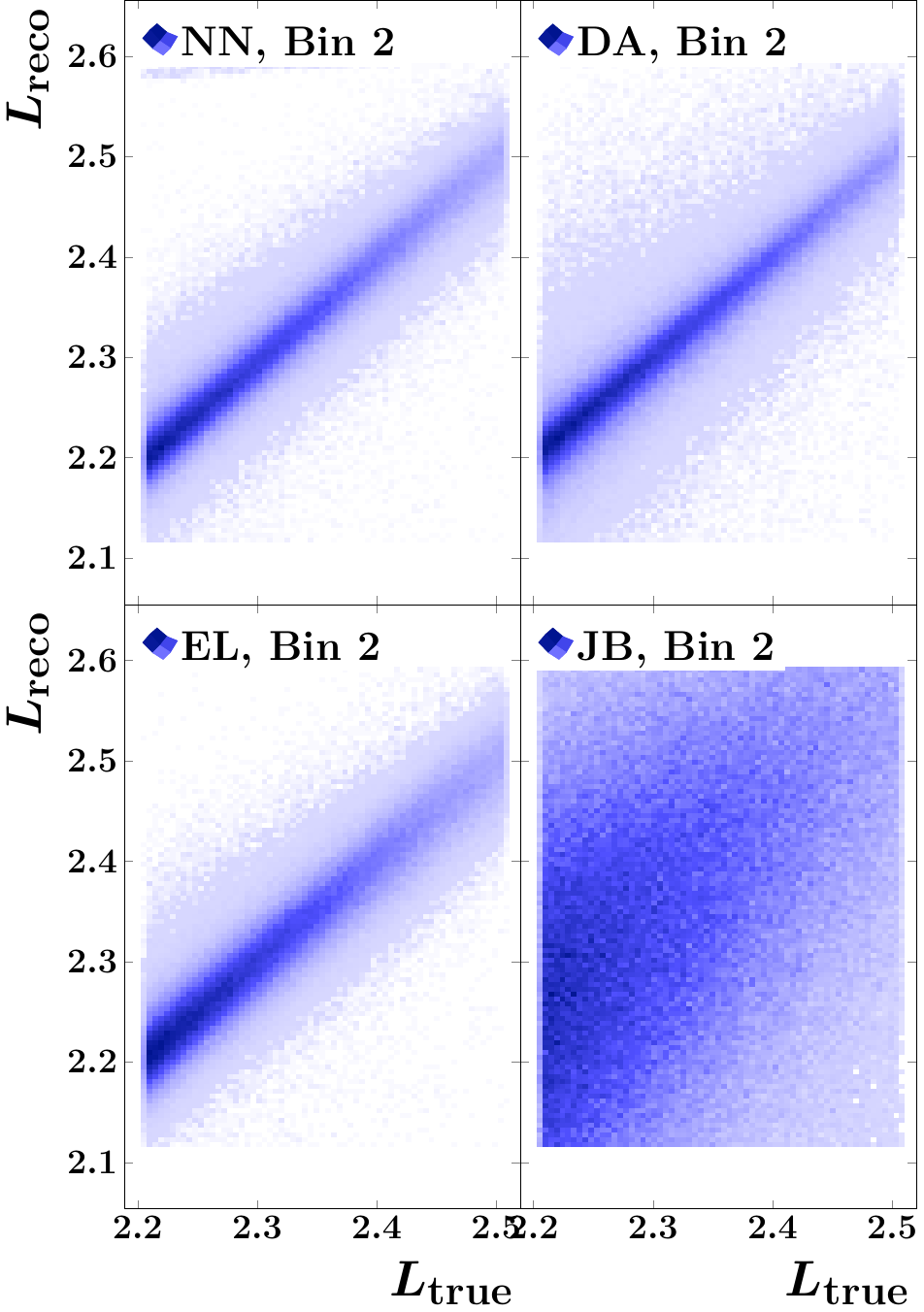}\includegraphics[width=\NARROWFIGWIDTH]{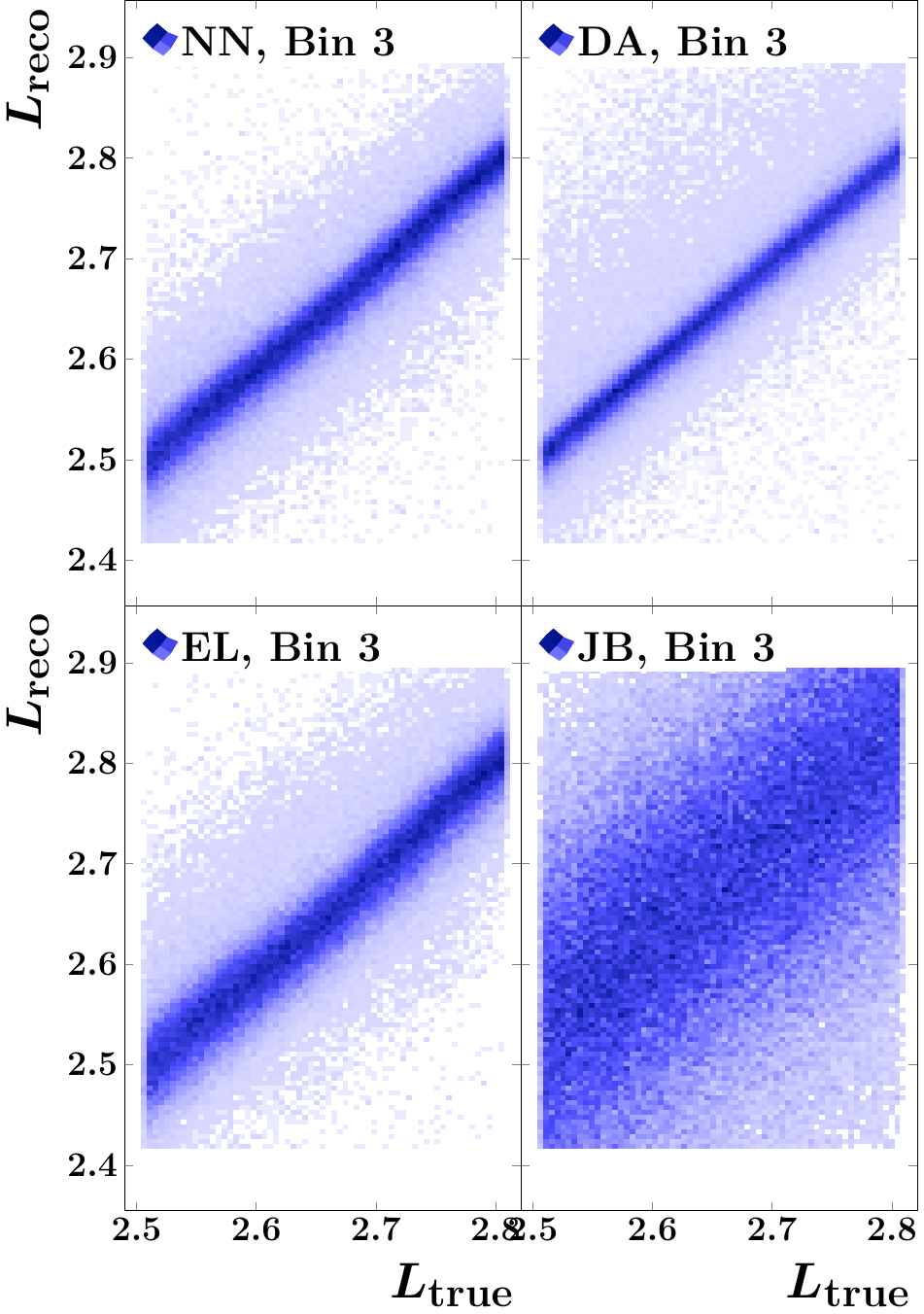}\\
\includegraphics[width=\NARROWFIGWIDTH]{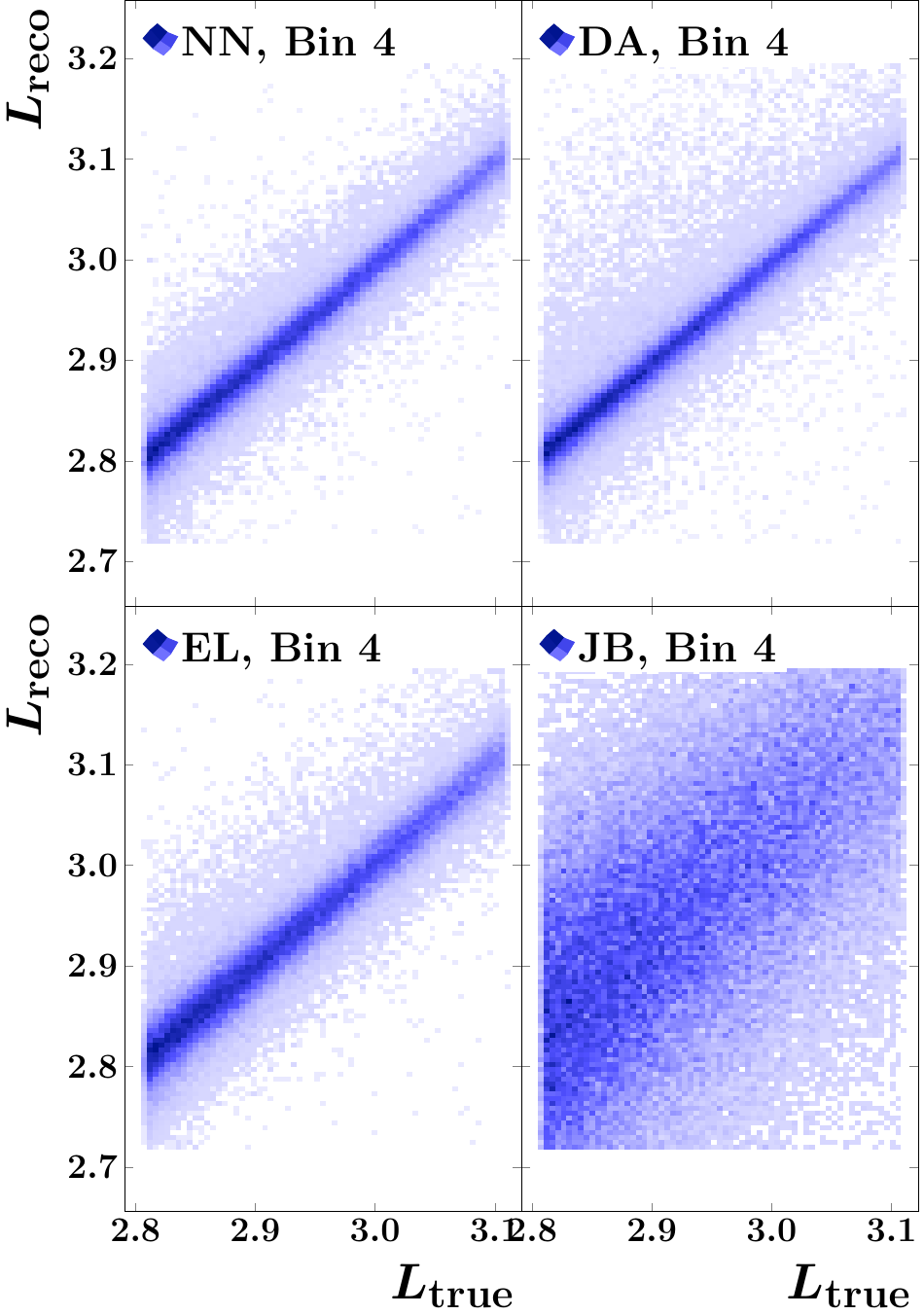}\includegraphics[width=\NARROWFIGWIDTH]{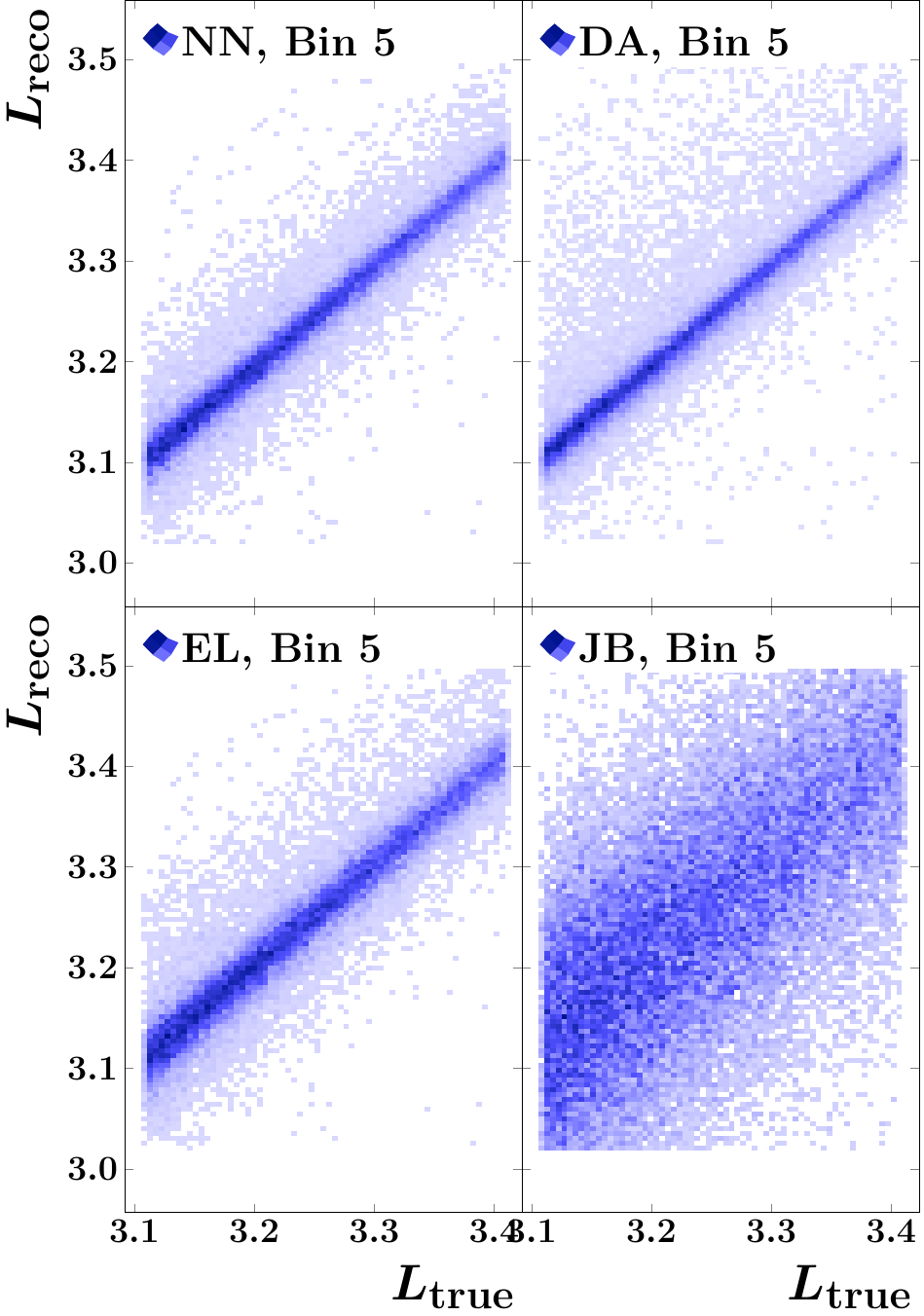}\includegraphics[width=\NARROWFIGWIDTH]{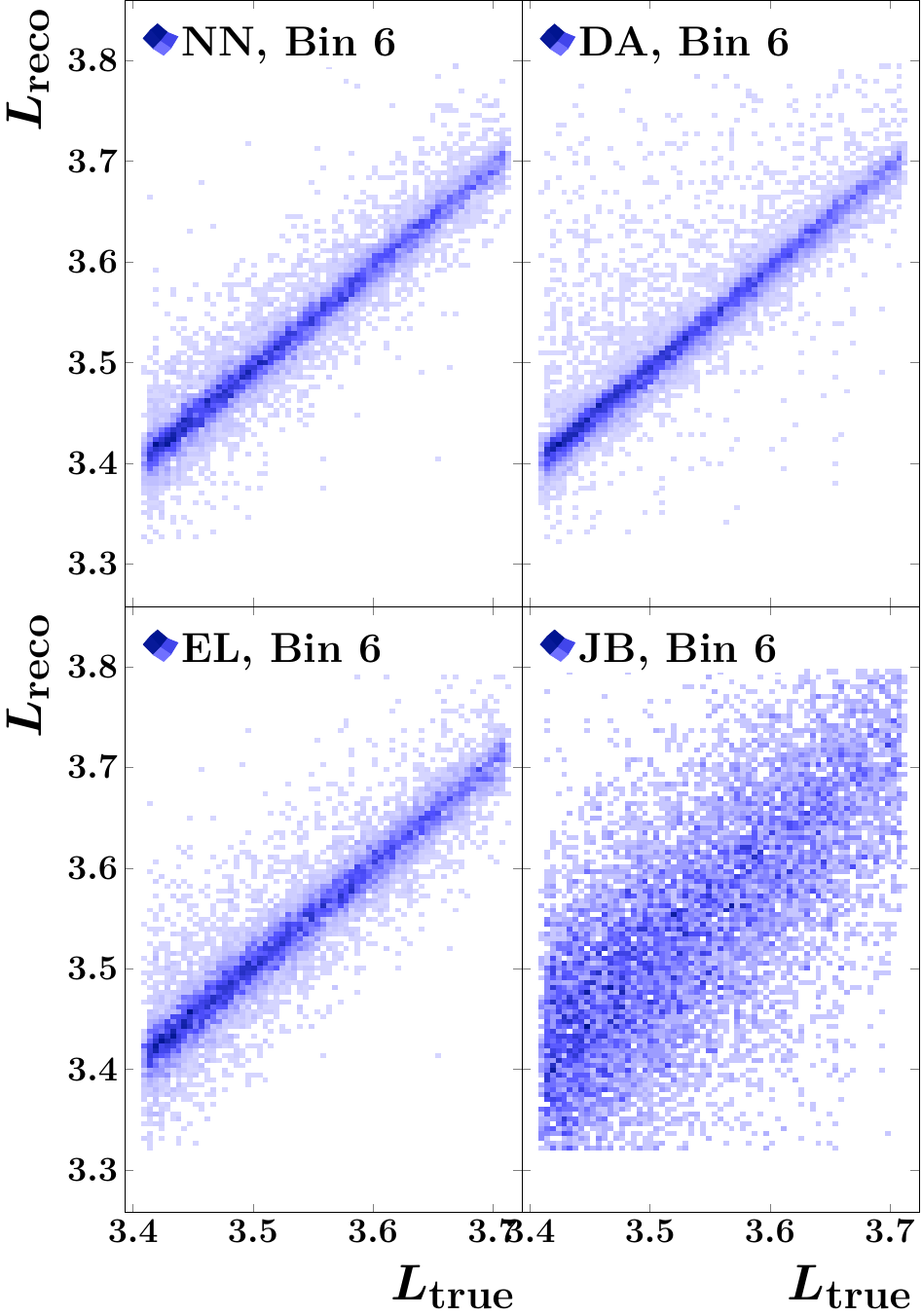}\\
\includegraphics[width=\NARROWFIGWIDTH]{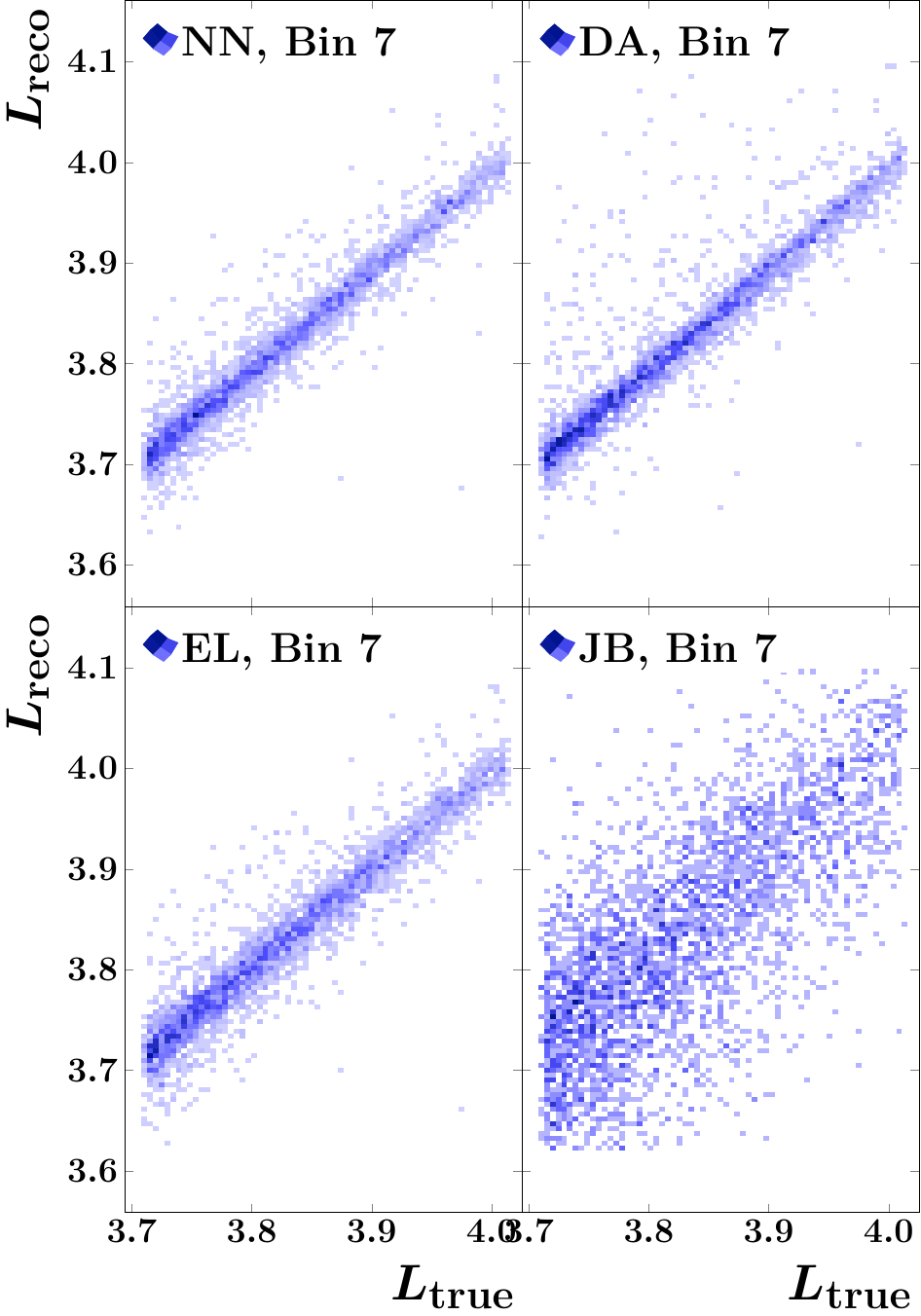}\includegraphics[width=\NARROWFIGWIDTH]{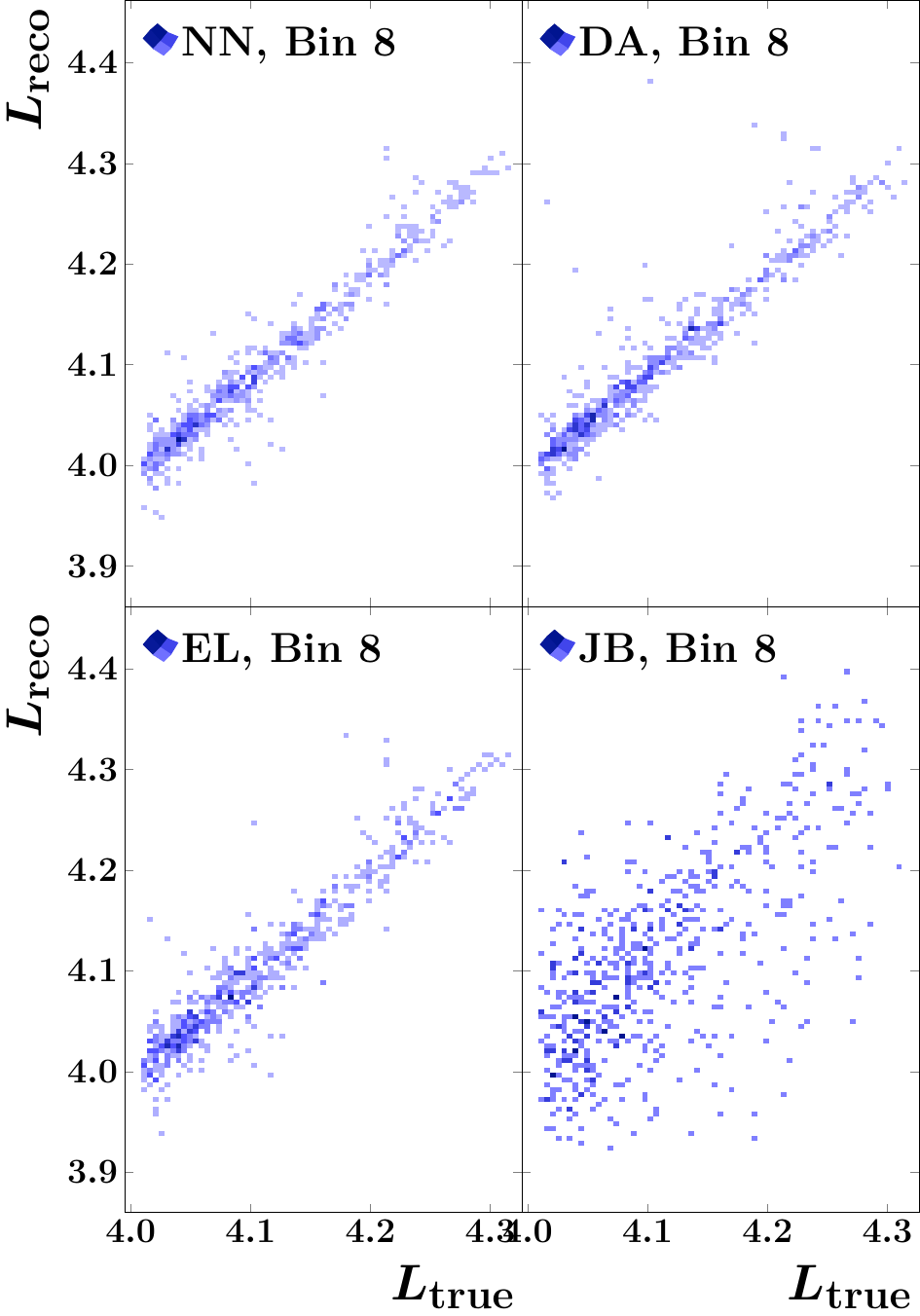}\\
\caption{#1}
\label{fig:qtwotwod}
\end{figure}
}
\newcommand{\FIGqtwotwodepjc}[1]{
\begin{figure}[htbp]\centering
\includegraphics[width=0.5\linewidth,height=0.235\textheight]{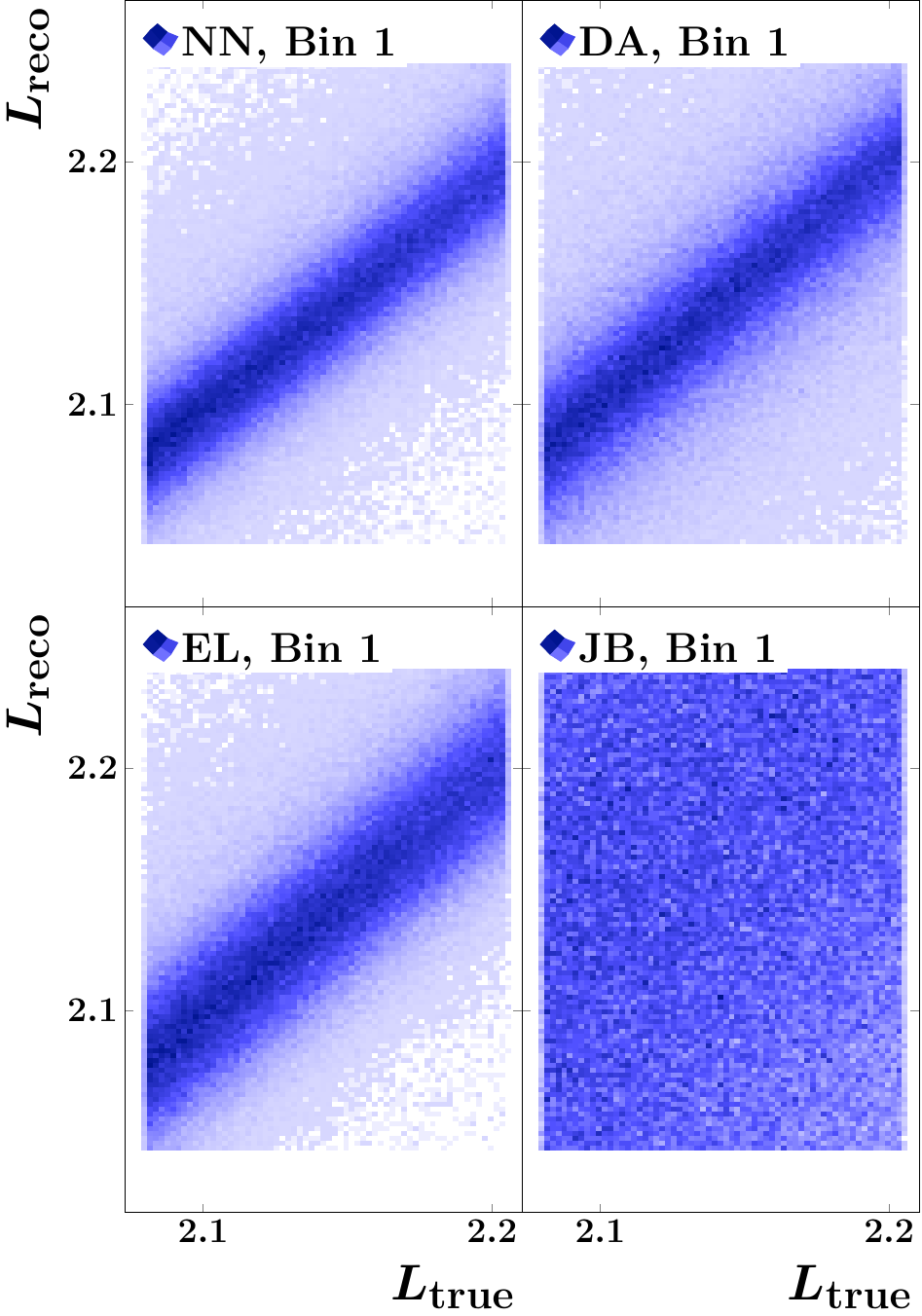}\includegraphics[width=0.5\linewidth,height=0.235\textheight]{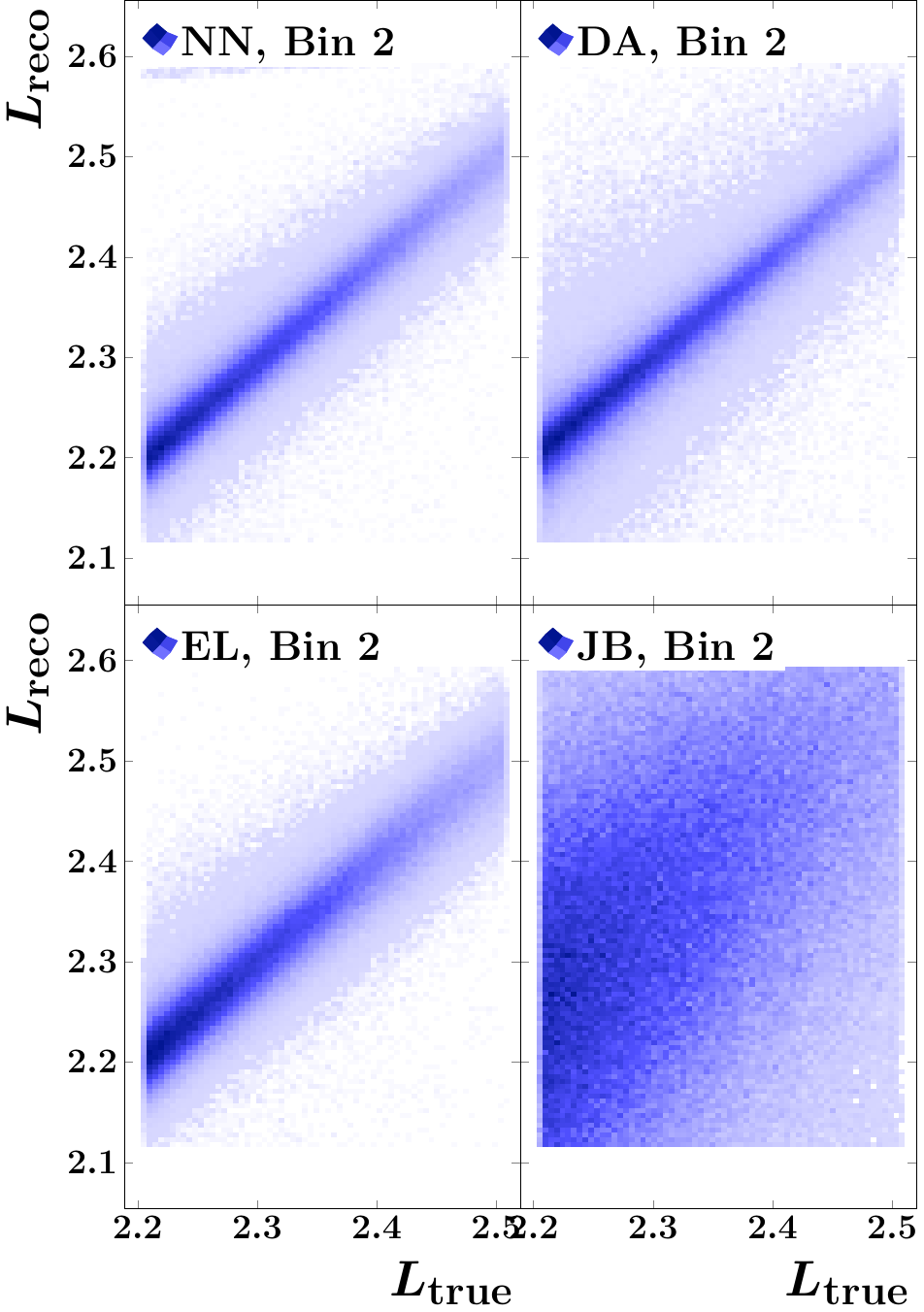}\\
\includegraphics[width=0.5\linewidth,height=0.235\textheight]{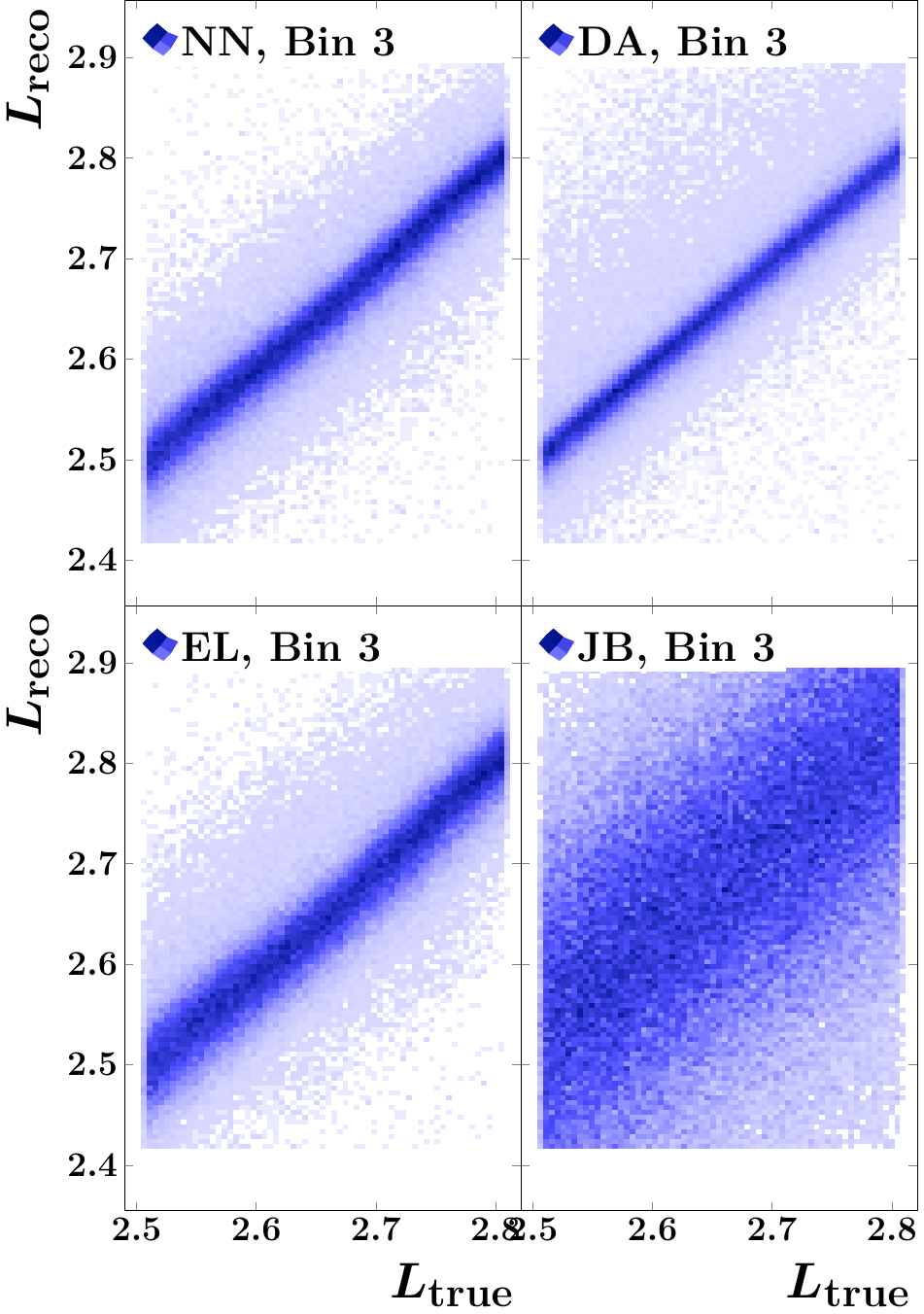}\includegraphics[width=0.5\linewidth,height=0.235\textheight]{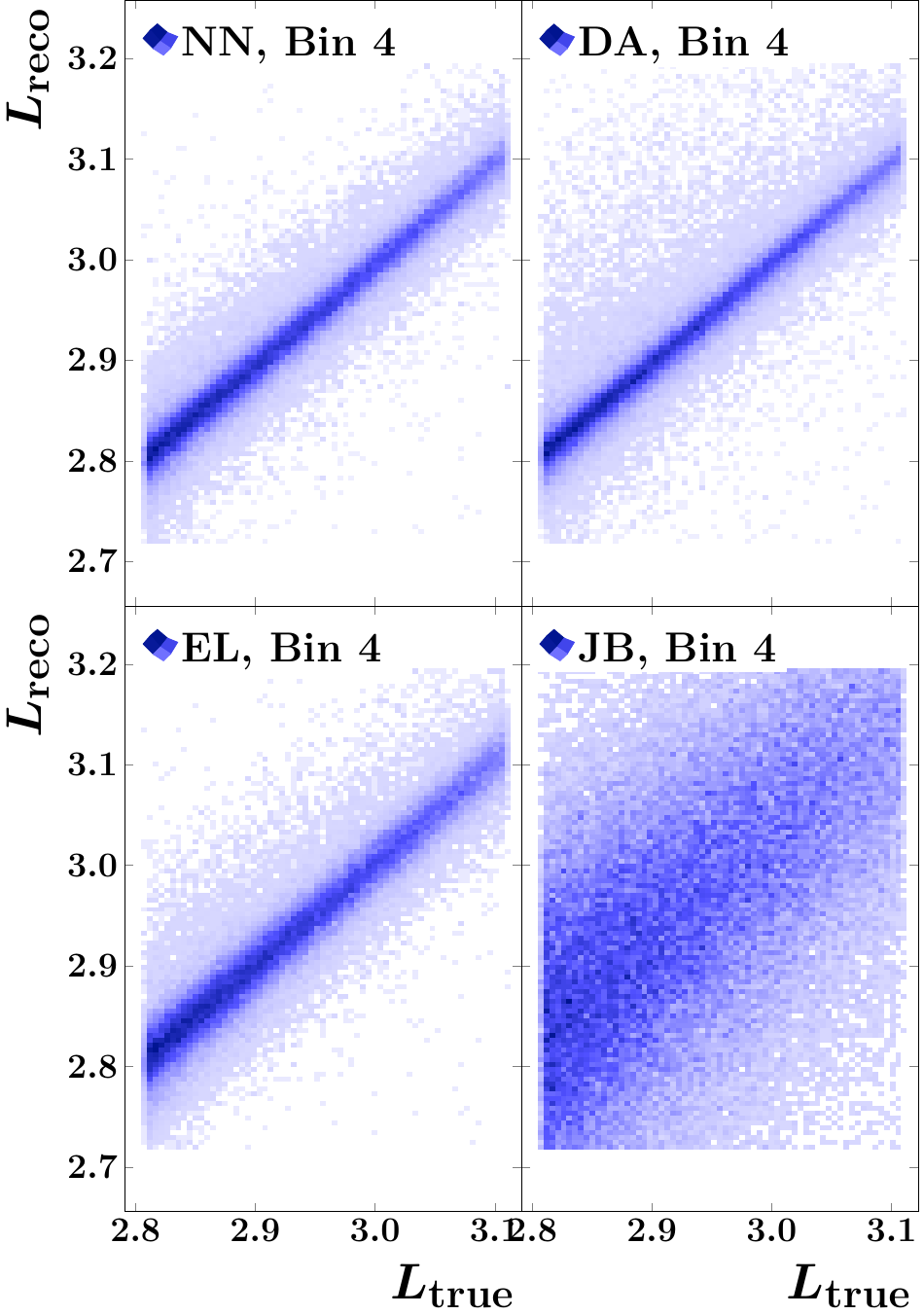}\\
\includegraphics[width=0.5\linewidth,height=0.235\textheight]{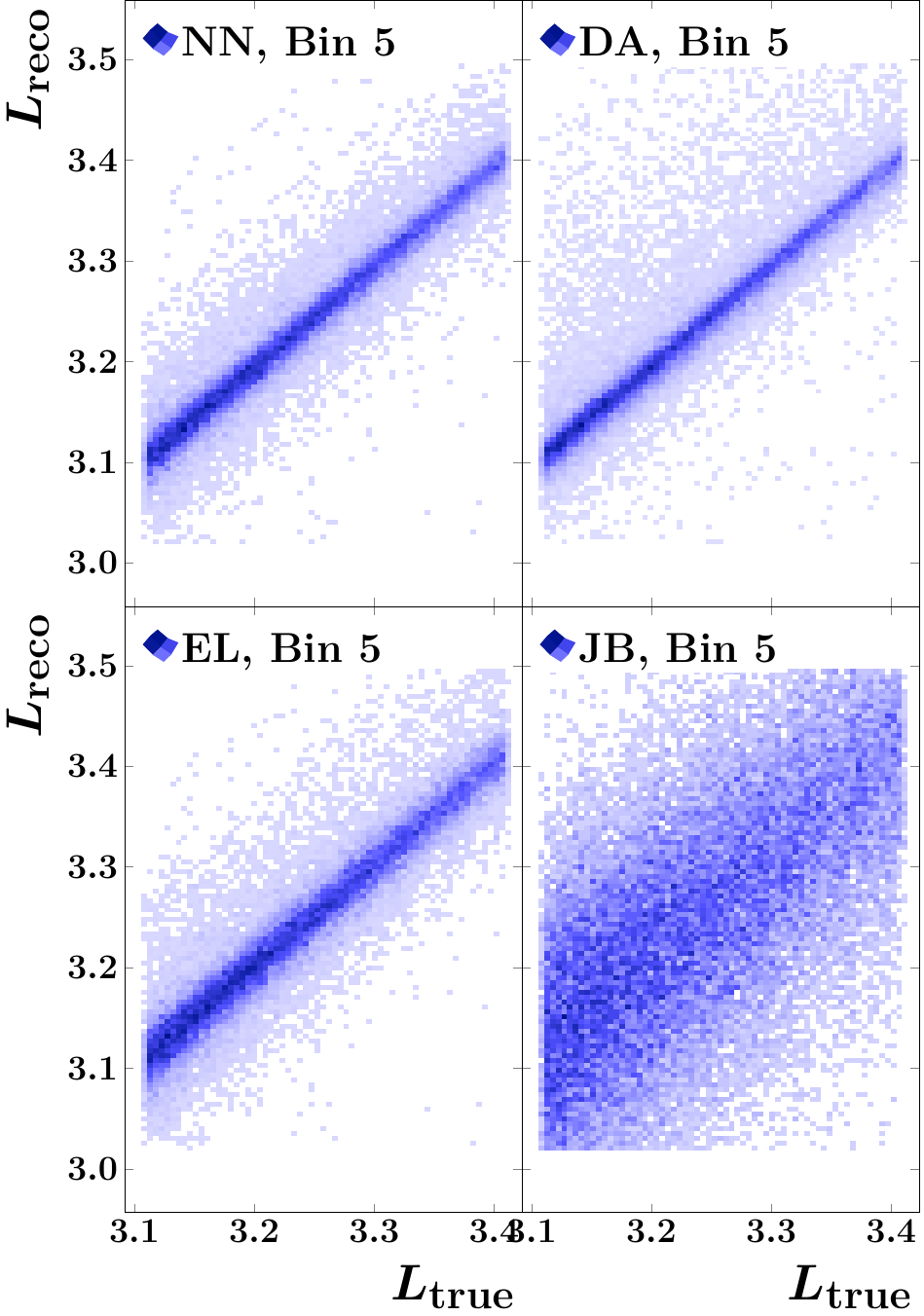}\includegraphics[width=0.5\linewidth,height=0.235\textheight]{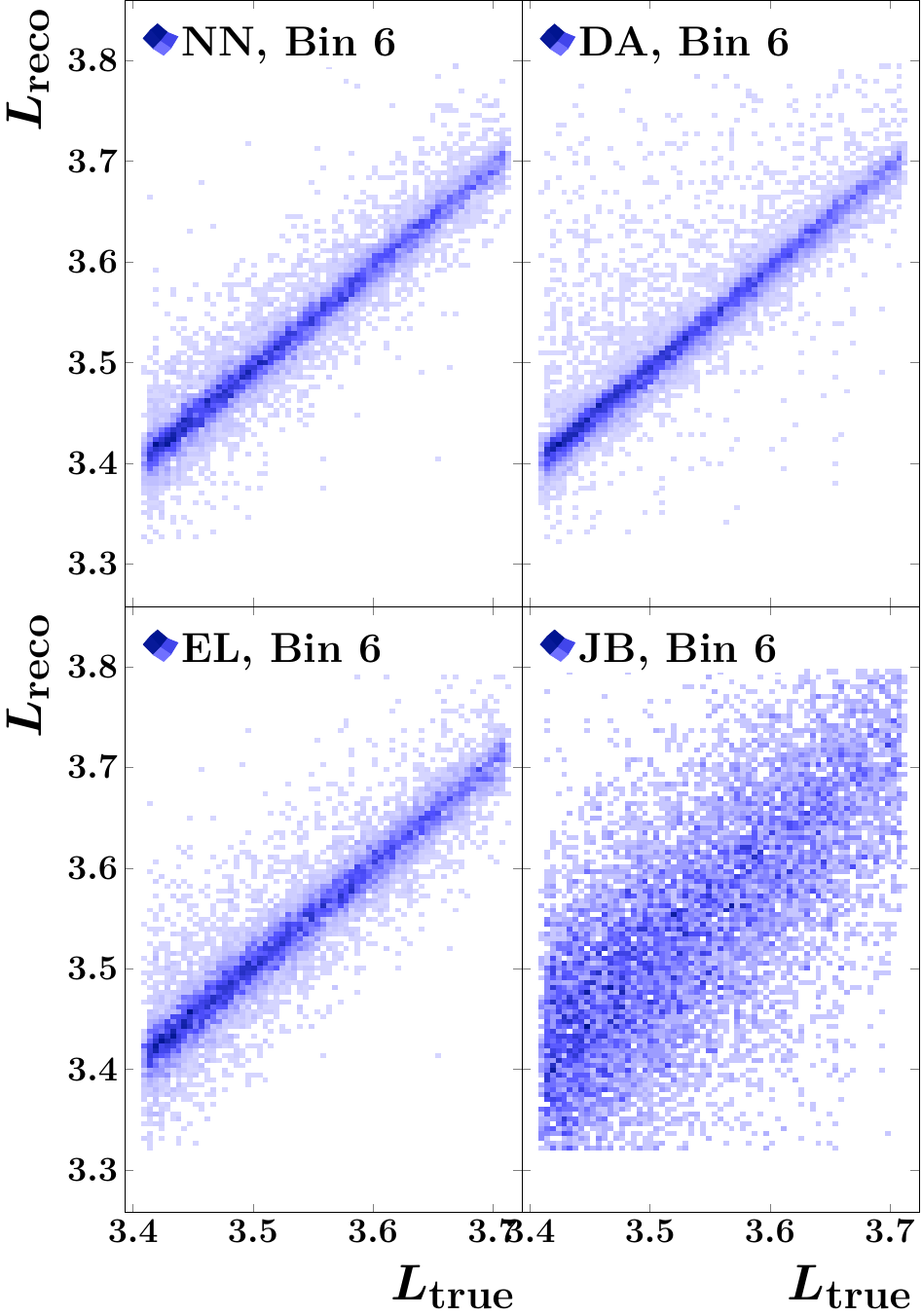}\\
\includegraphics[width=0.5\linewidth,height=0.235\textheight]{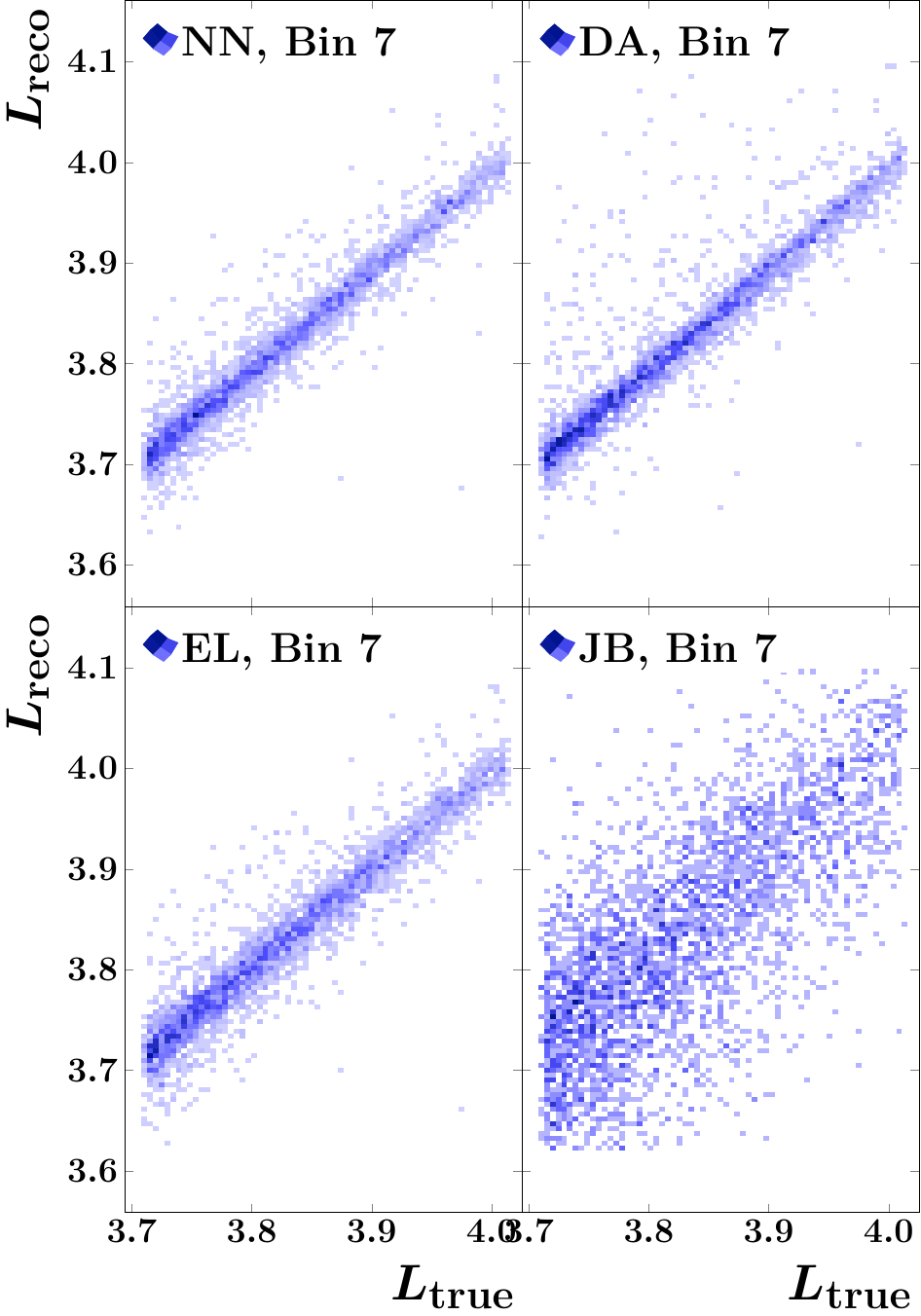}\includegraphics[width=0.5\linewidth,height=0.235\textheight]{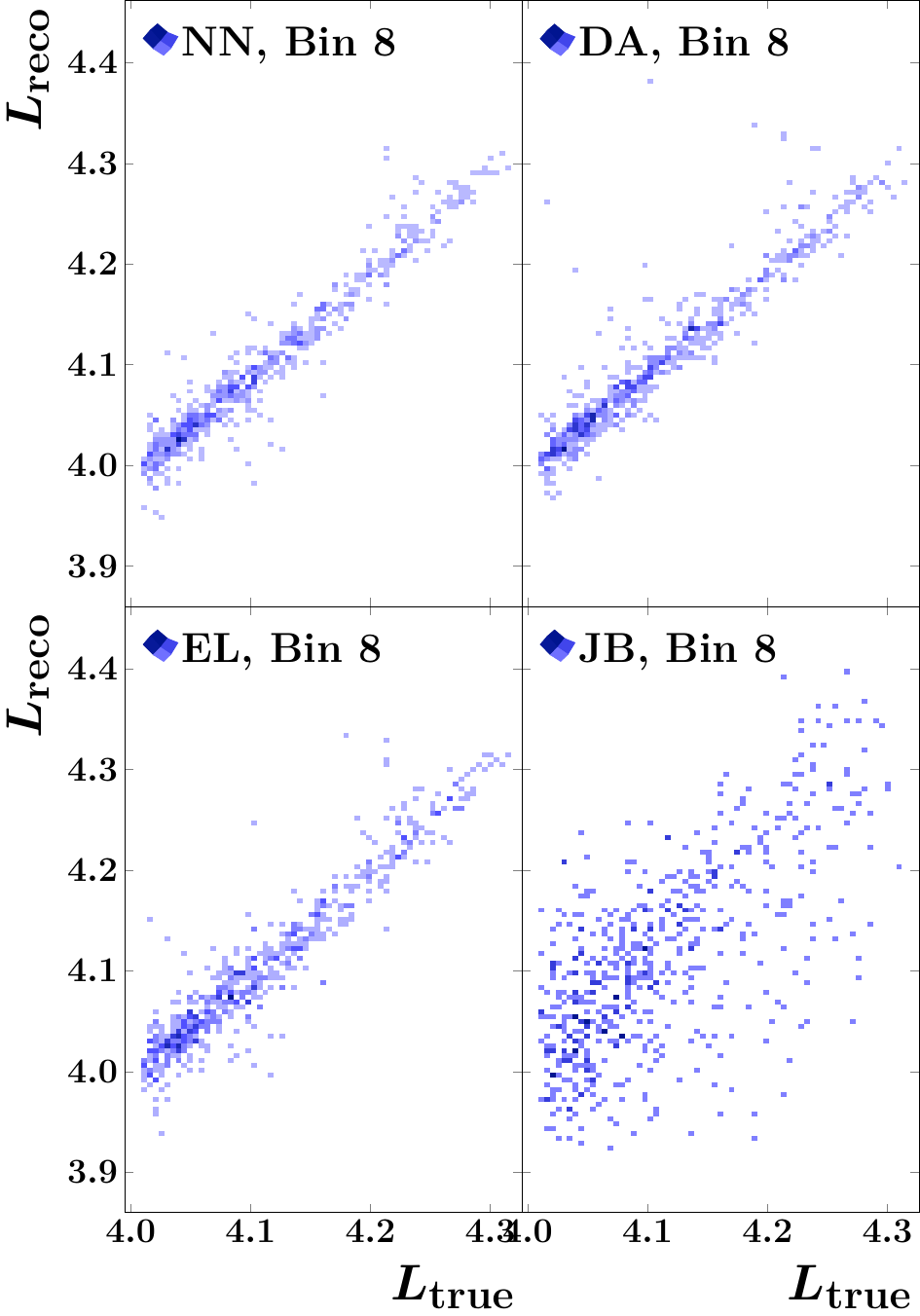}\\
\caption{#1}
\label{fig:qtwotwod}
\end{figure}
}
\newcommand{\FIGxtwod}[1]{
\begin{figure}[htbp]\centering
\includegraphics[width=\NARROWFIGWIDTH]{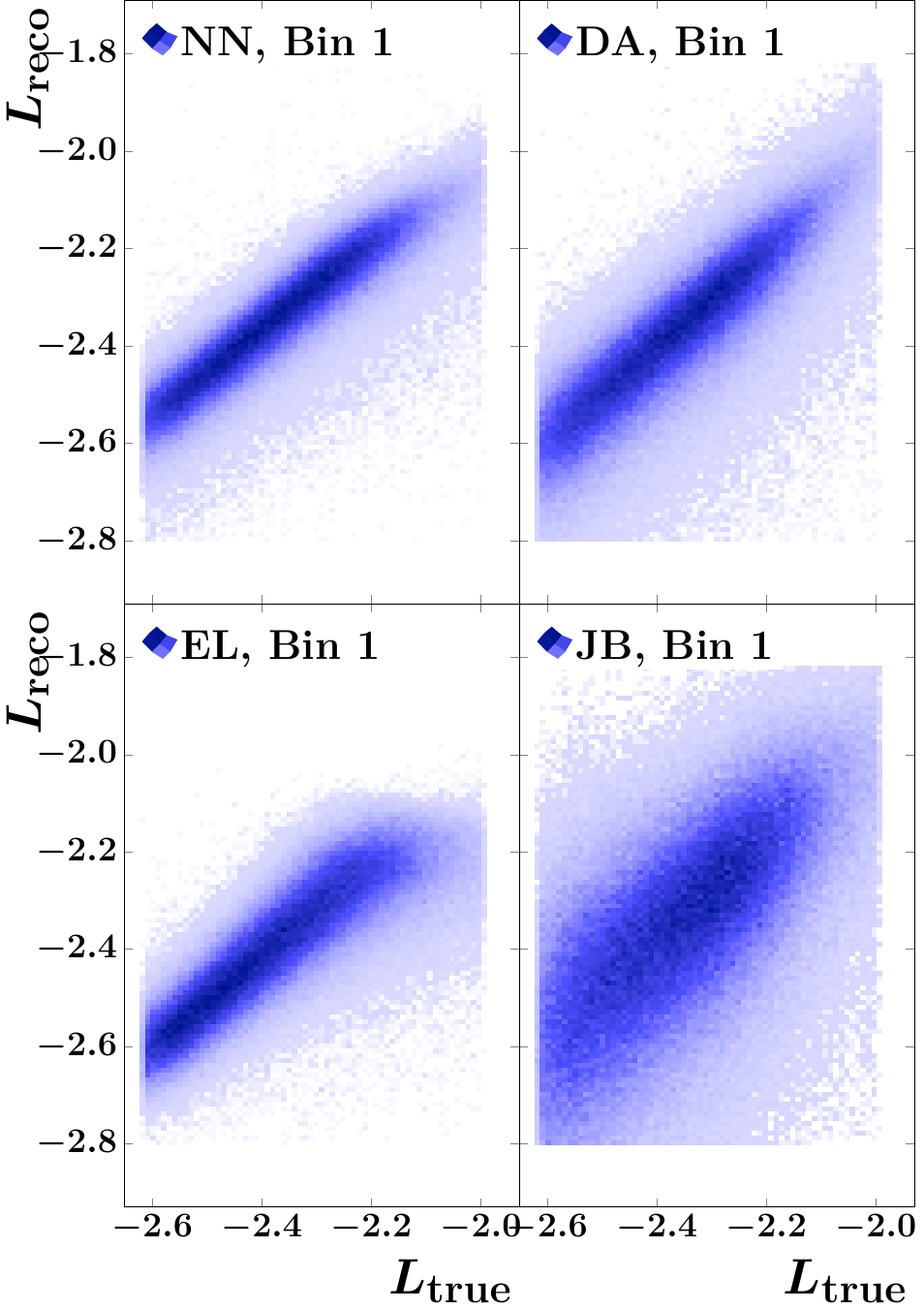}\includegraphics[width=\NARROWFIGWIDTH]{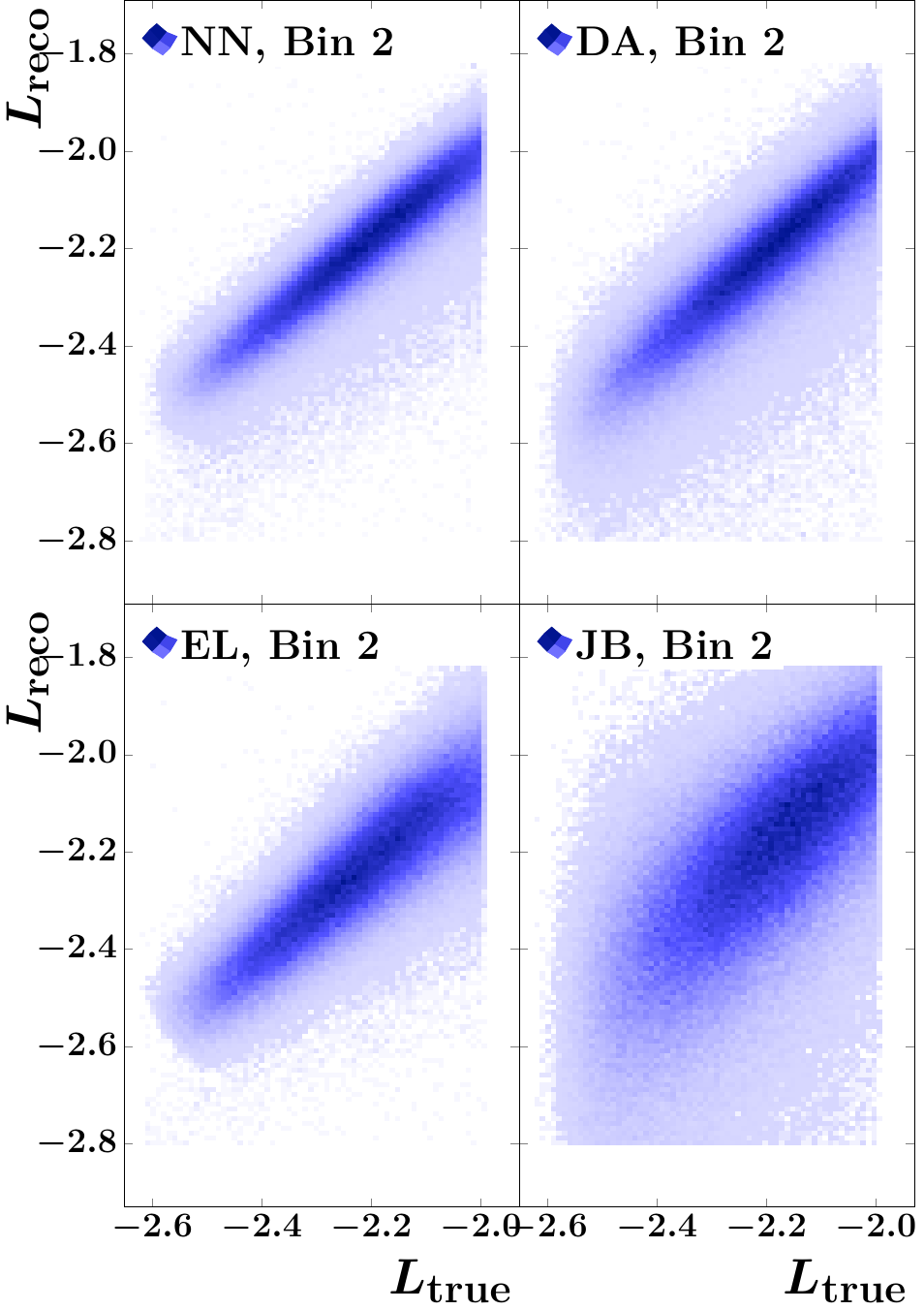}\includegraphics[width=\NARROWFIGWIDTH]{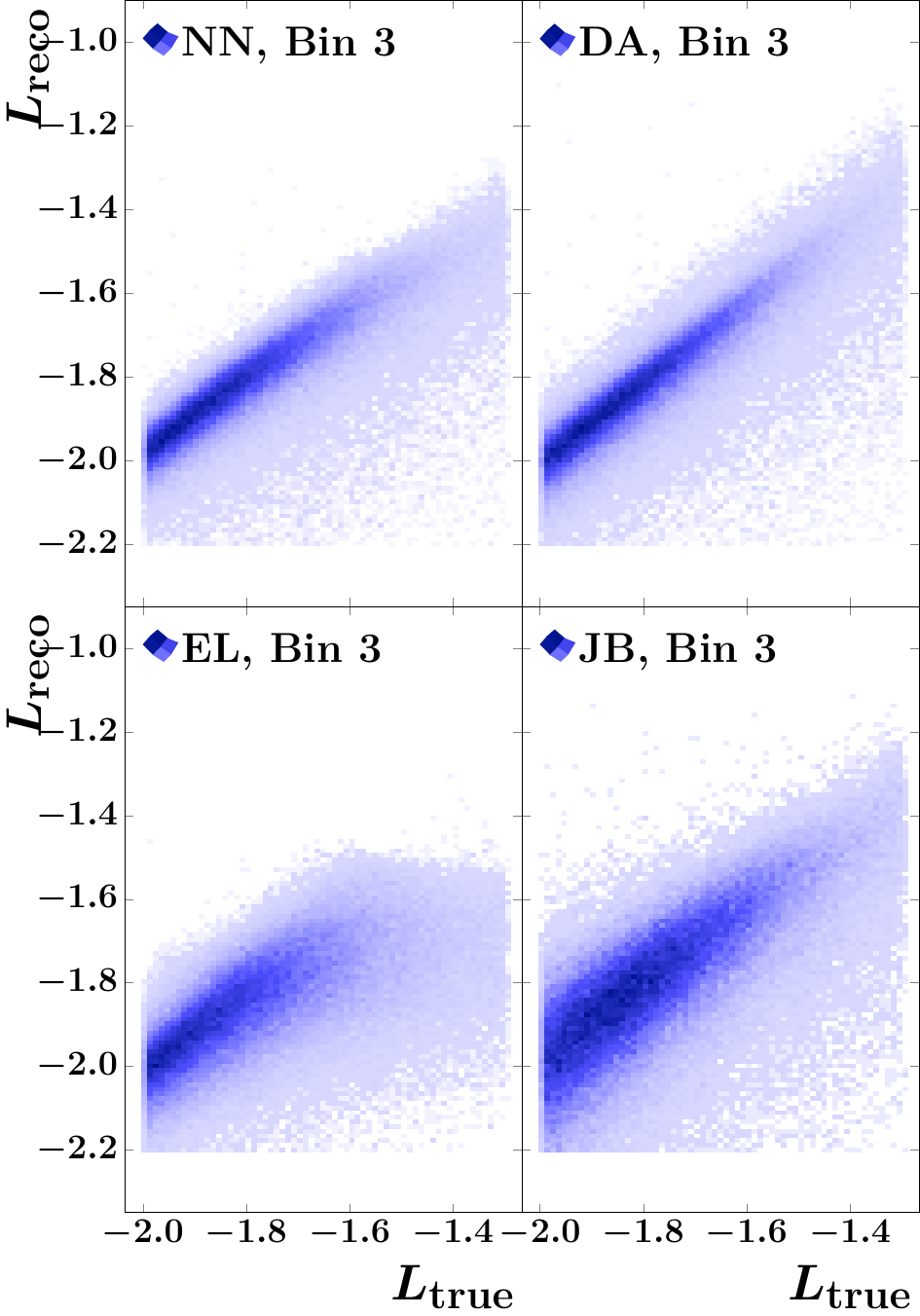}\\
\includegraphics[width=\NARROWFIGWIDTH]{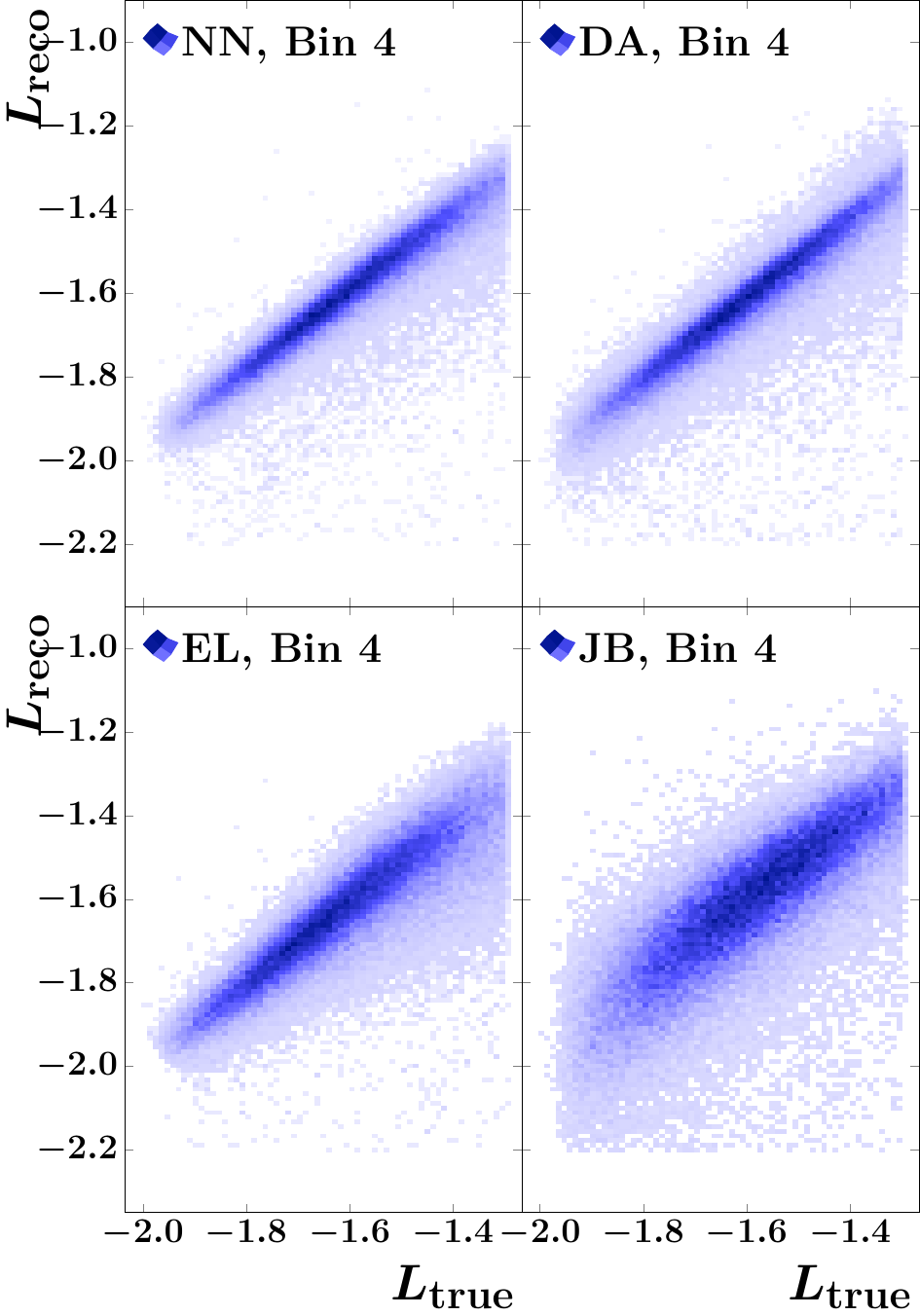}\includegraphics[width=\NARROWFIGWIDTH]{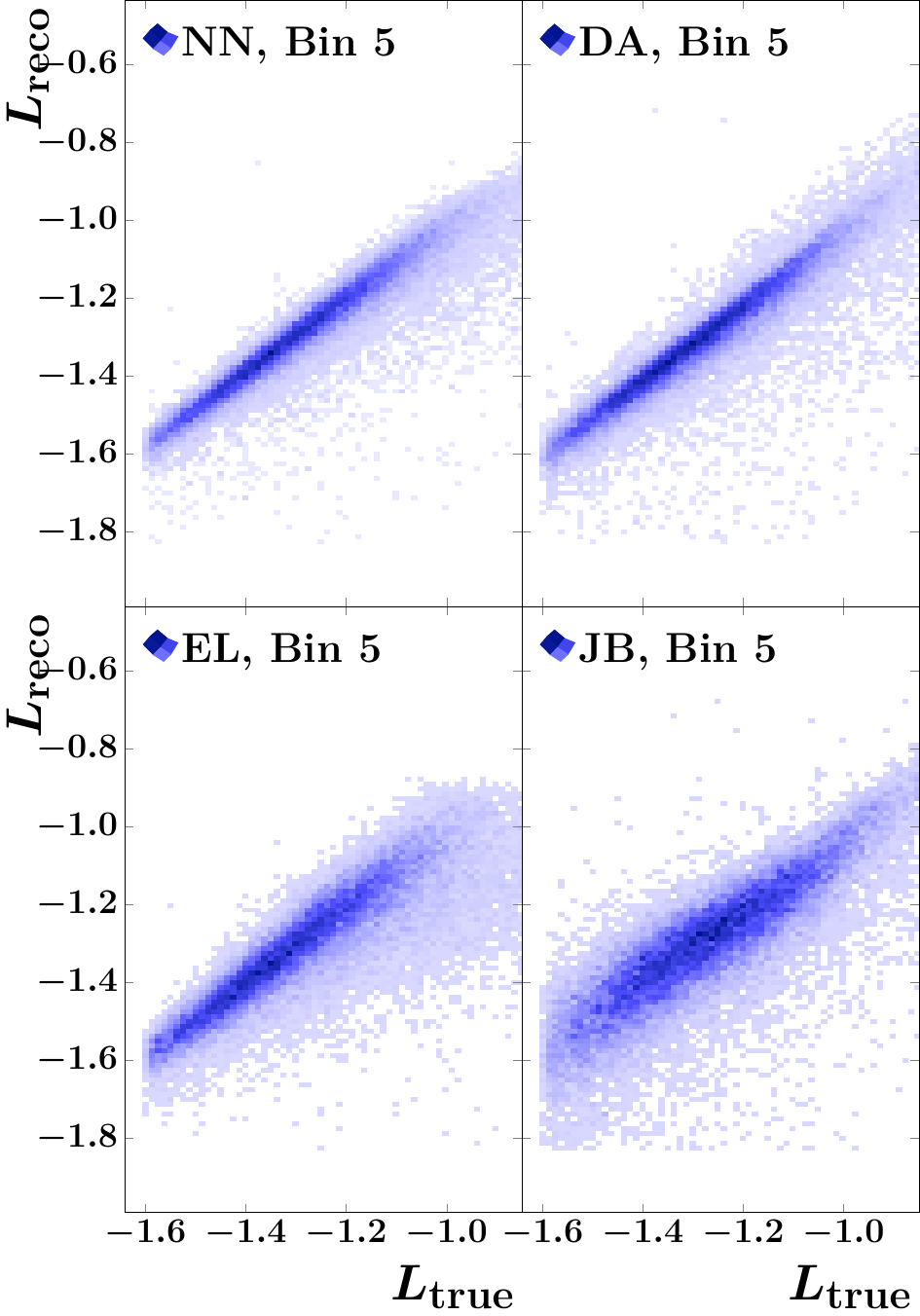}\includegraphics[width=\NARROWFIGWIDTH]{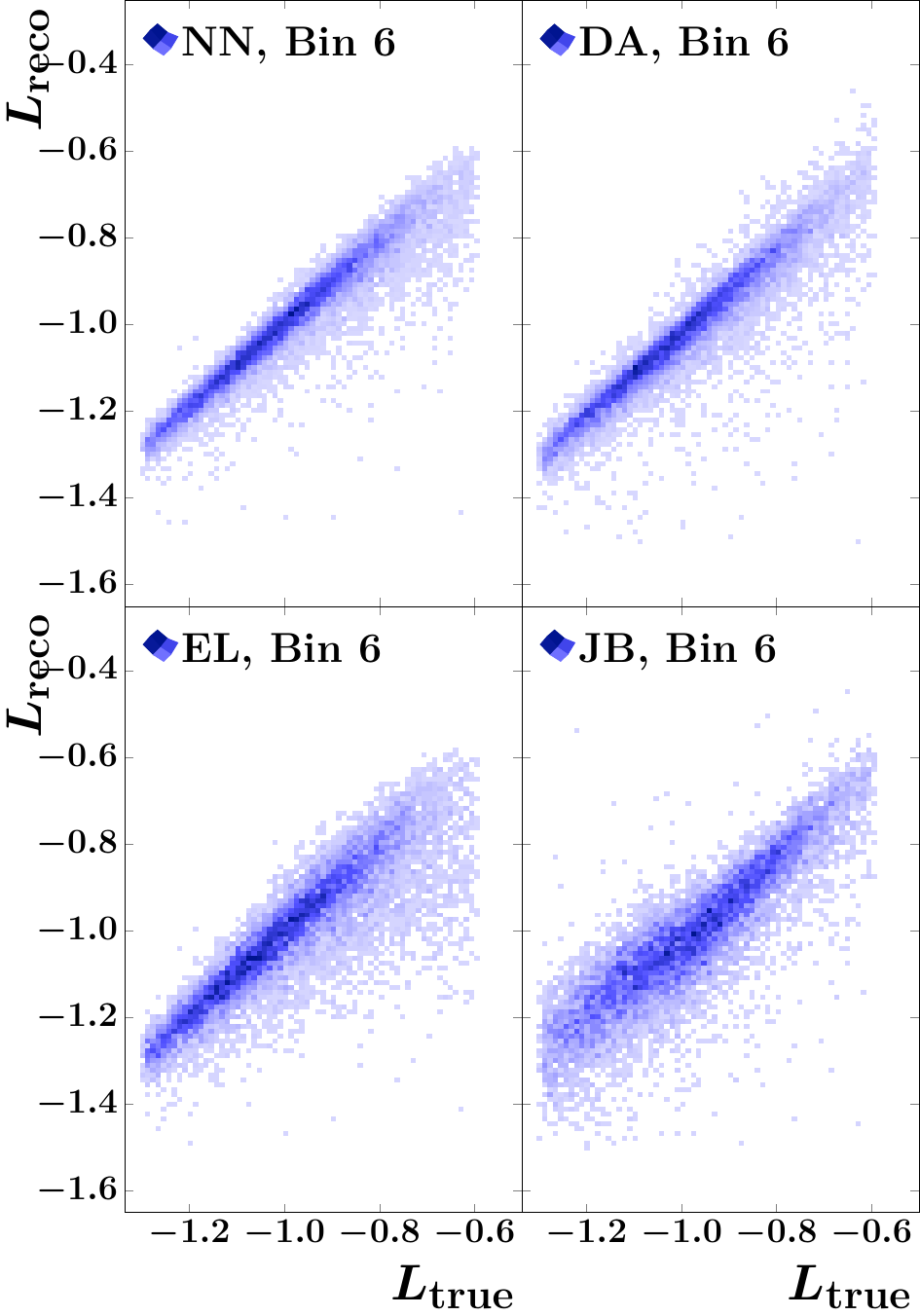}\\
\includegraphics[width=\NARROWFIGWIDTH]{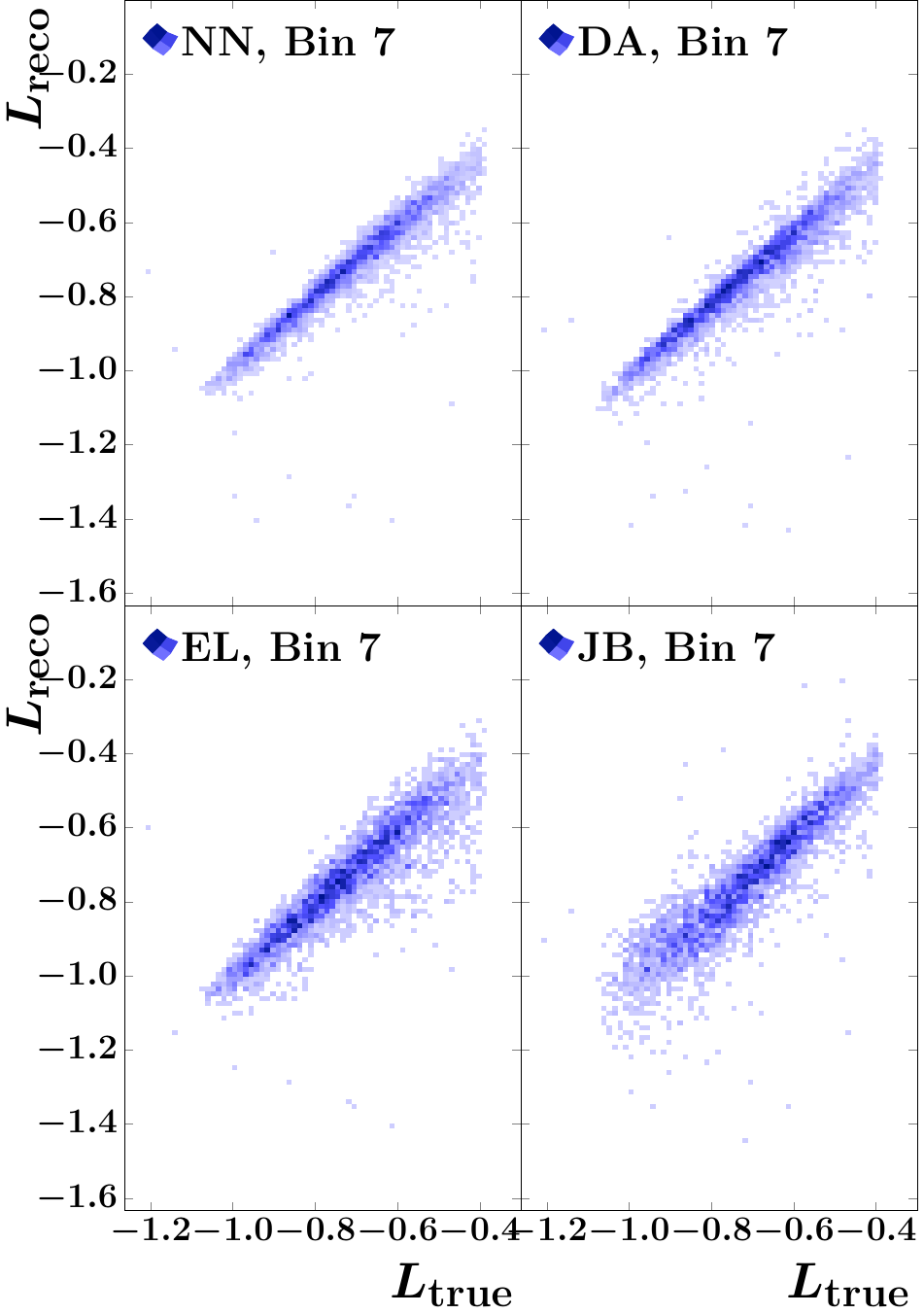}\includegraphics[width=\NARROWFIGWIDTH]{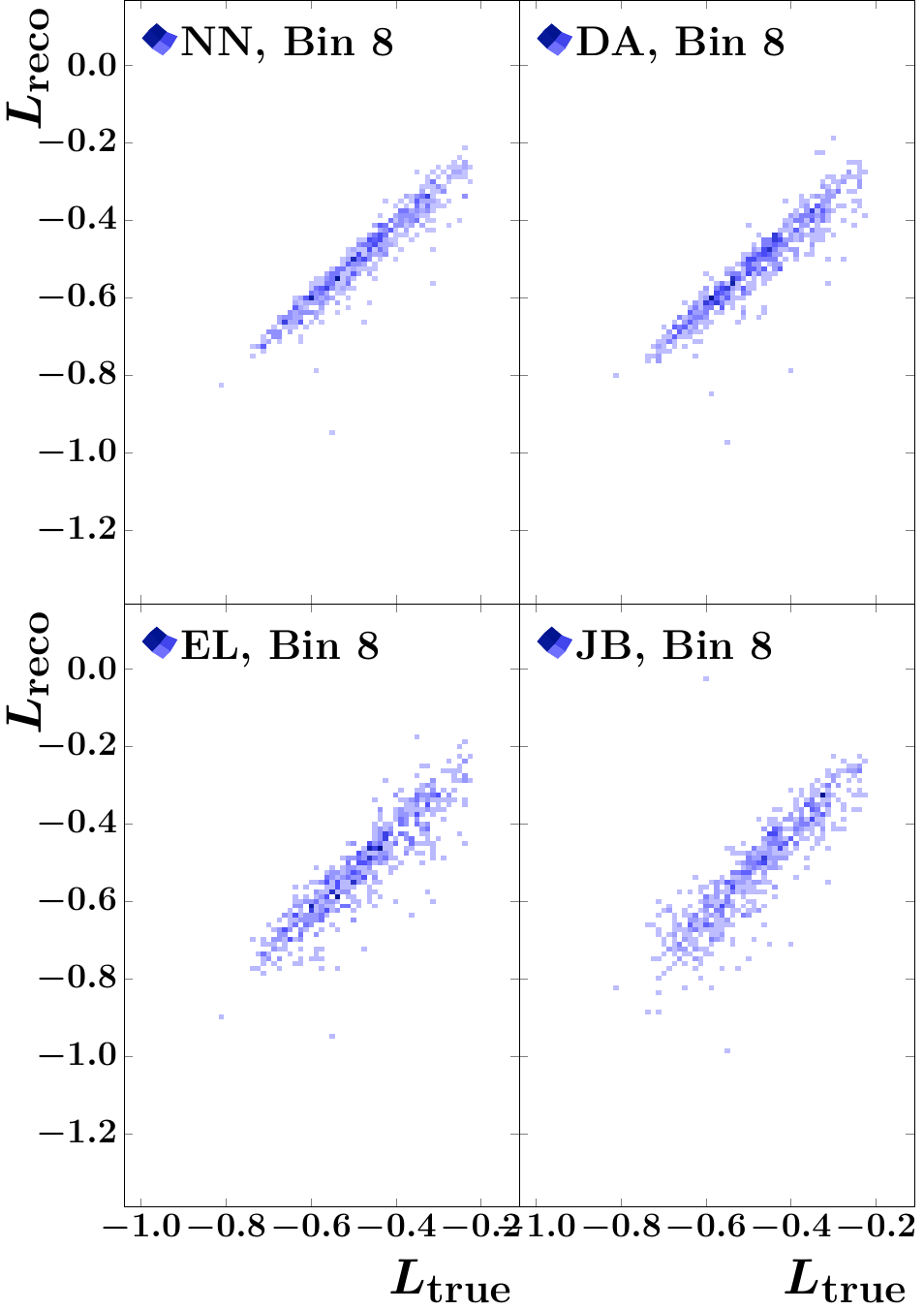}\\
\caption{#1}
\label{fig:xtwod}
\end{figure}
}

\newcommand{\FIGxtwodepjc}[1]{
\begin{figure}[htbp]\centering
\includegraphics[width=0.5\linewidth,height=0.235\textheight]{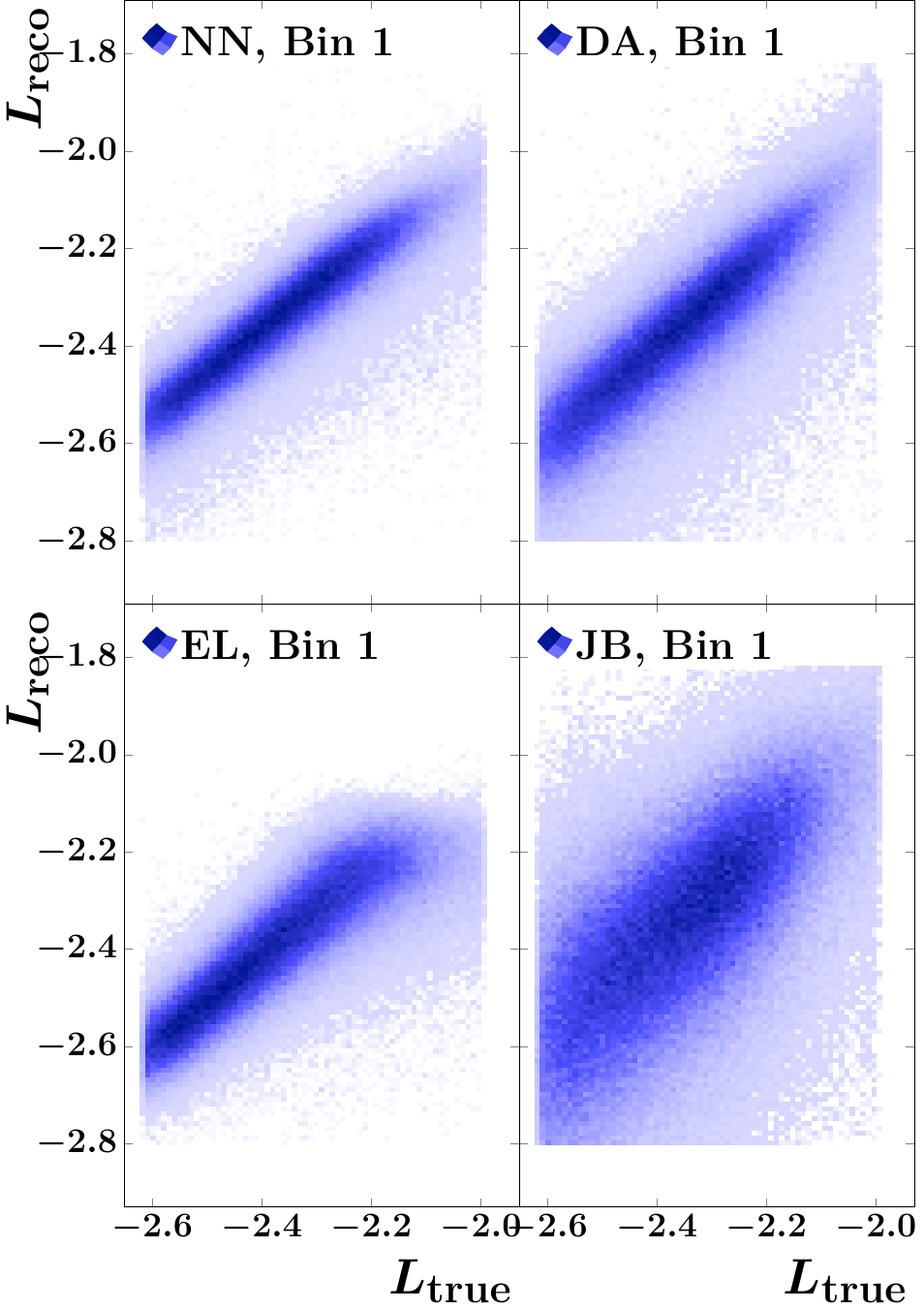}\includegraphics[width=0.5\linewidth,height=0.235\textheight]{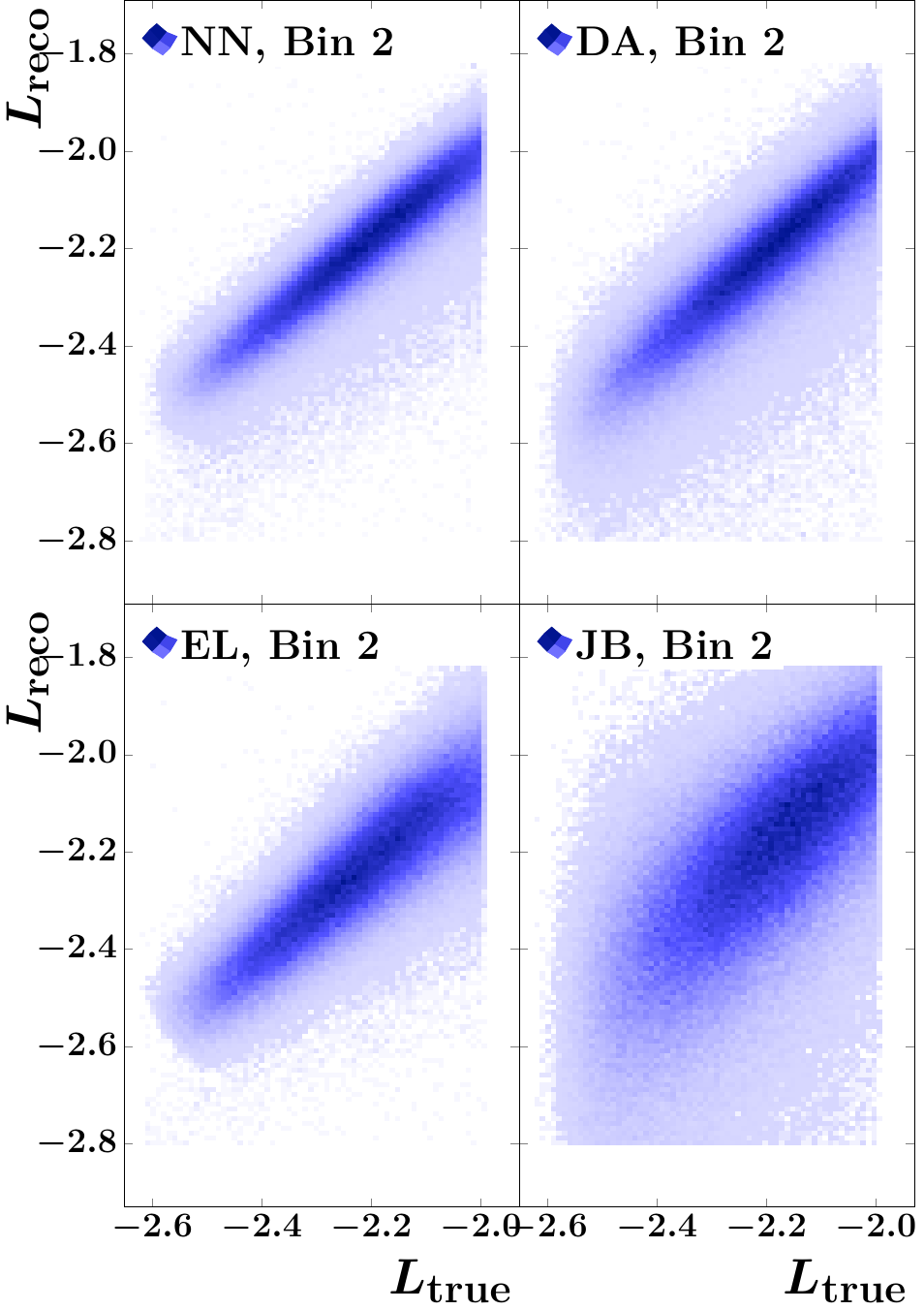}\\
\includegraphics[width=0.5\linewidth,height=0.235\textheight]{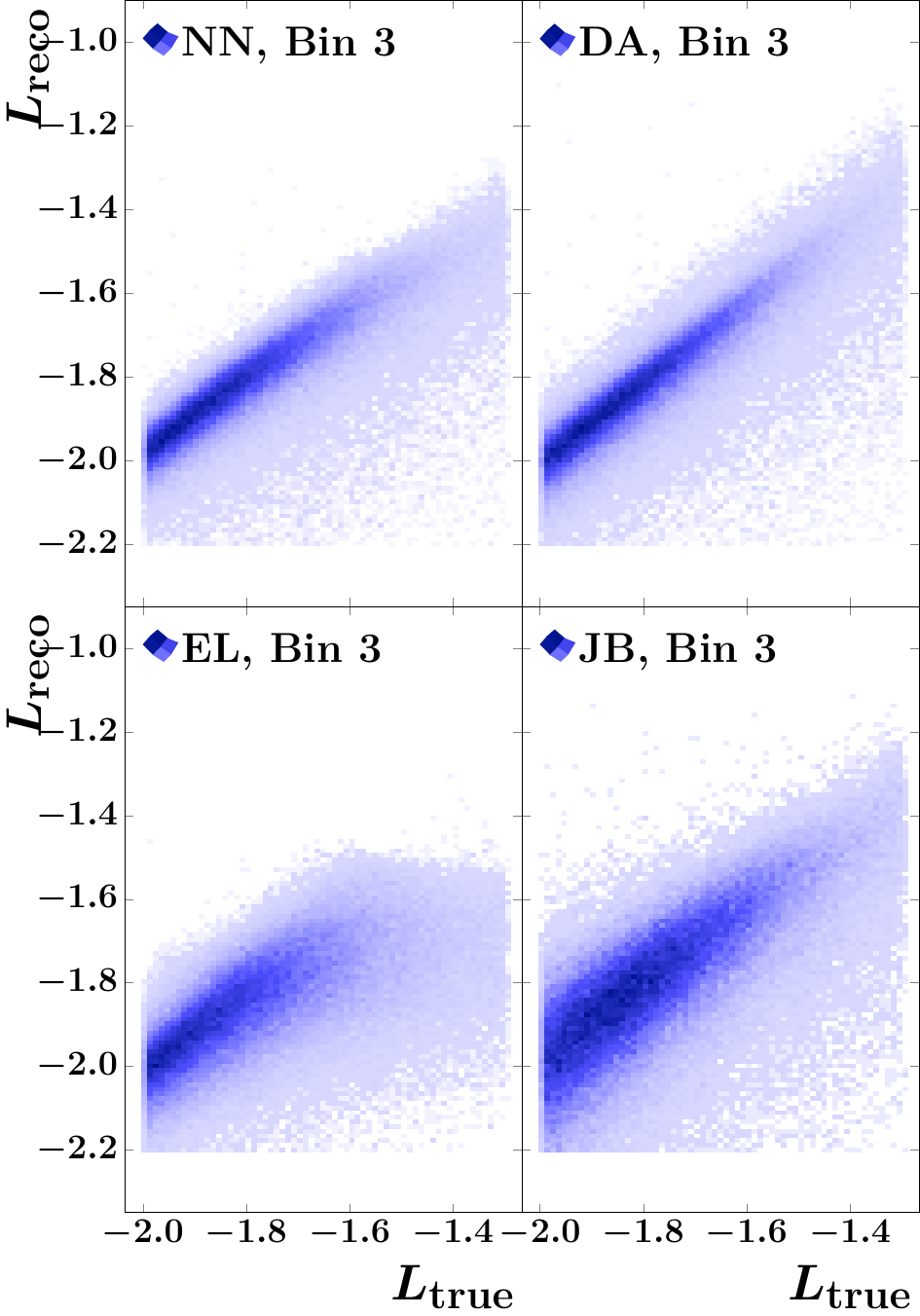}\includegraphics[width=0.5\linewidth,height=0.235\textheight]{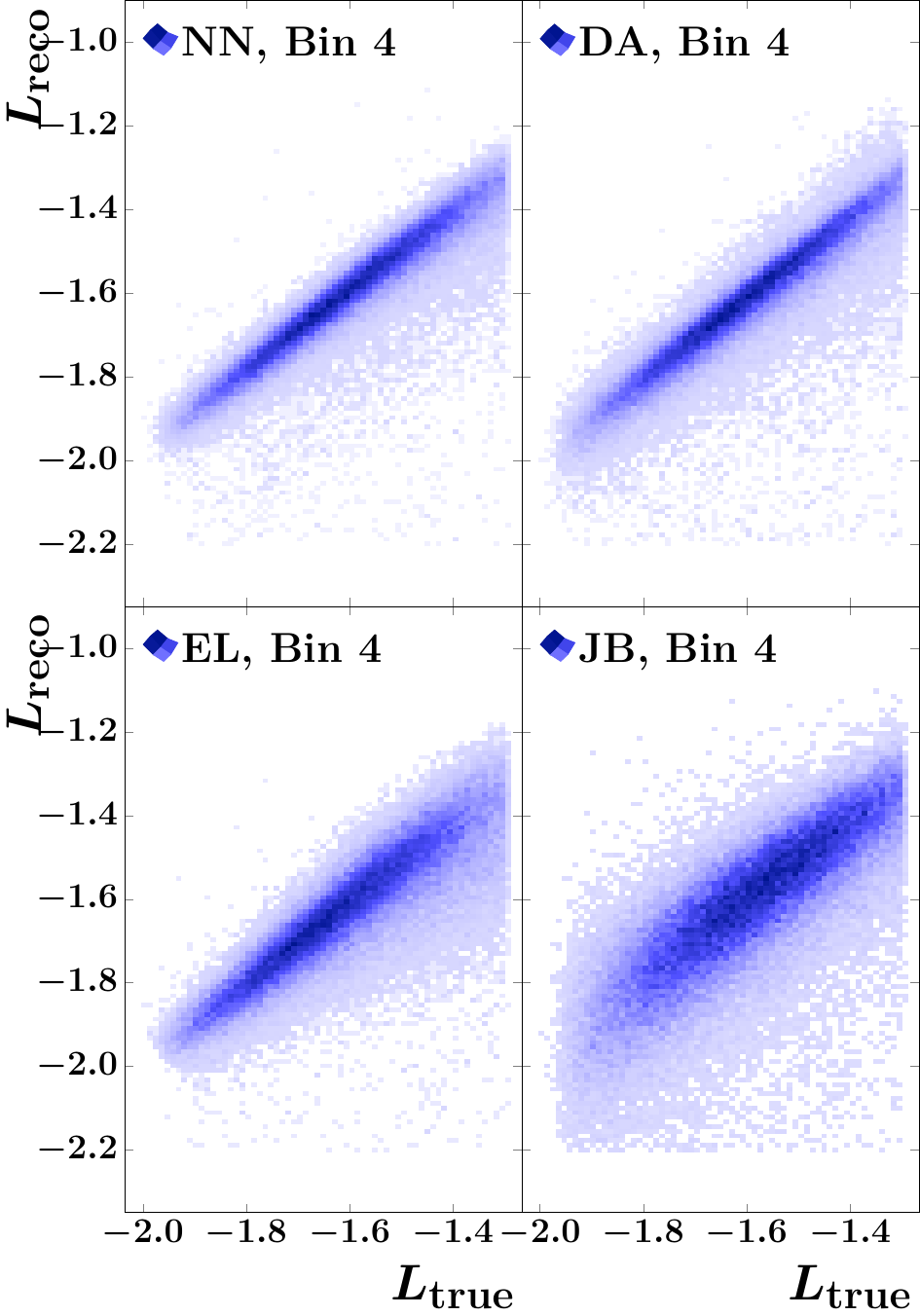}\\
\includegraphics[width=0.5\linewidth,height=0.235\textheight]{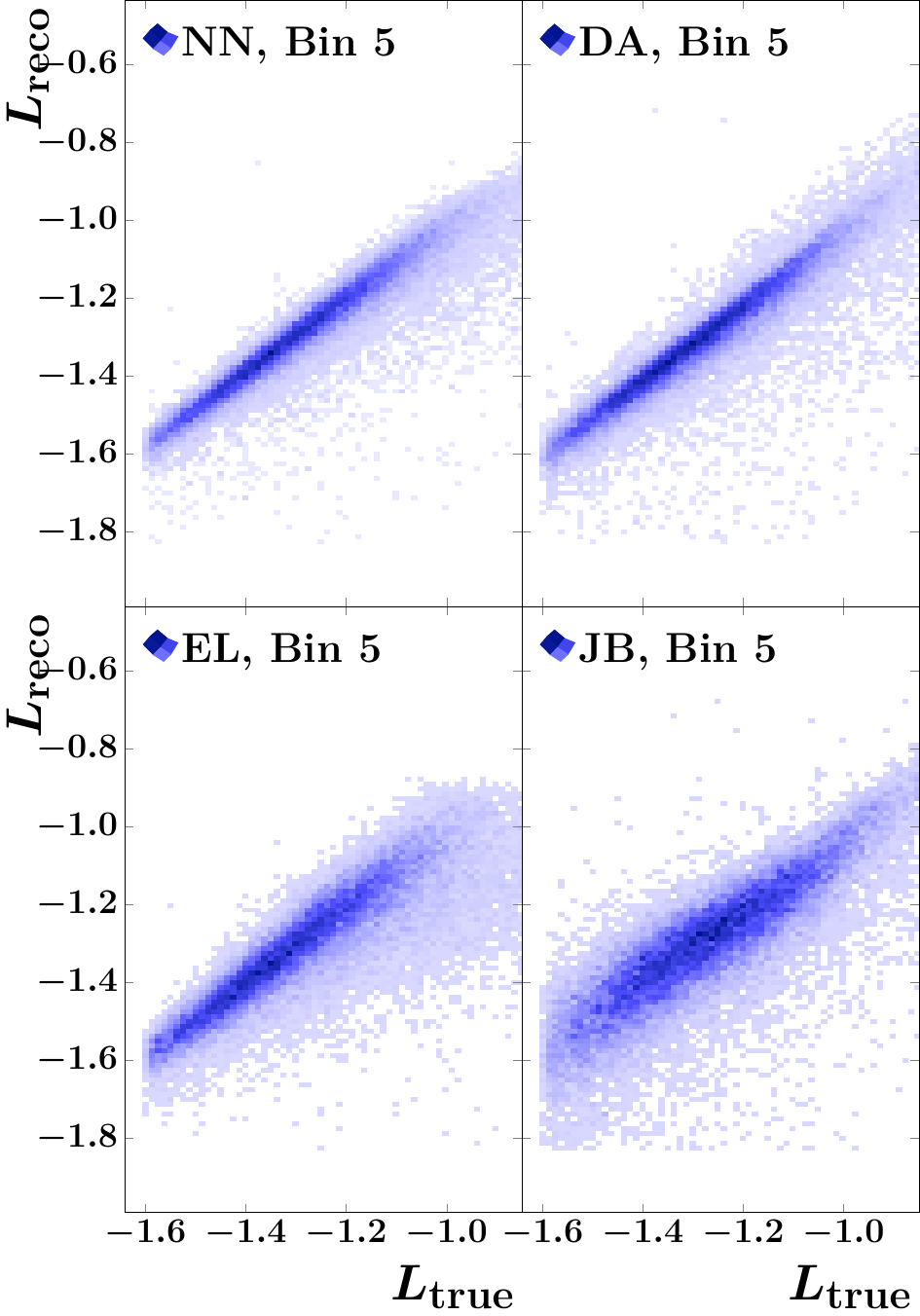}\includegraphics[width=0.5\linewidth,height=0.235\textheight]{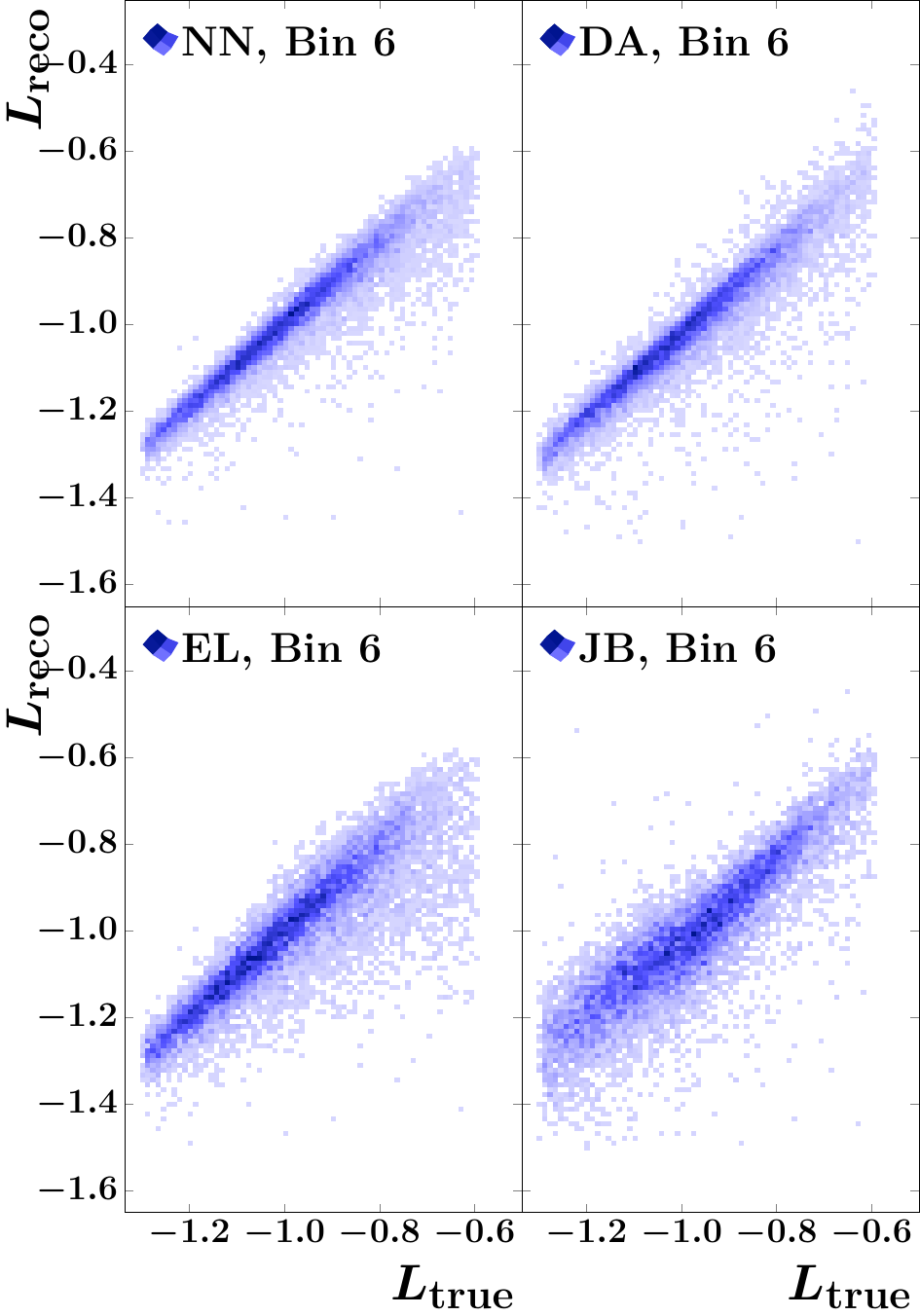}\\
\includegraphics[width=0.5\linewidth,height=0.235\textheight]{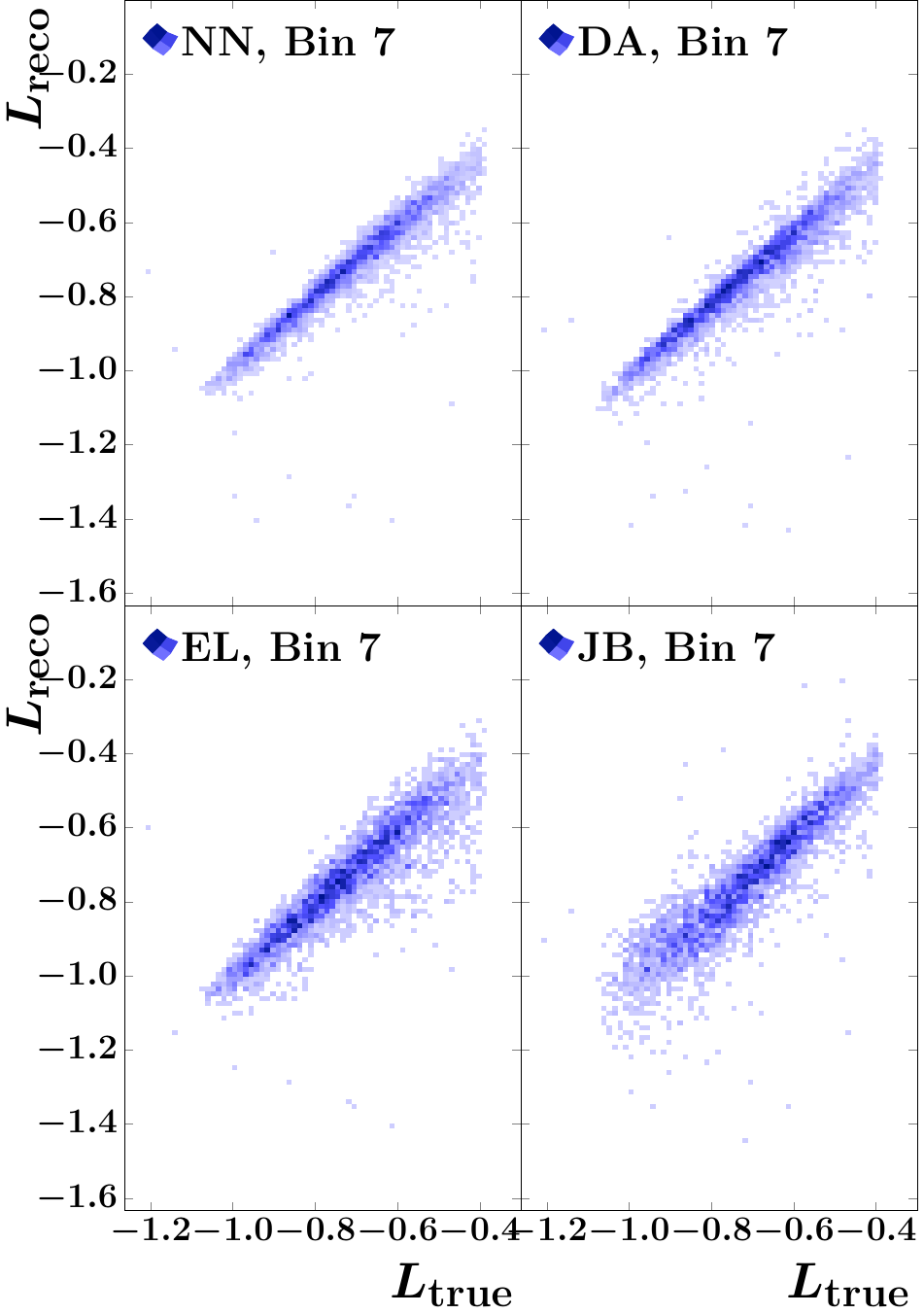}\includegraphics[width=0.5\linewidth,height=0.235\textheight]{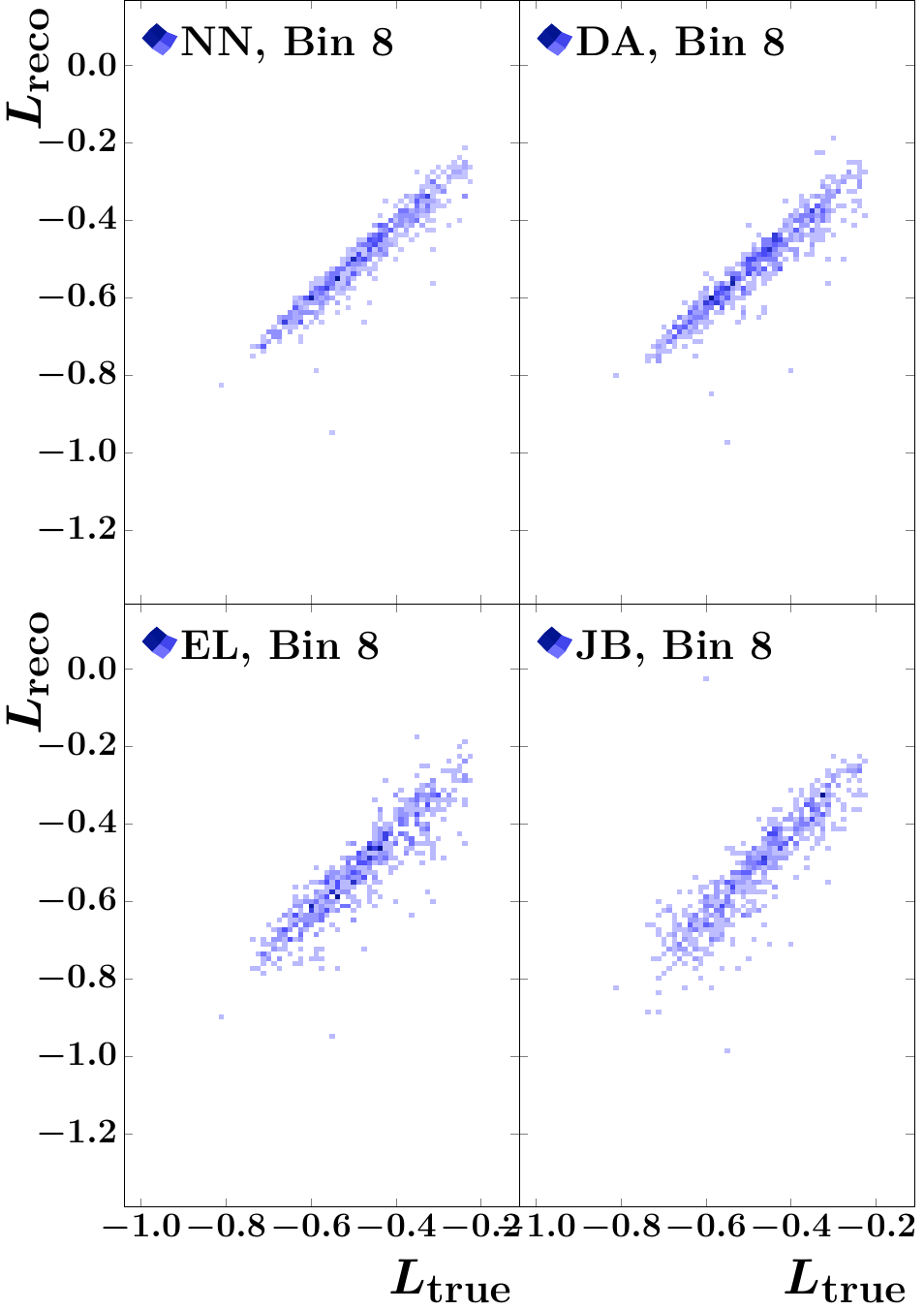}\\
\caption{#1}
\label{fig:xtwod}
\end{figure}
}

\newcommand\nospelling[1]     {#1}
\newcommand\prog[1]     {{\tt #1}}

\newcommand{\eVdist}{\kern-0.06667em}
\newcommand{\GeV}{{\,\text{Ge}\eVdist\text{V\/}}}

\newcommand{\bR}{\mathbb{R}}
\newcommand{\bN}{\mathbb{N}}

\renewcommand{\vec}[1]{#1}

\begin{document}
\maketitle 
\newpage
%\authorlist
\pagenumbering{arabic}
\pagestyle{plain}

\newpage

%%%%%%%%%%%%%%%%%%%%%%%%%%%%%%%%%%%%%%%%%%%%%%%%%%%%%%%%%%%%%%%%%%%%%%%%

\section{Introduction}
\label{sec:introduction}
Measurements of deep-inelastic scattering (DIS) by a multitude of experiments all-over the world~\cite{HERAPROPOSAL,Abi:2020wmh,Gautheron:2010wva, Arrington:2021alx,PDB2021} and the study of these measurements by the theoretical and experimental communities~\cite{PDB2021} have revealed information on the quark-gluon structure of nuclear matter and established quantum chromodynamics (QCD) as the theory of the strong interaction. The experiments at the HERA collider facility at the DESY research centre in Hamburg, Germany have played 
an important role in these studies. HERA has been the only electron-proton collider built so far~\cite{Voss:1994sr}.
 The data collected in the years 1992 - 2007 have provided truly unique
information on the internal structure of the proton and other hadrons~\cite{Abramowicz:2015mha}.

The key component in these studies has been a precise reconstruction of the DIS kinematics,
using information from the accelerator and the detectors. Multiple methods have been applied
at the HERA experiments~\cite{Bassler:1994uq} to reach an optimal precision in each particular measurement.
Each of the classical reconstruction methods uses only partial information from the
DIS event and is subject to specific limitations, either arising from the detector or the
assumptions used in the method.

With the DIS measurements at the upcoming Electron-Ion Collider in mind~\cite{AbdulKhalek:2021gbh}, 
we present in this work a novel method for the reconstruction of DIS kinematics based on
supervised machine learning and study its performance using Monte Carlo simulated data from the ZEUS experiment~\cite{ZEUS:1993aa} at HERA. 
For our approach, we develop deep neural network (DNN) models that
are optimised for the problem and are allowed to take full information from the DIS event into
account. We train the DNN models on simulated data from the ZEUS experiment and compare
the results from our trained model with the results from the classical reconstruction
methods.

We show that the reconstruction of the DIS kinematics using deep neural networks provides a rigorous, data-driven  
method to combine and outperform classical reconstruction 
methods over a wide kinematic range. In the past, neural networks had already been  
used in the context of DIS experiments~\cite{Abramowicz:1995zi} and we 
expect that our novel method and similar approaches will play an even more 
important role in ongoing and future DIS experiments~\cite{Accardi:2012qut,Liu:2020pzv}.

%%%%%%%%%%%%%%%%%%%%%%%%%%%%%%%%%%%%%%%%%%%%%%%%%%%%%%%%%%%%%%%%%%%%%%%%
\section{Deep inelastic scattering}
\label{sec:dis}
Deep inelastic scattering is a process in which a high-energy lepton ($l$) 
scatters off a nucleon or nucleus target ($h$) with large momentum transfer (the momentum of each entity is given in parenthesis): 
\begin{equation}
l(\vec{k}) + h(\vec{P}) \to l^\prime(\vec{k^\prime}) +  \mathcal{H}(\vec{P^\prime}) +{\rm remnant}.
 \label{eq:dis}
\end{equation}
The detectors in collider experiments are designed to measure the final state of the DIS process, consisting of 
the scattered lepton $l^\prime$ and the hadronic final state (HFS) $\mathcal{H}$. The latter consists of hadrons with a 
relatively long lifetime as well as some photons and leptons but does not include the hadron remnant. The H1~\cite{H1:1996prr} and ZEUS experiments were not able to register 
the remnant of the target due to its proximity to the proton beam pipe.

\subsection{Deep inelastic scattering at Born level}
\label{sec:dis:born}
In the leading order (Born) approximation, the leptons interact 
with quarks in the hadrons by the exchange of 
a single virtual $\gamma$ or $Z$ boson in the neutral current (NC) 
reaction, and the exchange of single $W^{\pm}$ boson in the charged 
current (CC) reaction. 
The kinematics of the leading order DIS process in a Feynman diagram-like form is shown in Fig.~\ref{fig:dis}. 
\FIGdis{Schematic representation of Deep Inelastic Scattering process at Born level.}{fig:dis} 
In this paper, we will only consider the neutral current electron scattering off a proton in a collider experiment.
 In this reaction, the final state lepton  is a charged particle (electron or positron) that can be easily
 registered and identified in the detector.

With a fixed centre-of-mass energy,
%\begin{equation} \label{eq:s}
$\sqrt{s}=\sqrt{(\vec{k}+\vec{P})^2}$,
%\end{equation}
two independent, Lorentz-invariant, scalar quantities are sufficient to 
describe the deep inelastic scattering event kinematics in the Born approximation. Typically, the used quantities are:
\begin{itemize}
\item  the negative squared four-momentum of the exchanged electroweak gauge boson: 
%the four-momentum  transferred to the hadronic system
\begin{equation} \label{eq:q2}
 Q^2 = -\vec{q}\cdot\vec{q} = -(\vec{k}-\vec{k^\prime})^2,
\end{equation}
\item the Bjorken scaling variable, interpreted in the frame of a fast moving nucleon as the fraction of incoming 
nucleon longitudinal momentum carried by  the  struck parton: 
\begin{equation} \label{eq:x}
x = \frac{Q^2}{2\vec{P} \cdot \vec{q}}.
\end{equation}
\end{itemize}
In addition to that, the inelasticity $y$ is used to define the kinematic region of interest. It is defined as 
the fraction of incoming electron energy taken  by the exchanged boson in the proton rest frame
\begin{equation} \label{eq:y}
y = \frac{\vec{P} \cdot \vec{q}}{\vec{P} \cdot \vec{k}}.
\end{equation}
Therefore, for the DIS an equation
\begin{equation} \label{eq:q2syx}
Q^2 = s y x
\end{equation} holds.
However, the Born-level  picture of the DIS process is not sufficient 
for the description of the  observed physics phenomena. A realistic  
description of DIS requires the inclusion of higher order QED and QCD processes~\cite{Kwiatkowski:1990es}.

%%%%%%%%%%%%%%%%%%%%%%%%%%%%%%%%%%%%%%%%%%%%%%%%%%%%%%%%%%%%%%%%%%%%%%%%
\subsection{Higher order corrections to deep-inelastic scattering process}
\label{sec:dis:rad}
The  DIS process with leading order QED corrections can be written as 
\begin{equation}
l(\vec{k}) + h(\vec{P}) \to l^\prime(\vec{k^\prime}) + \gamma(\vec{P_{\gamma}}) + X(\vec{P^\prime})   +{\rm remnant},
\label{eq:disrad}
\end{equation}
so the kinematics is defined not only by the 
kinematic of the scattered electron and the struck parton,  but also by 
the momentum of the radiated photon, $\vec{P_{\gamma}}$.  The lowest order electroweak radiative 
corrections can be depicted in a form of Feynman diagrams as shown in Fig.~\ref{fig:distwo} a)-d) and should be considered 
together with the virtual corrections e)-g). 
\epjconly{\FIGdistwoepjc{Feynman diagrams for Deep Inelastic Scattering process with some leading order 
electroweak corrections a)-g) and QCD corrections h)-k). The proton remnant is omitted.}{\nospelling{fig:distwo}}}
\draftonly{\FIGdistwo{Feynman diagrams for Deep Inelastic Scattering process with some leading order 
electroweak corrections. The proton remnant is omitted.}{\nospelling{fig:distwo}}}
The Fig.~\ref{fig:distwo} a),d) correspond to the initial state radiation (ISR) and the 
Fig.~\ref{fig:distwo} b),c) to the final state radiation (FSR).  

With the virtual corrections taken into account in the DIS process, Eq.~\ref{eq:q2} no longer holds, i.e.\ 
\begin{equation} 
\vec{q}\cdot\vec{q} \neq -(\vec{k}-\vec{k^\prime})^2.
\label{eq:q2not}
\end{equation}
%The presence of higher order QCD corrections makes the 
The presence of higher order QCD processes, (e.g.\  the boson-gluon fusion in Fig.~\ref{fig:distwo} h)  and QCD Compton Fig.~\ref{fig:distwo} i),j)) makes the 
kinematic description of the DIS process even more complicated. Therefore, 
the exact definitions of the kinematic observables used in the analysis 
of the DIS events and the corresponding simulations are essential for the correct physics interpretation. 

%%%%%%%%%%%%%%%%%%%%%%%%%%%%%%%%%%%%%%%%%%%%%%%%%%%%%%%%%%%%%%%%%%%%%%%%
\subsection{The simulation of the DIS events and  the kinematic variables in the simulated events }
\label{sec:dis:mc}
The simulation of the inclusive DIS events in the Monte Carlo event 
generators (MCEGs) starts from the simulation at the parton level, i.e.\  the simulation of the hard scattering process 
and the kinematics of the involved partons, e.g.\  given by parton distribution functions (PDFs)~\cite{Altarelli:1977zs} 
for the given hadron and considering processes with all types of partons in the initial state. 
The modelling of the hard scattering process combines the calculations of the perturbative QED and/or QCD matrix elements for the 
$2\rightarrow n $ processes at parton level with the different QCD parton cascade algorithms  designed to
take into account at least some parts of the higher order perturbative QCD corrections not present in the calculations of matrix elements.

The simulated collision events on the particle (hadron) level are obtained using  the parton level 
simulations as input and applying phenomenological hadronisation and particle decay models to them.

As of 2022, multiple MCEG programs are capable of simulating the inclusive 
DIS process at the hadron level with  different levels of theory precision and  sophistication of modelling of hadronisation, beam remnant and parton cascades,
e.g.\ \prog{Pythia6}~\cite{Sjostrand:2006za}, 
\prog{Pythia8}~\cite{Sjostrand:2014zea}, \prog{SHERPA-MC}~\cite{Bothmann:2019yzt}, 
\prog{WHIZARD}~\cite{Kilian:2007gr} and \prog{Herwig7}~\cite{Bellm:2015jjp}. 
In addition to that, the \prog{Lepto}~\cite{Ingelman:1996mq}, 
\prog{Ariadne}~\cite{Lonnblad:1992tz}, \prog{Cascade}~\cite{Jung:2010si} 
and \prog{Rapgap}~\cite{Jung:1993gf} programs can simulate the DIS process 
using parts of the \prog{Pythia6} framework for the simulation of hadronisation processes and decays of particles.

As it was discussed above, DIS beyond the Born approximation has a 
complicated structure which involves QCD and QED corrections~\cite{AbdulKhalek:2021gbh}. The 
most recent  MCEG programs, e.g.\  \prog{Herwig7}, \prog{Pythia8},  \prog{SHERPA-MC}, or \prog{WHIZARD}, contain these corrections as a particular case of their own general-purpose frameworks or 
are able to use specialised packages like 
\prog{OpenLoops}~\cite{Cascioli:2011va}, \prog{blackhat}~\cite{Bern:2013pya} 
or \prog{MadGraph}~\cite{Alwall:2011uj}. 
In general, the modern MCEGs do not specify their definitions of the 
DIS kinematic observables, but in some cases they can be calculated from the kinematics of the initial and final state, both for the true and reconstructed kinematics. For instance, under some assumptions, 
the $Q^{2}$ in the event could be calculated according to Eq.~\ref{eq:q2}.
 The total elimination of the ambiguities 
for such calculations is not possible, as the final state kinematics 
even at the parton level depend on the kinematics of all the emitted partons.
% for such calculations is not possible, as the final state kinematics
% even at the parton level depends not only on the hard process kinematics but also on the subsequent hard and soft QCD or QED radiation. 
 The calculations of the kinematic observables from the momenta of particles at hadron level add an additional ambiguity related to the identification of the scattered lepton and the distinction of that lepton from the leptons produced in the hadronisation and decay processes.

Contrary to the approaches adopted in modern MCEG programs, the MCEGs used for the HERA experiments relied upon generator-by-generator implementations of the higher order QED and QCD corrections specific for DIS or alternatively applied \prog{HERACLES}~\cite{Kwiatkowski:1990es} for the corrections.

%Some of these programs provided specific definitions of the
%DIS kinematic observables taking or not taking the radiative corrections into account.
%In this way all the ambiguities related to the determination of the kinematic 
%observables are contained in these specific definitions.

The way to get a simulation of the DIS collision even in a specific detector
is the same as for any other type of particle collision event. It involves  simulation of
the particle transport through the detector material, simulation of  
detector response and is typically performed in 
\prog{Geant}~\cite{Brun:1987ma,Agostinelli:2002hh} or similar tools. 
The simulated detector response is passed to the experiment-specific
reconstruction programs and should be indistinguishable from the 
real data recorded by the detector and processed in the same way.

%%%%%%%%%%%%%%%%%%%%%%%%%%%%%%%%%%%%%%%%%%%%%%%%%%%%%%%%%%%%%%%%%%%%%%%%
\subsection{Reconstruction of the kinematic variables at the detector level}
\label{sec:dis:reco}
The kinematics of the DIS events are reconstructed in collider experiments by 
identifying and measuring the momentum of the scattered lepton 
$l^\prime$ and/or the measurements of the hadronic final state  ($\mathcal{H}$). 
The identification of the scattered lepton is 
ambiguous even at the particle level of the simulated DIS collision events. 
The same ambiguity is present in the reconstructed real and simulated data at the detector level.
Therefore, the identification of the scattered electron candidate for the purposes of 
physics analyses is a complicated task on itself and was a subject of 
multiple studies in the past, some of which also involved neural 
network-related techniques~\cite{Abramowicz:1995zi}. 
In our paper, we rely on the standard method of the electron identification at the ZEUS experiment~\cite{Abramowicz:1995zi}
and discuss solely the reconstruction of the kinematic variables using the identified electron and other quantities measured in the detector.

The physics analyses performed in the experiments at HERA relied on the following quantities for the calculation
of the kinematic observables $x$ and $Q^2$: 
\begin{itemize}
\item The energy ($E_{l^\prime}$) and polar angle ($\theta_{l^\prime}$) of the 
scattered electron.  Most of the DIS experiments are equipped to register 
the scattered electron using the tracking and calorimeter detector 
subsystems. While the tracking system is able to provide information 
on the momentum of the scattered electron, the calorimeter system can be 
used to estimate the energy of the electron and the total energy of the collinear 
radiation emitted by the electron. At the detector level, the estimation of 
these energies can be done by comparing the momentum of the electron reconstructed 
in the tracking system and the energy deposits registered in the calorimeter 
system around the extrapolated path of the electron in the calorimeter.
\item The energy of the HFS expressed in terms of the following convenient 
variables:
\begin{equation}
\delta_{\mathcal{H}} =\sum_{i\in\mathcal{H}} E_i-P_{Z,i}
 \label{eq:deltah}
\end{equation}
and
\begin{equation}
P_{T,\mathcal{H}}=\sqrt{\left( \sum_{i\in\mathcal{H}} P_{X,i} \right) + \left( \sum_{i\in\mathcal{H}} P_{Y,i} \right) },
\label{eq:PTh}
\end{equation}
\end{itemize}
where the sums run over the registered objects $i$ excluding the scattered 
electron. Depending on the analysis requirements, the used objects could 
be either registered tracks, energy deposits in the calorimeter system, 
or a combination of both.

The measurements of the quantities listed above overconstrain the 
reconstruction of the DIS kinematics.  Therefore, in the simplest case, any subset of two observables 
in $E_{l^\prime}$, $\theta_{l^\prime}$, $\delta_{\mathcal{H}}$, and $P_{T,\mathcal{H}}$, 
can be used for the reconstruction.

In our analysis, we consider three specific classical reconstruction methods  based on 
these observables which were used by the ZEUS collaboration in the past: 
the electron, the Jacques-Blondel, and the double-angle methods. 
We briefly provide some details on the methods in this section, while a more detailed
  description can be found elsewhere~\cite{Bentvelsen:1992fu}. 

The electron (EL) method uses only measurements of the scattered lepton, 
$E_{l^\prime}$ and $\theta_{l^\prime}$, to do the reconstruction of $Q^2$ and $x$. 
The kinematic variables calculated from these measurements are given by:
\begin{equation}
Q^2_{EL} = 2 E_l E_{l^\prime}(1+\cos\theta_{l^\prime})
 \label{eq:q2el}
\end{equation}
and 
\begin{equation}
x_{EL}  = \frac{E_l E_{l^\prime}(1+\cos\theta_{l^\prime})}{E_P(2E_l-E_{l^\prime}(1-\cos\theta_{l^\prime}))}.
 \label{eq:xel}
\end{equation}
The electron method provides precise reconstruction of kinematics, but, 
since it uses only information from the scattered lepton, this method is 
affected by initial and final state QED radiation. Namely, the QED 
radiation registered in the detector separately from the scattered 
electron will not be taken into account in the calculations with this 
method. Practically, the reconstruction with this method gives reasonable 
results when $E_l$ and $E_{l^\prime}$ are significantly different 
from one another, but the resolution and stability becomes poor otherwise.

The Jacques-Blondel (JB) method uses only measurements of the final state hadronic 
system, $\delta_{\mathcal{H}}$ and $P_{T,\mathcal{H}}$, for the reconstruction.
The kinematic variables are calculated from these by:
\begin{equation} \label{eq:q2jb}
Q^2_{JB} = \frac{2 E_l P_{T,\mathcal{H}}^2}{2 E_l - \delta_{\mathcal{H}}},
\end{equation}
\begin{equation} \label{eq:xjb}
x_{JB}  = \frac{2 E_l Q^2_{JB}}{s \delta_{\mathcal{H}}}.
\end{equation}
%The Jacques-Blondel method reconstructs the DIS kinematic variables using 
%only information measured from the HFS. 
The JB method
is resistant to possible biases because of unaccounted QED FSR, but requires precise  measurements 
of the particles momenta in the hadronic final state. The factors that 
limit the precision of the measurements are the uncertainties in the particle identification, 
the finite resolution of the calorimeter and tracking detectors,
the inefficiencies of these detectors, the impossibility of the particle detection around the beampipe, and the presence of objects that avoid detection (e.g.\  neutrinos from particle decays).

The double-angle (DA) method combines measurements from the scattered lepton and 
the final-state hadronic system, $\theta_{l^\prime}$ and $\gamma_\mathcal{H}$, 
to perform the kinematic reconstruction as follows:
\begin{equation}
Q^2_{DA} = \frac{4E_l^2\sin\gamma_{\mathcal{H}}(1+\cos\theta_{l^\prime})}{\sin\gamma_{\mathcal{H}} +\sin\theta_{l^\prime}-\sin(\gamma_{\mathcal{H}} + \theta_{l^\prime})},
 \label{eq:q2da}
\end{equation}
\begin{equation}
x_{DA}  = \frac{E_l\sin\gamma_{\mathcal{H}}(1+\cos\theta_{l^\prime})}{E_P\sin\theta_{l^\prime}(1-\cos\gamma_{\mathcal{H}})},
 \label{eq:xda}
\end{equation}
where the angle $\gamma_\mathcal{H}$ is
defined as 
\begin{equation}
\cos\gamma_\mathcal{H}=\frac{P_{T,\mathcal{H}}^2-\delta_{\mathcal{H}}^2}{P_{T,\mathcal{H}}^2+\delta_{\mathcal{H}}^2}.
 \label{eq:gammah}
\end{equation}
The angle $\gamma_\mathcal{H}$ depends on the ratio of the measured quantities $\delta_{\mathcal{H}}$ 
and $P_{T,\mathcal{H}}$, and thus, uncertainties in the hadronic energy measurement 
tend to cancel, leading to good stability of the reconstructed kinematic variables. 
Similar to the electron method, when $E_l$ and $E_{l^\prime}$ are significantly 
different from one another, the double-angle method provides reliable results, 
but the resolution and stability  are poor otherwise.

%%%%%%%%%%%%%%%%%%%%%%%%%%%%%%%%%%%%%%%%%%%%%%%%%%%%%%%%%%%%%%%%%%%%%%%%
\subsection{The methodology of measurements in the deep inelastic $ep$ collisions}
\label{sec:dis:meas}
The methodology of measurements at lepton-hadron colliders in general is 
similar to the methodologies used at $e^+ e^-$ and hadron-hadron colliders.
Briefly, in the most cases the quantity of interest is measured from the 
real data registered in the detector corrected for detector effects. 
The corrections are estimated by comparing the analysis at the detector level with the same analysis 
at particle level using detailed simulations of the collision events with the inclusion of higher order QED and QCD processes. 
 
The main difference between the measurements at lepton-hadron colliders and elsewhere,
 is in the way the measurements involve collision  kinematics.
At $e^+ e^-$ colliders, the initial kinematics of the interactions is given by the lepton energies 
that are known  parameters of the accelerators. 
Therefore,  it is straightforward for most of the measurements in the 
$e^+ e^-$ experiments to estimate the centre-of-mass energy of the hard 
collision process. In hadron-hadron collider experiments, there is no way to 
measure the kinematic properties of the partons initiating the collision 
process, as the involved partons cannot be observed in a free state and 
most measurements in the hadron-hadron collisions  are inclusive in the 
kinematics of the initial state. The  DIS collisions at 
electron-proton colliders take a middle stance between these cases. The kinematic observables of the DIS process 
are measured on an event-by-event basis at the detector level using the methods described above. 

In an experiment, the measurements of event kinematics is affected by various effects. 
%in particular radiative and detector smearing effects that cannot be separated 
%on an event-by-event basis.  
For a 
proper comparison of the measurements of HFS, e.g. jet cross-sections or event shape observables to corresponding perturbative QCD (pQCD) predictions, the 
detector-level measurements are unfolded for detector effects 
while hadronisation correction factors are calculated using MCEGs or specialised programs
 and applied to the pQCD predictions~\cite{Currie:2017tpe,H1:2021wkz,Arbuzov:1995id,Liu:2021jfp}.
The prescription for calculation of those correction factors  vary 
depending on the HFS quantities measured and the used definitions of the kinematic observables.
Typically, at ZEUS and other experiments at 
HERA, after the unfolding of the detector effects, the measurements were also
scaled by radiation correction factors to facilitate a comparison to
theoretical calculations available  at Born level in QED, see e.g.\ Ref.~\cite{ZEUS:2010fxa}. 
The factors were obtained from separate high-statistics MC simulations. 
This is a well-understood Monte Carlo approach and our deep learning 
technique can be used with it in exactly the same 
way as the classical methods, both for the experimental and the MC data.  
For future DIS measurements, e.g.\  at the upcoming Electron-Ion Collider, it is 
expected that the effects of QED and QCD radiation can be treated in an unified 
formalism~\cite{Liu:2021jfp}, with the QED effects taken into account into a factorised approach. 
Our DNN-based reconstruction of  DIS kinematics is compatible with such a 
factorised approach as well.

Therefore, in this analysis, we keep the calculation of correction factors for QED and hadronisation effects out of scope 
and limit the discussion to detector-level measurements. 
At particle-level in the MC generated events, 
we use a single definition of ``true'' kinematic 
observables that is based on the kinematics of the exchanged boson extracted from the MC event information.

Namely, we use the definitions of the true-level observables  $Q^{2}_{\rm true}$ and  $x_{\rm true}$  that follow the 
definitions implemented in ZEUS Common NTuples~\cite{Malka:2012hg,AndriiVerbytskyifortheZEUS:2016zup} 
and were used in many ZEUS analyses.  With these definitions the  $Q^{2}_{\rm true}$ is calculated directly 
from the squared four-momentum of the exchange boson $\vec{q}_{\rm boson}$, 
\begin{equation}
Q^{2}_{\rm true} = - |\vec{q}_{\rm boson}|^2.
\end{equation}
The $x_{\rm true}$ is calculated according to the formula
\begin{equation}
x_{\rm true} = \frac{Q^{2}_{\rm true}}{y_{\rm after\ ISR} s_{\rm after\ ISR}},
\end{equation}
with
\begin{equation}
y_{\rm after\ ISR} = \frac{\vec{q}_{\rm boson}\cdot \vec{P}_{\rm proton\ beam}} {  ( \vec{P}_{\rm lepton\ beam} - \vec{P}_{\rm ISR\ radiation} )\cdot \vec{P}_{\rm proton\ beam}} 
\end{equation}
and
\begin{equation}
s_{\rm after\ ISR} = | \vec{P}_{\rm proton\ beam} +  \vec{P}_{\rm lepton\ beam} - \vec{P}_{\rm ISR\ radiation} |^2,
\end{equation}
where $\vec{P}_{\rm proton\ beam}$, $\vec{P}_{\rm lepton\ beam}$ and $\vec{P}_{\rm ISR\ radiation}$ 
represent the four-momenta of the proton beam particles, lepton beam particles and the 
momenta lost by the lepton beam particles to ISR.  Thus, $y_{\rm after\ ISR}$ corresponds to the 
fraction of energy of the bare lepton (i.e.\ without the ISR) transferred to the HFS in 
the centre of mass of the proton and $s_{\rm after\ ISR}$ to the centre-of-mass-energy squared of 
the proton and the incoming bare electron.

%%%%%%%%%%%%%%%%%%%%%%%%%%%%%%%%%%%%%%%%%%%%%%%%%%%%%%%%%%%%%%%%%%%%%%%%

\section{Machine learning methods}
\label{sec:ml}
The reconstruction of DIS event kinematics is overconstrained by the previously 
mentioned measurements, $E_{l^\prime}$, $\theta_{l^\prime}$, $\delta_{\mathcal{H}}$, 
and $P_{T,\mathcal{H}}$. We trained an ensemble of deep neural networks (DNNs) to reconstruct $x$ 
and $Q^2$ by correcting results from classical reconstruction methods  using the information on the 
scattered lepton and the final-state hadronic system.

The ensemble DNN method presented here is a new approach designed specifically to address this 
reconstruction problem. The remainder of this section discusses in detail specifics of the DNN architectures,
the specific optimisation methods used to find the optimal parameters defining the DNNs, and the specific 
structure of the ensemble models. Shown rigorously in what follows, the main motivation for using the unique model is:
\begin{itemize}
\item the universal approximation capabilities of the DNN models with our specific architectures
\item the necessary reduction in the approximation error with the increase in the depth of our DNN models
\item avoidance of a degradation problem~\cite{He:2015wrn} in training due to the residual structure of our DNN models.
\end{itemize}

\noindent Further details on the DNN-based reconstruction method will be provided in Ref.~\cite{Farhat:2022}.

\subsection{Neural networks architecture}
\label{subsec:universality}
The problem of reconstruction of DIS event kinematics can be posed as a task to construct a 
functional relationship  between a set of input variables ${\rm x}$ and a set 
of response variables ${\rm y}$. In many cases, such a relation is a continuous function, 
and can be approximated to arbitrary accuracy with a neural network. 
The corresponding neural network can have various architectures. The simplest architecture
 is a sequence of fully-connected, feed-forward (hidden) layers. For 
an input ${\rm x} \in \bR^m$, the output of the first hidden layer is:
\begin{equation}
h^1 ({\rm x}) = \alpha \left( A_0 {\rm x} \right), \quad h^1 \in \bR^{d_1},
\label{eq:layer1}
\end{equation}
where $A_0 : \bR^m \to \bR^{d_1}$ is an affine transformation, the composition of a translation and a 
linear mapping, and $\alpha : \bR \to \bR $ is a nonlinear function, called an activation function.
Define the application of $\alpha$ on a vector ${\rm x}:=\left( {\rm x}_1,{\rm x}_2,...,{\rm x}_m \right)^T \in \bR^m$, by
$\alpha({\rm x}) := \left( \alpha({\rm x}_1),\alpha({\rm x}_2),...,\alpha({\rm x}_m) \right)^T$.

For $A_1 : \bR^{d_1} \to \bR^{d_2}$, the output $h^1$ is 
passed into a second hidden layer:
\begin{equation}
h^2 ({\rm x})= \alpha \left( A_1 h^1({\rm x}) \right), \quad h^2 \in \bR^{d_2},
\label{eq:layer2}
\end{equation}
and so on, for a desired number of hidden layers. An affine transformation 
on the output of the final hidden layer is the output of the network.
For $i\in\bN$, call $d_i$ the number of nodes in hidden layer $i$.

Define $\mathbf{\Psi}^D_{m,n}(\alpha)$ to be the set of neural networks 
with $D$ hidden layers that map $\bR^m \to \bR^n$ with continuous, 
nonlinear activation function $\alpha$. If $\psi \in \mathbf{\Psi}^D_{m,n}$, 
then there exists a set of natural numbers 
$\{d_i\}_{i=0}^{D+1} \subset \bN$ with $d_0 = m$ and $d_{D+1} = n $, 
a set affine transformations $\{A_i : \bR^{d_{i}} \to \bR^{d_{i+1}} \}_{i=0}^D$, 
such that, for ${\rm x} \in \bR^m$, 

\begin{equation} \label{eq:mlp}
\begin{cases}
h^0 ({\rm x})= {\rm x}, \\
h^{i+1} ({\rm x})= \alpha \left( A_{i} h^i ({\rm x}) \right) \text{ for } 0 \leq i < D, \\
\psi ({\rm x}) = A_D h^{D} ({\rm x}).
\end{cases}
\end{equation}

Specific universality properties of $\mathbf{\Psi}^D_{m,n}(\alpha)$ are 
proven in Refs.~\cite{Hornik:1989, Leshno:1993}. In particular, 
there exists some element in the class 
of neural networks  that is arbitrarily close to any continuous function. 
For this reason, we say that $\mathbf{\Psi}^D_{m,n}(\alpha)$ is dense in 
the space of continuous functions.

We consider a particular subclass of neural networks that also maintain the universal approximation property. 
We call it $\mathbf{\Phi}^D_{m,n}(\alpha)$. If a function $\phi \in \mathbf{\Phi}^D_{m,n}(\alpha) $, there exists 
a set of natural numbers  $\{d_i\}_{i=0}^{D+1} \subset \bN$ with $d_0 = m$ and $d_{D+1} = n $, 
a set of affine transformations $\{A_i : \bR^{d_{i}} \to \bR^{d_{i+1}} \}_{i=0}^{D-1}$, 
a set of matrices  $\{W_i \in \bR^{d_i \times n}\}_{i=0}^{D}$,
such that, for ${\rm x} \in \bR^m$, 

\begin{equation} \label{eq:nn}
\begin{cases}
h^1 ({\rm x})= \alpha \left( A_{0} {\rm x} \right) , \\
h^{i+1} ({\rm x})= \alpha \left( A_{i} h^i ({\rm x}) \right) \text{ for } 0 \leq i < D, \\
\phi ({\rm x}) = W_0 {\rm x} + \sum_{i=1}^D W_i h^i ({\rm x}).
\end{cases}
\end{equation}
Here the affine transformations $A_i$, $i=0,1,\dots, D-1$, have the form
$$
A_i{\rm x}=M_i{\rm x}+{\rm b}_i, \ \ \mbox{for}\ \ {\rm x}\in \bR^{d_i} 
$$
for matrices $M_i\in R^{d_i\times d_{i+1}}$ and vectors ${\rm b}_i\in R^{d_{i+1}}$. The function $\phi$ defined by \eqref{eq:nn} is determined by matrices $M_i$ and $W_i$, and vectors ${\rm b}_i$ whose entries/components will be embraced in a single notation $\omega$ later. That is, $\phi=\phi_\omega$. When $\omega$ is chosen via an optimisation problem, $\phi({\rm x})$ is a function of the input variable ${\rm x}\in \bR^{d_0}$.

Let $\mathbf{\Phi}^D_{m,n}(\alpha)$ be the function class in which we 
search for an optimal kinematic reconstruction method. In addition to the universality property, 
it can be shown that this class is good for searching for an appropriate reconstruction function because 
there the necessary reduction in the approximation error with the increase in the depth and the residual 
structure avoids a degradation problem in the weights during training~\cite{Farhat:2022}.

\subsection{Optimisation methods}
\label{subsec:optimization}
In any given problem to be solved with the DNN,
the task is to choose an optimal function from the class 
$\mathcal{F}=\mathbf{\Phi}^D_{m,n}(\alpha)$ to be a functional relationship 
between a set of input variables ${\rm x}$ and a set of response variables ${\rm y}$.

Each function $\phi\in\mathcal{F}$ is completely determined by the 
elements of the matrices and vectors that determine the values of the 
hidden layers.  Let $\omega \in \bR^p$ be the collection of all these 
elements, where $p$ is the total number of them. We can then write for 
any ${\rm x}\in\bR^m$: $\phi({\rm x})=\phi_\omega({\rm x})$.

Choosing an optimal function from this class entails finding the 
collection of parameters $\omega$ in $\bR^p$ that minimises  $\mathcal{L}_\mathcal{D}$, 
the expected discrepancy or generalisation error between ${\rm y}$ and $\phi_\omega({\rm x})$, 
over the joint probability distribution of ${\rm x}$ and ${\rm y}$, $\mathcal{D}$, defined by:
\begin{equation} \label{eq:optimal}
\mathcal{L}_\mathcal{D}(\omega) = 
\mathop{\mathbb{E}}_{{\rm x},{\rm y}\sim\mathcal{D}} \ell (\omega,{\rm x},{\rm y}) =
\int \ell (\omega,{\rm x},{\rm y}) d \mathcal{D}({\rm x},{\rm y}),
\end{equation}
where $\ell$ is some loss function measuring the discrepancy between ${\rm y}$ 
and $\phi_\omega({\rm x})$.

With a randomly sampled data set $\{{\rm x}_i,{\rm y}_i\}_{i=1}^N$, if $N$ is 
sufficiently large, then the generalisation error can  be approximated 
by an empirical error:
\begin{equation} \label{eq:sampleerror}
\mathcal{L}_\mathcal{D} \approx \frac{1}{N} \sum_{i=1}^N \ell (\omega,{\rm x}_i,{\rm y}_i).
\end{equation}

This summation is sometimes called the fidelity term, as it measures the 
discrepancy between the data and model predictions. A commonly used fidelity term is the mean 
square error, with:
\begin{equation} \label{eq:mseloss}
\ell (\omega,{\rm x}_i,{\rm y}_i) = \|{\rm y}_i - \phi_\omega({\rm x}_i)\|^2_2.
\end{equation}

The fidelity term used in this study is a modification of this, the mean square logarithmic error:
\begin{equation} \label{eq:loss}
\ell (\omega,{\rm x}_i,{\rm y}_i) = \|\log({\rm y}_i) - \log(\phi_\omega({\rm x}_i))\|^2_2.
\end{equation}

Due to the universality of neural networks, there 
is a model that can achieve zero empirical error. However, in the presence of 
noise this means the model can overfit to the data sample and loose its
generalisability. This problem can be addressed by adding a regularisation 
term to the optimisation problem as a penalty for certain irregular behaviour. 
Common regularisation terms include the $\ell^2$-norm, used to limit the 
size of the parameters, and the $\ell^1$-norm, which can induce sparse 
solutions.  For out analysis, we choose the  $\ell^1$-regularisation.
The Theorem $1.3$ in~\cite{Candes:2006} 
and Proposition 27 of~\cite{Zhang:2009} provide an evidence that minimising the $\ell^1$ norm 
provides sparse optimal solutions with a minimal number of nonzero elements.
Well-constructed models should be able to generalise the information from one given 
sample to any possible event. Thus, regularisation with the $\ell^1$ 
norm produces a model determined by a minimal number of parameters so 
that the optimal solution does not fit completely to the training set 
and loose its generalisability.

Therefore, the determination of the optimal neural network model consists of 
minimising this final loss function, the sum of the sample fidelity term 
and  a weighted regularisation term:
\begin{equation} \label{eq:optimization}
\min_{\omega\in\bR^p} \frac{1}{N} \sum_{i=1}^N \ell (\omega,{\rm x}_i,{\rm y}_i) + \mathcal{R} \cdot ||\omega||_{1},
\end{equation}
where $\ell$ is defined in Eq.~\ref{eq:loss} and $\mathcal{R}$ is a hyperparameter to determine 
the magnitude of the regularisation.

Expression~\ref{eq:optimization} can be minimised using stochastic gradient methods 
on batches of the data sample~\cite{Bertsekas:2000}. The training is 
accelerated using classical momentum methods~\cite{Sutskever:2013}. In 
particular, randomly select an initial set of parameters $\omega^0$.
Select a sequence of step sizes, or learning rates, ${\mathcal{L}}^k$ that diminish to zero.  
Randomly selecting a batch of data with indices $I \subset \{ 1,...,N \}$.
Choosing a momentum parameter $\mu$. By defining:

\begin{equation}
v^{k+1} = \mu v^k - {\mathcal{L}}^k \nabla_\omega L_I(\omega^k),
\label{eq:velocity}
\end{equation}

\begin{equation}
\omega^{k+1} = \omega^k + v^{k+1}.
\label{eq:sgd}
\end{equation}
Then the sequence $\omega^k$ converges to a set of parameters defining 
the optimal neural network with the minimal generalisation error, in the 
sense described here.

\subsection{The model}
\label{subsec:model}
We construct a model to rigorously weight classically derived 
reconstructions of $x$ and $Q^2$ with corrections based on measurements 
from the final state lepton and hadronic system. The final reconstruction 
of these observables with the neural network approach are
labelled below as $Q^2_{NN}$ and $x_{NN}$ respectively.  

The constructed model is an ensemble of networks from the previously 
defined function class $\mathbf{\Phi}^D_{m,1}(\alpha)$, where $\alpha$ 
is the rectified linear unit (\prog{ReLU}). The values of $D$ and $m$ vary with each 
network in the ensemble.

The NC DIS events studied in this analysis  are by definition the events containing the scattered 
electron in the final state, therefore we aim to reconstruct the  $Q^2_{NN}$ primarily from the 
related observables, i.e.\ using the properties of the electron  directly measured in the experiment.
In particular, we use as inputs three set of variables:   
the classically reconstructed kinematic observables \epjcbreak{} $\left(Q^2_{EL},Q^2_{DA},Q^2_{JB}\right)$, 
 measurements on the scattered lepton $\left(E_{l^{\prime}},\theta_{l^{\prime}}\right)$, 
 and measurements on the final-state hadronic system $\left(\delta_{\mathcal{H}},P_{T,\mathcal{H}}\right)$. 
We reconstruct the $Q^2$  in the form:
\begin{equation}
\begin{split}
Q^2_{NN} & =  A_{Q^2}\left(Q^2_{EL},Q^2_{DA},Q^2_{JB}\right) + L_{Q^2}\left(A_{Q^2},E_{l^{\prime}},\theta_{l^{\prime}}\right) +\epjconly{\\& +} H_{Q^2}\left(A_{Q^2},\delta_{\mathcal{H}},P_{T,\mathcal{H}}\right),
\label{q2nn}
\end{split}
\end{equation}
in which $A_{Q^2}$ could be understood as a rigorous average of classically 
derived reconstructions, $L_{Q^2}$ is a correction term computed from 
the scattered lepton, and $H_{Q^2}$ is another correction term computed 
from the final-state hadronic system. In our analysis $A_{Q^2}$, $L_{Q^2}$, $H_{Q^2}$ are simultaneously trained networks in 
$\mathbf{\Phi}^5_{3,1}(\alpha)$, with each hidden layer of the networks containing 2000 nodes.

The $x$ observable for the NC DIS events is actually calculated from the electron-related observables as well.
Therefore,  we reconstruct $x_{NN}$, with $Q^2_{NN}$ also as an input, for a total of eight inputs
 ($x_{EL},x_{DA},x_{JB},E_{l^{\prime}},\theta_{l^{\prime}},\delta_{\mathcal{H}}\draftonly{,}\epjconly{$, $} P_{T,\mathcal{H}},Q^2_{NN}$), in the form:
\begin{equation}
\begin{split}
x_{NN} & =  A_{x}\left(x_{EL},x_{DA},x_{JB}\right) + L_x\left(A_{x},Q^2_{NN},E_{l^{\prime}},\theta_{l^{\prime}}\right) +\epjconly{\\& +} H_{x}\left(A_{x},Q^2_{NN},\delta_{\mathcal{H}},P_{T,\mathcal{H}}\right),
\label{xnnwq2}
\end{split}
\end{equation}
where $A_{x}$, $L_x$, and $H_x$ are defined similarly to $A_{Q^2}$, $L_{Q^2}$, and $H_{Q^2}$, but 
$A_{x}$ is a network in $\mathbf{\Phi}^{20}_{3,1}(\alpha)$ with each hidden layer containing 1000 nodes, 
and $L_{x}$, $H_{x}$ are networks in $\mathbf{\Phi}^{10}_{4,1}(\alpha)$, with each hidden layer containing 500 nodes.

The number of hidden layers and the number of nodes in each hidden layer were each progressively increased until sufficiently desirable results were achieved. Smaller networks provided good results on average, but larger networks were needed to find best results in small, specific regions of the kinematic space. A further increasing of the numbers of hidden layers and nodes per layer beyond the chosen values was found to not significantly alter the performance of the kinematic reconstruction, due to the convergence results of deep neural networks. The convergence theorems of deep neural networks with the ReLU activation function as the number of layers increases were recently established in \cite{xu2021, xu2021conv}.

In the ensemble neural network model, we used  the \prog{ReLU} 
\begin{equation}
\alpha({\rm x}) := \max({\rm x},0), \ \ {\rm x}\in \bR
\label{eq:relu}
\end{equation}
as the nonlinear activation function
It has been shown~\cite{Krizhevsky:2012} that with a gradient descent algorithm, 
using the \prog{ReLU} function as the activation function provides a smaller training time 
compared to that with the use of functions with saturating nonlinearities, 
such as a sigmoid or hyperbolic tangent function. 
The reduced training time enabled us to experiment with more sophisticated networks. 
In addition to this, the \prog{ReLU} functions do not need input 
normalisation to prevent them from saturating, which is a desirable 
property for the present analysis. 
%In addition, suitable convergence results on deep \prog{ReLU} networks 
%have been demonstrated in Refs.~\cite{xu2021,xu2021conv}. 
%In particular,
Moreover, it was shown in \cite{xu2021} that the selection of the \prog{ReLU} 
activation function produces a neural network as a piecewise linear function 
over a nonuniform partition, of the domain, determined by parameters involved in the affine transformations $A_i$. Such a structure ensures a good representability of the neural network and can overcome the problem of underfitting.

To accommodate for the large range of the $Q^{2}$ and $x$ variables in the analysis, 
we select the loss function defined in Eq.~\ref{eq:loss} for the training of the DNN models.
%%%%%%%%%%%%%%%%%%%%%%%%%%%%%%%%%%%%%%%%%%%%%%%%%%%%%%%%%%%%%%%%%%%%%%%%

\section{Experimental setup}
\label{sec:zeusexp}
The Monte Carlo simulated events used to train our deep neural networks were specifically generated using
 the conditions of $e^{\pm}p$ scattering in the ZEUS detector at HERA.
A detailed description of the ZEUS detector can be found elsewhere~\cite{ZEUS:1993aa}. 

In our analysis, we have used two samples of Monte Carlo simulated $e^{+}p$ DIS   
events that are provided by the ZEUS collaboration. These samples were generated 
with an inclusion of QED and higher order QCD radiative effects
using the \prog{HERACLES 4.6.6}~\cite{Kwiatkowski:1990es} package with \prog{DJANGOH 1.6}~\cite{Schuler:1991yg} interface
and the \prog{ARIADNE 4.12} and \prog{LEPTO 6.5.1} packages for the simulation of the parton cascade. For both samples the same
 set of kinematic cuts was applied during the generation, the same set of PDFs were used, CTEQ5D~\cite{Lai:1999wy} and the same hadronisation 
settings were used to model the hadronisation with the \prog{Pythia6} program.
Therefore, the essential difference between the two samples is the way 
the higher order corrections are partially modelled with the corresponding algorithms (QCD cascades).
Namely, the \prog{LEPTO} MCEG utilises the parton shower approach~\cite{Bengtsson:1987rw},
 while \prog{ARIADNE} implements a color-dipole model~\cite{Gustafson:1987rq}.
Accordingly, we label the data-sets produced by the \prog{LEPTO} generator as ``CDM'' data sets and those with 
\prog{ARIADNE} as ``MEPS'' data sets.

The generated particle-level events were passed 
 through the ZEUS detector and trigger simulation programs based on \prog{Geant} 3.21~\cite{Brun:1987ma},
  assuming the running conditions of the ZEUS experiment in the year 2007 with a proton beam energy of 920\GeV.
The simulated detector response was processed and reconstructed using exactly the same software chain
 and the same procedures as being used for real data.
The results of the processing were saved in ROOT~\cite{Antcheva:2011zz} files in a form of ZEUS 
Common NTuples~\cite{Malka:2012hg,AndriiVerbytskyifortheZEUS:2016zup}, 
a format that can be easily used for physics analysis without any ZEUS-specific software. 
%%%%%%%%%%%%%%%%%%%%%%%%%%%%%%%%%%%%%%%%%%%%%%%%%%%%%%%%%%%%%%%%%%%%%%%%%%%%%%%%%%%%

\section{Event selection}
\label{sec:eventSelection}
The selection of events for the neural network training is motivated 
by the selection procedure applied in the previous ZEUS analyses~\cite{Breitweg:1999id,Chekanov:2006yc,Chekanov:2006hv,Chekanov:2006xr,Abramowicz:2010ke,Abramowicz:2010cka}.
Even though the presented analysis performed on the Monte Carlo simulated events, 
the selection cuts are choosen and applied as if the analysis is performed on real data for the purpose of being
 as close as possible to the measurements. The general motivation for these cuts is the same as in many
  analyses performed by the ZEUS collaboration: to define unambiguously the phase space of the measurement,
   ensure low fraction of background events, and a reasonable description of the detector acceptance by Monte Carlo simulations.

%%%%%%%%%%%%%%%%%%%%%%%%%%%%%%%%%%%%%%%%%%%%%%%%%%%%%%%%%%%%%%%%%%%%%%%%
\subsection{Phase space selection}
The phase space for the training of the neural networks in this analysis is selected as
  $100\GeV^{2}<Q^{2}<20480\GeV^{2}$, being close to the phase space of the physics analysis in Ref.~\cite{Abramowicz:2010ke}. 

In the phase-space region at low $x$ and very low inelasticity $y$, the QED predictions 
from the Monte Carlo simulations are not reliable because of a limit of higher orders 
in the calculations~\cite{Lontkovskyi:2015}. To avoid these phase-space 
regions, the events are required to have $y_{JB}\cdot(1-x_{DA})^2>0.004$ and $y_{JB} > 0.04$~\cite{Lontkovskyi:2015}. 
To ensure optimal electron identification and electron energy resolution, similar to the previous physics analyses, 
a kinematic cut $0.2<y_{EL}< 0.7$ is used.

%%%%%%%%%%%%%%%%%%%%%%%%%%%%%%%%%%%%%%%%%%%%%%%%%%%%%%%%%%%%%%%%%%%%%%%%
\subsection{Event selection}
The deep inelastic scattering events of interest are those characterised 
by the presence of a scattered electron in the final state and a significant deposit of energy in
 the calorimeter from the hadronic final state. 
The scattered electrons are 
registered in the detector as localised energy depositions primarily in the electromagnetic 
part of the calorimeter, with little energy flow into the hadronic part of 
the calorimeter. On the other hand, hadronic showers propagate in the detector much more 
extensively, both transversely and longitudinally.

In addition to the DIS process, there are also background processes which leave similar signatures in the detector as
those described above. Therefore, the correct and efficient 
identification of the scattered lepton is crucial for the selection of the NC DIS events.
For this analysis, the \prog{SINISTRA} algorithm is used to identify 
lepton candidates~\cite{Abramowicz:1995zi}.
Based on the information from the detector and the results of the \prog{SINISTRA} algorithm,
 the following selection criteria are applied to select the 
events for the further analysis:
\begin{itemize}
\item \textbf{Detector status:} It is required that for all the events the detector was fully functional. 

\item \textbf{Electron energy:} At least one electron candidate with energy greater than  
$10\GeV$~\cite{Abramowicz:2010ke} is identified in the event. 

\item \textbf{Electron identification probability:} 
The \prog{SINISTRA}~\cite{Abramowicz:1995zi} probability of a lepton candidate 
being the DIS lepton was required to be greater than 90\%. If several 
lepton candidates satisfy this condition, the candidate with the highest 
probability is used. In addition to this, there must be no problems reported by the 
\prog{SINISTRA} algorithm.

\item \textbf{Electron isolation:} To assist in removing events where the energy deposits 
from the hadron system overlap with those of the scattered lepton, the 
fraction of the energy not associated to the lepton is required to be 
less than 10\% over the total energy deposited within a cone around the 
lepton candidate. The cone is defined with a radius of $0.7$ units in the 
pseudorapidity-azimuth plane around the lepton momentum direction~\cite{Abramowicz:2010ke}.

\item \textbf{Electron track matching:} The tracking system covers the region of 
polar angles restricted to $0.3 < \theta < 2.85$ rad. Electromagnetic clusters 
within that region that have no matching track are most likely photons. If 
the lepton candidate is within this region, the presence of a matched track is required. This track must have a distance of closest 
approach between the track extrapolation point at the front surface of the 
CAL and the cluster centre-of-gravity-position of less than 10 cm. The track 
energy must be greater than 3\GeV~\cite{Abramowicz:2010ke}.

\item \textbf{ Electron position:} To  minimise the impact of imperfect simulation of some detector regions, 
additional requirements on the position of the electromagnetic 
shower are imposed. The events in which the lepton is found in the following 
regions ($x$,$y$ and $z$ being the cartesian coordinates in the ZEUS detector) are rejected~\cite{Perrey:2011hga}:
\begin{itemize}
\item $\sqrt{x^2 + y^2} < 18 \text{ cm}$, regions close the beam pipe
\item $z < -148 \text{ cm}$ and $y > 90 \text{ cm}$ and $-14 < x < 12 \text{ cm}$, 
a part of the RCAL where the depth was reduced due to the cooling pipe for 
the solenoid (chimney), 
\item $-104 < z < -98.5 \text{ cm}$ or $164 < z < 174 \text{ cm}$, regions 
in-between calorimeter sections (super-crack).
\end{itemize}

\item {\bf  Primary vertex position:} It was required that the reconstructed primary vertex position is close to
 the central region of the detector, applying the selection $-28.5<Z_{\rm vtx}<26.7\text{ cm}$~\cite{Abramowicz:2010ke}.

\item {\bf  Energy-longitudinal momentum balance:} To suppress photoproduction and beam-gas interaction background events 
and imperfect Monte Carlo simulations of those, restrictions are put on the 
energy-longitudinal momentum balance. This quantity is defined as:
\begin{equation}
\delta = \delta_l + \delta_{\mathcal{H}} = (E_{l^\prime}-P_{z,l^\prime}) + (E_{\mathcal{H}}-P_{z,\mathcal{H}}) = \sum_{i} (E_i-P_{z,i}),
\end{equation}
where the final summation index runs over all energy deposits in the detector. 
In this analysis, we apply a condition of $38 < \delta < 65 \GeV$~\cite{Abramowicz:2010ke}.

%\item {\bf  Removal of elastic scattering events :} #CHECK
\item {\bf  Missing transverse energy:}  To remove beam-related background 
and cosmic-ray events, a cut on the missing energy is imposed.
$P_{T,\rm miss}/\sqrt{E_{T}}<2.5\GeV^{1/2}$,  where  $E_{T}$ is the total transverse energy in the CAL and 
$P_{T,\rm miss}$ is  the  missing  transverse  momentum, the transverse component of the vector sum of the 
hadronic final state and scattered electron momenta.
\end{itemize}

%%%%%%%%%%%%%%%%%%%%%%%%%%%%%%%%%%%%%%%%%%%%%%%%%%%%%%%%%%%%%%%%%%%%%%%%
\section{Training the DNN models}
\label{sec:trainingMethods}
We use the neural network models defined in Sec.~\ref{subsec:model} 
and consider them as optimisation problem in terms of Eq.~\ref{eq:optimization}, i.e., we minimise the loss
 function across the selected training set and satisfy sparse regularity conditions.
 Every neural network making the ensemble model for $x$ and $Q^2$ are trained 
 simultaneously. 
 The optimal regularity 
condition depends on the selection of the training set, the events batch size, 
the initial learning rate, and the regularisation parameter.
We aim to select these in a way that minimises overfitting in 
particular regions of the kinematic space while still maximising the 
mean accuracy of the model.

We select a  set of events from the ``MEPS'' data sets for training,  as described  in Sec.~\ref{sec:eventSelection}, and define the
true values of the $x$ and $Q^{2}$ as described in Sec.~\ref{sec:dis:mc}.
Fig.~\ref{fig:xq2original} shows a  distribution of selected events in the $(x_{\rm true},Q^2_{\rm true})$ plane and the boundaries of the chosen analysis bins. 

\FIGtraining{Distribution of events from the training set in $(x_{\rm true},Q^2_{\rm true})$ plane and the 
boundaries of the analysis bins from Tab.~\ref{table:bins}.}{\nospelling{fig:xq2original}}

First, we train the network 
to reconstruct $Q^2$ by optimising Eq.~\ref{eq:optimization}  
with an initial learning rate of $\mathcal{L} = 1.0 \times 10^{-5}$ and a regularisation parameter of ${\mathcal{R} =1.0\times 10^{-5}}$. We select a batch size of 10000.

The reconstruction of $x$ is more complex. We fix the regularisation parameter to $\mathcal{R} = 1.0\times 10^{-5}$ and select optimal parameters for the
 learning rate ${\mathcal L}$ and batch size ${\mathcal B}$
experimentally by varying the learning rate against the batch size. 
The appropriate selection of these parameters ensures fast convergence 
of the stochastic gradient method by balancing noise in the gradients 
with the stability of the algorithm. For each set of parameters, ten attempts are made 
and the best result in terms of the mean square error of the $x$ reconstruction 
model over the training set is chosen. The results are listed in Tab.~\ref{table:lrBatch}. 
The smaller learning rate assures a higher stability of the results than a larger learning rate. It prolongs
 the training process but avoids a poor convergence of the learning process as observed for larger learning rates.
  The larger batch size does not offer any advantages in our analysis as shown in Tab.~\ref{table:lrBatch}. 
We, therefore, select an initial learning rate of $10^{-5}$  with a minimal batch size.

\begin{table}
\centering
\begin{tabular}{| c | c | c | c |}
    \hline
    & \multicolumn{3}{c |}                  { RMS of $\log{x}-\log{x_{\rm true}}$  }       \\ \hline
	 ${\mathcal{L}}$           & ${\mathcal{B}}$ & ${\mathcal{B}}$ & ${\mathcal{B}}$ \\ 
	                                & 10000      & 50000      & 100000 \\ \hline\hline
\input{Tables/tablecdmcentral.dat}
\\\hline
\end{tabular}
\caption{Resolution of $\log{x}$ reconstruction after 200 epochs of training with different values of initial learning rate ${\mathcal{L}}$ and batch size ${\mathcal{B}}$.}
\label{table:lrBatch}
\end{table}

To choose the regularisation parameter close to the optimal one, we vary its value   
with constant batch size of $10000$ and initial learning rate of $10^{-5}$ 
and again observe the mean square error of the $x$ reconstruction model 
over the training set. For each set of parameters, ten attempts are made 
and the best result is chosen. The results are presented in Tab.~\ref{table:regularization}.
Accordingly, we choose a regularisation parameter of $10^{-6}$. 
Using this regularisation, the neural network models for both $x$ and $Q^2$ are 
defined by weight parameters, of which 50\% are effectively zero, or less than $10^{-8}$

\begin{table}
\centering
\begin{tabular}{| c | c |}
        \hline
${\mathcal{R}}$  & RMS of $\log{x}-\log{x_{\rm true}}$     \\ \hline\hline
\input{Tables/table2cdmcentral.dat}
\\\hline
\end{tabular}
\caption{Resolution of $\log{x}$ reconstruction after 200 epochs of training with different values of regularisation parameter ${\mathcal{R}}$.}
\label{table:regularization}
\end{table}

Following the suggestions in Ref.~\cite{Smith:2017}, we start
with a small batch size, and increase it in initial training epochs. 
We test this approach by comparing the mean square error of the $x$ reconstruction 
model over the training set over the first 200 epochs of training over 
three different training regimes. The results are summarised in 
Fig.~\ref{fig:xhistory} and imply to use a gradually increasing batch 
size up to a maximum batch size of $1000$.
\FIGxhistory{Training history for $x$ reconstruction model using different training parameters but the same 
$Q^2$ reconstruction model obtained with 
${\mathcal L}=1.0\times10^{-5}$, ${\mathcal R}=1.0\times10^{-6}$ 
and ${\mathcal B}=10000$. In each of the cases, the initial learning rate was set to ${\mathcal L}=1.0\times10^{-5}$ and 
the regularisation parameter to ${\mathcal R}=1.0\times10^{-6}$. 
}{\nospelling{fig:xhistory}}

%%%%%%%%%%%%%%%%%%%%%%%%%%%%%%%%%%%%%%%%%%%%%%%%%%%%%%%%%%%%%%%%%%%%%%%%

\section{Results}
\label{sec:results}
We evaluate the performance of our approach for the reconstruction of DIS kinematics by applying it to detailed Monte Carlo simulations from the ZEUS experiment and by comparing our results to the results from the electron, double-angle, and Jacques-Blondel reconstruction methods. For the comparison, we use various combinations of statistically independent data sets, one for the training, and another for the evaluation. In our systematic studies, we have found no signs of overtraining and also no indication that the results depend on the selected Monte Carlo simulations. For the results presented in this section, we use the ``MEPS'' data set for the training and the ``CDM'' data set for evaluation. The main quantities of the comparison are the resolutions of the reconstructed variables $\log{Q^2/1\GeV^{2}}$  and $\log{x}$ 
as measured in selected $x-Q^{2}$ regions (bins).
The resolutions are defined as \epjconly{\\} $\sqrt{\sum^{N}_{i} \left( \log{Q^2_{i}/1\GeV^{2}} - \log{Q^2_{i, \rm true}/1\GeV^{2}} \right)^{2} /N}$ and  \epjconly{\\}  $\sqrt{ \sum^{N}_{i} \left( \log{x_{i}}-\log{x_{i,\rm true}} \right)^{2}/N}$,
 where $N$ stands for the number events in the corresponding bin.
The boundaries of the bins are given in Tab.~\ref{table:bins} and are chosen to be close to 
the bins used in ZEUS DIS analyses, e.g.\ in Ref.~\cite{Chekanov:2006hv}.
\TABQxbins

The distributions of the  $\log{Q^2/1\GeV^{2}} - \log{Q^2_{\rm true}/1\GeV^{2}}$ and $\log{x}-\log{x_{\rm true}}$  quantities are given 
in the Fig.~\ref{fig:qtworesolution} and  Fig.~\ref{fig:xresolution}, respectively.
The numerical values for the resolution 
%\footnote{The NN optimisation procedure minimises the value of $\int^{+\infty}_{-\infty} \frac{d N}{d \delta}\delta^{2}d\delta + {\rm regularisation\ terms}$, where $\frac{d N}{d \delta}$ are the distributions in  Figs.~\ref{fig:qtworesolution} and ~\ref{fig:xresolution} and therefore the $\frac{d N}{d \delta}$ does not necessarily peak at $\delta = 0$.} 
are summarised for all the bins and methods in Tab.~\ref{table:resolution}. 
The DNN optimisation procedure minimises the generalisation error described 
in Eq.~\ref{eq:optimal} plus the regularisation penalty, so 
distributions in  Figs.~\ref{fig:qtworesolution} and ~\ref{fig:xresolution}
do not necessarily peak at zero.
In addition to that, Fig.~\ref{fig:qtwotwod} and
Fig.~\ref{fig:xtwod} show the two dimensional distributions of events in
 $\log{Q^2/1\GeV^{2}}$ vs. $\log{Q^2_{\rm true}/1\GeV^{2}}$ and $\log{x}$  vs.  $\log{x_{\rm true}}$ planes.

\epjconly{
\FIGqtworesolution{Distributions of $\log{Q^2/1\GeV^{2}} - \log{Q^2_{\rm true}/1\GeV^{2}}$  for various
 reconstruction methods in individual analysis bins. 
For better visibility, the data points for each reconstruction method are connected with straight lines.}
}

\draftonly{
\clearpage
\FIGqtworesolution{Distributions of $\log{Q^2/1\GeV^{2}} - \log{Q^2_{\rm true}/1\GeV^{2}}$  for various
 reconstruction methods in individual analysis bins. 
For better visibility, the data points for each reconstruction method are connected with straight lines.}
\clearpage
\FIGxresolution{Distributions of $\log{x}-\log{x_{\rm true}}$  for various reconstruction methods in indivdual analysis bins.
For better visibility, the data points for each reconstruction method are connected with straight lines.}
\clearpage
\FIGqtwotwod{Distribution of events in $L_{\rm reco}=\log{Q^2/1\GeV^{2}}$ versus $L_{\rm true}=\log{Q^{2}_{\rm true}/1\GeV^{2}}$
 plane for various  reconstruction methods in individual analysis bins. }
\clearpage
\FIGxtwod{Distribution of events in $L_{\rm reco}=\log{x}$  versus $L_{\rm true}=\log{x_{\rm true}}$ plane for various reconstruction methods in individual analysis bins. }
\FloatBarrier

\begin{table}
\centering
\begin{tabular}{| c | c | c  c c  c}\hline
 Bin &  Events & \multicolumn{2}{c |}{Resolution of $\log{x}$ }& \multicolumn{2}{c |}{  Resolution of  $\log{Q^2/1\GeV^2}$ }\\ \hline\hline
1  &  301780   &  \bf NN: 0.070 &      EL: 0.083  &  \bf NN: 0.035 &      EL: 0.035 \\        
   &          &      JB: 0.180 &      DA: 0.103  &      JB: 0.203 &      DA: 0.062 \\\hline 
2  &  350530   &  \bf NN: 0.069 &      EL: 0.082  &  \bf NN: 0.040 &      EL: 0.043 \\        
   &          &      JB: 0.167 &      DA: 0.096  &      JB: 0.192 &      DA: 0.064 \\\hline 
3  &  138456   &  \bf NN: 0.098 &      EL: 0.130  &      NN: 0.055 &  \bf EL: 0.053 \\        
   &          &      JB: 0.138 &      DA: 0.100  &      JB: 0.150 &      DA: 0.077 \\\hline 
4  &   74844   &  \bf NN: 0.067 &      EL: 0.084  &  \bf NN: 0.044 &      EL: 0.046 \\        
   &          &      JB: 0.117 &      DA: 0.077  &      JB: 0.138 &      DA: 0.063 \\\hline 
5  &   31043   &  \bf NN: 0.064 &      EL: 0.091  &  \bf NN: 0.036 &      EL: 0.041 \\        
   &          &      JB: 0.102 &      DA: 0.073  &      JB: 0.117 &      DA: 0.053 \\\hline 
6  &   11475   &  \bf NN: 0.053 &      EL: 0.079  &  \bf NN: 0.033 &      EL: 0.036 \\        
   &          &      JB: 0.083 &      DA: 0.061  &      JB: 0.100 &      DA: 0.045 \\\hline 
7  &    3454   &  \bf NN: 0.050 &      EL: 0.069  &  \bf NN: 0.036 &      EL: 0.038 \\        
   &          &      JB: 0.074 &      DA: 0.055  &      JB: 0.093 &      DA: 0.042 \\\hline 
8  &     624   &  \bf NN: 0.036 &      EL: 0.055  &  \bf NN: 0.033 &      EL: 0.037 \\        
   &          &      JB: 0.067 &      DA: 0.045  &      JB: 0.095 &      DA: 0.041 \\\hline 

\end{tabular}
\caption{Resolution of the reconstructed kinematic variables in 
bins of $x$ and $Q^2$. The resolution  for $x$ and $Q^{2}$ is defined as 
the RMS of the distributions  $\log(x)-\log(x_{\rm true})$ and 
$\log(Q^{2})-\log(Q^{2}_{\rm true})$ respectively.
}
\label{\nospelling{table:resolution}}
\end{table}

}

The comparison of the DNN-based approach with the classical methods
demonstrates that the DNN-based approach is well suited for 
the reconstruction of DIS kinematics. Specifically, for most of the bins, our approach provides 
the best resolution as measured by the standard deviation of the 
logarithmic differences of true and reconstructed variables. The better performance of the DNN-based approach in most of the bins is a consequence of using
 additional available information about the final state. In this respect, the DNN-based approach is similar to averaging of
  the values provided by the classical methods with some weights or to alternative 
  approaches for the same task, e.g. see kinematic fitting in Ref.~\cite{Aggarwal:2012}. 
However, in addition to a better resolution, 
the reconstruction with the DNN  has an important advantage over the classical methods or any simple combination of them. It allows to combine the methods without the intrinsic biases of each method and enables an extension of the model with additional physics observables in a robust way. 
\epjconly{
\FIGxresolution{Distributions of $\log{x}-\log{x_{\rm true}}$  for various reconstruction methods in individual analysis bins.
For better visibility, the data points for each reconstruction method are connected with  straight lines.}

\begin{table}[hbp]
\centering
\begin{tabular}{| c | c | c  c | c  c}\hline
 Bin &  Events & \multicolumn{2}{c |}{Resolution  of   }& \multicolumn{2}{c |}{  Resolution of  }\\
     &         & \multicolumn{2}{c |}{$\log{x}$, $\times10^{3}$ }         & \multicolumn{2}{c |}{ $\log{Q^2/1\GeV^2}$, $\times10^{3}$ }\\ \hline\hline
1  &  301780   &  \bf NN: 70 &      EL: 83  &  \bf NN: 35 &      EL: 35 \\        
   &          &      JB: 180 &      DA: 103  &      JB: 203 &      DA: 62 \\\hline 
2  &  350530   &  \bf NN: 69 &      EL: 82  &  \bf NN: 40 &      EL: 43 \\        
   &          &      JB: 167 &      DA: 96  &      JB: 192 &      DA: 64 \\\hline 
3  &  138456   &  \bf NN: 98 &      EL: 130  &      NN: 55 &  \bf EL: 53 \\        
   &          &      JB: 138 &      DA: 100  &      JB: 150 &      DA: 77 \\\hline 
4  &   74844   &  \bf NN: 67 &      EL: 84  &  \bf NN: 44 &      EL: 46 \\        
   &          &      JB: 117 &      DA: 77  &      JB: 138 &      DA: 63 \\\hline 
5  &   31043   &  \bf NN: 64 &      EL: 91  &  \bf NN: 36 &      EL: 41 \\        
   &          &      JB: 102 &      DA: 73  &      JB: 117 &      DA: 53 \\\hline 
6  &   11475   &  \bf NN: 53 &      EL: 79  &  \bf NN: 33 &      EL: 36 \\        
   &          &      JB: 83 &      DA: 61  &      JB: 100 &      DA: 45 \\\hline 
7  &    3454   &  \bf NN: 50 &      EL: 69  &  \bf NN: 36 &      EL: 38 \\        
   &          &      JB: 74 &      DA: 55  &      JB: 93 &      DA: 42 \\\hline 
8  &     624   &  \bf NN: 36 &      EL: 55  &  \bf NN: 33 &      EL: 37 \\        
   &          &      JB: 67 &      DA: 45  &      JB: 95 &      DA: 41 

\\\hline
\end{tabular}
\caption{Resolution of the reconstructed kinematic variables in 
bins of $x$ and $Q^2$. The resolution  for $x$ and $Q^{2}$ is defined as 
the RMS of the distributions  $\log(x)-\log(x_{\rm true})$ and 
$\log(Q^{2})-\log(Q^{2}_{\rm true})$ respectively.
}
\label{\nospelling{table:resolution}}
\end{table}

}

\epjconly{
\clearpage
\FloatBarrier
\FIGqtwotwodepjc{Distribution of events in $L_{\rm reco}=\log{Q^2/1\GeV^{2}}$ versus $L_{\rm true}=\log{Q^{2}_{\rm true}/1\GeV^{2}}$
 plane for different reconstruction methods in individual analysis bins. }
\FIGxtwodepjc{Distribution of events in $L_{\rm reco}=\log{x}$  versus $L_{\rm true}=\log{x_{\rm true}}$ plane for different reconstruction methods in individual analysis bins. }
\FloatBarrier
}
The resolution is improved by the DNN-based approach in two ways. 
For the first couple of bins, the main improvement is caused by the more precise 
estimation of the reconstructed observables. This is clearly seen in 
Figs.~\ref{fig:qtworesolution} and~\ref{fig:xresolution}.
For the bins with higher $Q^2$ and $x$ the main improvement is due to the rejection of outliers,
 which can be seen in Figs.~\ref{fig:qtwotwod} and~\ref{fig:xtwod}.
The bins with higher $Q^2$ and $x$ also demonstrate another, very specific advantage on the DNN approach.
Due to the low number of training events in the  high $Q^2$ and $x$ region, it would not be possible to train a DNN model
or combine the $Q^{2}$ and $x$ observables with other methods using information from this region only. 
However, the DNN training process benefits from the constraints from the higher number of events available elsewhere in the kinematical space and 
delivers models that perform well even in the bins with highest $Q^{2}$ and $x$.

\section{Conclusions}

We have presented the use of DNN to reconstruct the
kinematic observables $Q^2$ and $x$ in the study of neutral current DIS events at the ZEUS experiment at HERA. The DNN models
 are specially designed to be effective in their universal approximation capability, robust in the sense that increasing 
the depth of the networks will necessarily reduce empirical error, and computationally 
efficient with a structure that avoids ``vanishing'' gradients arising in the backpropagation algorithm.

Compared to the classical reconstruction methods, the DNN-based approach enables significant improvements in the resolution
 of $Q^{2}$ and $x$. At the same time, it is evident that the usage of the DNN approach allows
  to match easily any definition of $Q^{2}$ and $x$ at the true level 
that is preferred for a given physics analysis.

The large samples of simulated data required for the training of the DNN can be generated rapidly at modern data centres.
 Also, DNN allow to effectively extract information from large data sets. This suggests that our new approach for the
  reconstruction of DIS kinematics can serve as a rigorous method to combine and outperform the classical reconstruction
   methods at ongoing or upcoming DIS experiments. We will extend the approach beyond inclusive DIS measurements and will
    study next the use of DNN for the reconstruction of event kinematics in semi-inclusive and exclusive DIS measurements.

\section*{Acknowledgement}
\label{sec:acknowledgement}
We are grateful to the ZEUS Collaboration for allowing us to use ZEUS Monte Carlo simulated samples in this analysis and the Data Preservation efforts~\cite{DPHEP:2015npg} in DESY and 
Max-Planck for Physics that made this analysis possible.
 We thank R.~Aggarwal, C.~Fanelli, Y.~Furletova, D.~Higinbotham, D.~Lawrence, S.~Menke, A.~Vossen, and R.~Yoshida for helpful discussions. 

The work of M.~Diefenthaler was supported by the US Department of Energy, Office of Science, Office of Nuclear Physics
 contract DE-AC05-06OR23177, under which Jefferson Science Associates, LLC operates the Thomas Jefferson National Accelerator Facility.

A.~Farhat and Y.~Xu were supported in part by National Science Foundation under grant DMS-1912958.\epjconly{\\}  A.~Farhat was also
 supported by an EIC Center Fellowship at the Thomas Jefferson National Accelerator Facility. 

\FloatBarrier

\appendix
\allowdisplaybreaks
%%%%%%%%%%%%%%%%%%%%%%%%%%%%%%%%%%%%%%%%%%%%%%%%%%%%%%%%%%%%%%%%%%%%%%%%
\section{Software used in the analysis}
\label{sec:software}

The \prog{ROOT} package~\cite{Antcheva:2011zz} of version 6.22 was used 
to read the ZEUS data, provided by the data preservation at Max-Planck for Physics~\cite{DPHEP:2015npg}, analyze it and prepare plain text (or ROOT 
files) with selected information to be used with the ML tools. The selected 
information from the plain text (or ROOT) files was piped using the \prog{pandas} 
package~\cite{mckinney-proc-scipy-2010} into 
\prog{Keras}~\cite{chollet2015keras} interface to 
\prog{tensorflow}~\cite{tensorflow2015whitepaper}  2.3.0 library to 
train the ML models. The  packages \prog{Eigen}~\cite{eigenweb}, 
\prog{frugally-deep}~\cite{fdeep}, 
\prog{JSON for Modern C++}~\cite{nlohmann} and \epjcbreak{} \prog{FunctionalPlus}~\cite{fplus} were used for execution of 
the trained models after these were converted into \prog{frugally-deep} 
model format~\cite{fdeep} to be used with C++ application. The dependencies for the \prog{tensorflow} 
were supplied from the \prog{PyPi} repository. The training 
was performed using libraries for computing on GPUs from the \prog{CUDA}~\cite{cuda} framework of version 10.1.
We are grateful to MPCDF\footnote{Max Planck Computing and Data Facility,
Gie{\ss}enbachstra{\ss}e 2,
85748 Garching} for the ability to compile and execute the codes on the HPC cluster ``Cobra'' in ~\cite{cobra}. 
The operation system used was Linux on ${\rm x}86\_64$ architecture using  \prog{gcc}~\cite{Stallman:2009:UGC:1593499} 
of version 7.3 and \prog{python}~\cite{python3} of version 3.6.8.

The figures with the final results were produced with the \prog{PGFPlots}~\cite{pgfplots} package. 

{\bibliographystyle{DISML}{\raggedright\bibliography{DISML.bib}}}\vfill\eject
\clearpage
\end{document}